\documentclass[prb,twocolumn,nopacs,amsmath,amssymb]{revtex4-1}

\usepackage{graphicx}
\usepackage{dcolumn}
\usepackage{bm}
\usepackage{comment}
\usepackage[percent]{overpic}
\usepackage{color}

\newcommand{\vk}{{\mbox{\boldmath$k$}}}

\newcommand{\ex}{{\mbox{$\mathrm{e}$}}}

\begin{document}

\preprint{APS/123-QED}

\title{Topological nodal superconducting phases and topological phase transition\\ in the hyperhoneycomb lattice}

\author{Adrien Bouhon}%
 \email{adrien.bouhon@physics.uu.se}
\author{Johann Schmidt}
\author{Annica M.~Black-Schaffer}
\affiliation{%
Department of Physics and Astronomy, Uppsala University, Box 516, SE-751 21 Uppsala, Sweden 
}

\date{\today}

\begin{abstract}
We establish the topology of the spin-singlet superconducting states in the bare hyperhoneycomb lattice and derive analytically the full phase diagram using only symmetry and topology in combination with simple energy arguments.
The phase diagram is dominated by two states preserving time-reversal symmetry. 
We find that the line-nodal state dominating at low doping levels is topologically nontrivial and exhibits surface Majorana flat bands, which we show perfectly match the bulk-boundary correspondence using Berry phase approach. At higher doping levels we find a fully gapped state with trivial topology. 
By analytically calculating the topological invariant of the line nodes, we derive the critical point between the line-nodal and fully gapped states as a function of both pairing parameters and doping. We find that the line-nodal state is favored not only at lower doping levels but also if symmetry-allowed deformations of the lattice is present. Adding simple energy arguments we establish that a fully gapped state with broken time-reversal symmetry likely appears covering the actual phase transition. We find this time-reversal symmetry broken state to be topologically trivial, while we find an additional point nodal state at very low doping levels to have nontrivial topology with associated Fermi surface arcs.
We eventually address the robustness of the phase diagram to generalized models also including adiabatic spin-orbit coupling, and show how all but the point nodal state are reasonably stable.

\end{abstract}

\maketitle

\section{Introduction}
In the last few years a plethora of new topological states have been predicted.~While numerous topological insulators, semimetals, and metals have been identified already, the discovery of bulk topological superconductors is still a big challenge. Particularly, nontrivial topological superconductivity typically requires unconventional pairing mechanisms for which no universal framework exists. 
Known unconventional pairing mechanisms are often strongly anisotropic which can easily favor nodal pairing states. \cite{Schnyder_nodal_0, Schnyder_nodal_0b, Schnyder_nodal_1,Schnyder_nodal_2} The nontrivial topological nature of several well known nodal superconductors has in fact been revealed \textit{a posteriori}, e.g.~the non-centrosymmetric heavy fermion systems and the $d_{x^2-y^2}$-wave state of high $T_c$ cuprate-based superconductors,\cite{Sato_nodal_2006,Beri_nodal,Sato11,Schnyder_nodal_0,Schnyder_nodal_0b,Schnyder_nodal_1,GoswamiBalicas_URu2Si2,Schnyder_nodal_2} not to mention the early discussion of the A-phase of liquid $^3$He by Volovik.\cite{Volovik_nodal_99,Volovik} However, as for the prediction and design of new topological superconductors a better understanding of the interaction between pairing mechanisms and the normal state band structure and thus the lattice is required. 

A very interesting system for intrinsic topological superconductivity is the hyperhoneycomb lattice, which has recently been synthesized in the strongly correlated lithium iridate, the so-called $\beta$-phase of Li$_2$IrO$_3$.\cite{Takagi15} This material has been considered as a Kitaev spin-liquid candidate,\cite{Kim_HHL_kitaev} even though stoichiometric $\beta$-Li$_2$IrO$_3$ seems to favor ordered magnetic phases in the undoped, half-filled case.\cite{Kim_HHL_magnetic_pd, Veiga_HHL} Moreover, the simplest possible normal state band structure on the hyperhoneycomb lattice features a nodal line at half-filling.\cite{Ezawa_HHL} It is this combination of a nontrivial nodal-line normal state and strong correlations in iridate hyperhoneycomb materials that opens for very exciting possibilities in terms of nontrivial topological superconductivity. 

In this work we study the possible superconducting states, their topology, and topological phase transitions in the iridate hyperhoneycomb materials under doping away from the magnetic ground state at half-filling. To most clearly elucidate the effect of the lattice, we concentrate on the superconducting phases supported by the bare hyperhoneycomb lattice structure, such that only the spin-singlet pairing channel is relevant. 
Several different stable spin-singlet states have previously been obtained from an effective $t-J$-model solved numerically at the mean-field level on the hyperhoneycomb lattice.\cite{SBBS_16} The previously established phase diagram is primarily composed of a fully gapped phase at high doping (here called $\Gamma_{1,a}^+$) and a nodal phase at lower doping ($\Gamma_{1,b}^+$) which both preserve time-reversal symmetry (TRS). Intermediary between these two states, largely hindering a direct phase transition, is a sliver of a fully gapped state ($\Gamma_{1,c}^+$) with spontaneously broken time-reversal symmetry (BTRS). At very low doping a stable nodal state with BTRS ($\Gamma_{d}^+$) has also been found that in addition spontaneously breaks point group symmetries. 

Here we first study and characterize the topology of all previously identified phases. In particular, we show that the line-nodal phase with TRS has topologically nontrivial nodal lines and exhibits surface Majorana flat bands. We are able to substantiate the bulk-boundary correspondence by showing a perfect match between the $\mathbb{Z}_2$ quantized Berry phases computed in the bulk and the positions of surface Majorana states computed for different slab geometries. We also show that the nodal phase with BTRS found at very low doping has topologically nontrivial nodal points characterized by Chern numbers and, in analogy with Weyl semimetals, exhibits surface Fermi arcs that can be traced out from the projected Berry flux lines. The two gapped phases we however find to be topologically trivial.

Secondly, by combining symmetry and topology, we are able to derive fully analytically the critical point of the topological phase transition between the fully gapped and the line-nodal states with TRS, even going beyond the parameters of the simplified $t-J$ model used previously.\cite{SBBS_16} The resulting phase diagram finds the line-nodal state preferred at lower doping levels but also under symmetry-allowed deformations of the hyperhoneycomb lattice. These results not only extend the previously found phase diagram, but also establish that the overall phase diagram can be constructed based on topological arguments alone. Moreover, combining the global topology of the normal state with the energy spectrum of the superconducting states, we find the same phase diagram and can in addition predict the BTRS state as a natural intermediary state covering the phase transition. We also verify the robustness of the phase diagram under the generalization to longer range hopping and pairing terms and including adiabatic spin-orbit coupling. These results show that symmetry and topological arguments are advantageous and versatile tools for establishing superconducting phase diagrams going beyond particular pairing mechanisms in the search for bulk topological superconductors. 

The remaining of the paper is organized in the following way.~In Section \ref{model} we introduce the hyperhoneycomb lattice and the tight-binding Bogoliubov-de Gennes (BdG) model for general pairing within the spin-singlet channel.~In Section \ref{TRS_states} we discuss the pairing states with TRS, where the bulk topology and the bulk-boundary correspondence in term of Berry phase of the line nodal state are presented. We also show the existence of surface Majorana flat bands. In Section \ref{BTRS_states} we discuss the pairing states with BTRS. We compute the Chern numbers of the nodal points through the flow of Berry phase and show the existence of surface Fermi arcs. In Section \ref{Top_PT} we find analytically the topological phase transition between the fully gapped and the line-nodal phases with TRS using topological arguments and derive the overall phase diagram. 
In Section \ref{conclusions} we conclude and also discuss the robustness of the phase diagram for generalized BdG models.

\section{Hyperhoneycomb lattice and BdG model}\label{model}
\begin{figure}[t]
\centering
\begin{tabular}{c} 
	\begin{overpic} [width=0.92\linewidth]{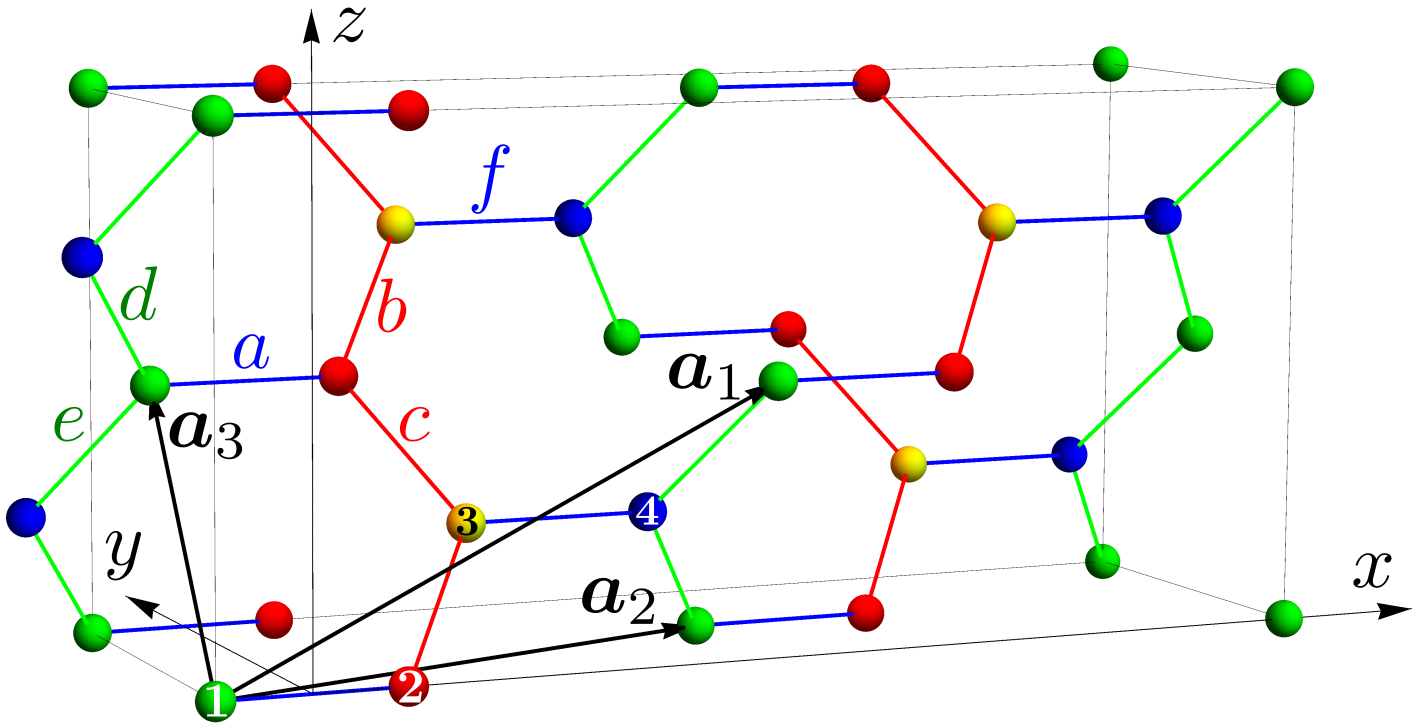}
 		\put (0,0) {(a)}
	\end{overpic}\\
	\begin{overpic} [width=0.6\linewidth]{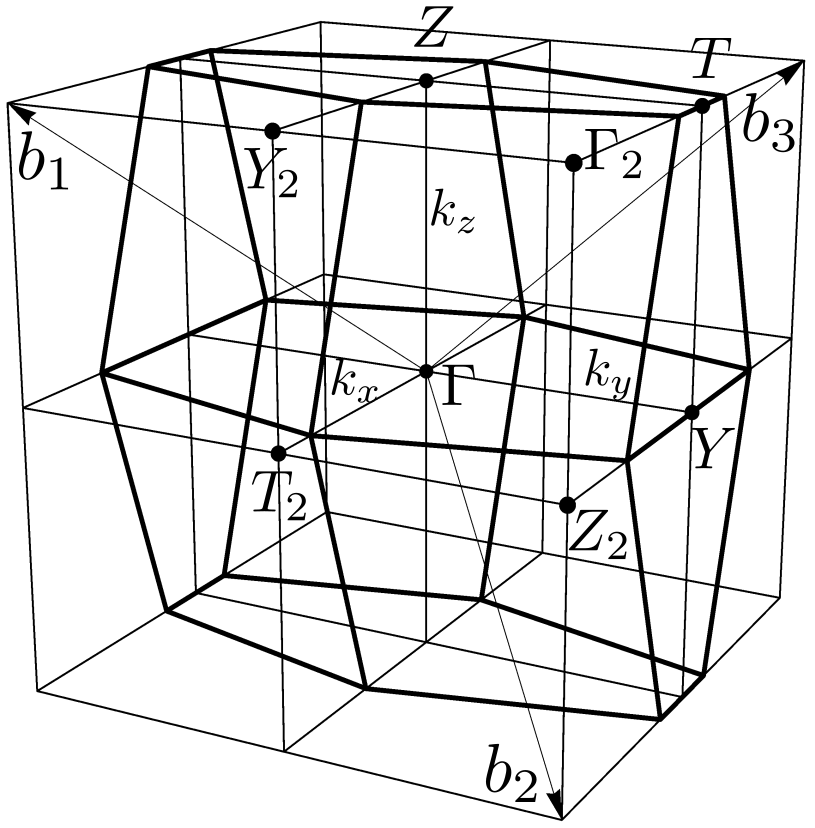}
 		\put (0,0) {(b)}
	\end{overpic}
\end{tabular}
\caption{\label{fig_HHL} (a) Hyperhoneycomb lattice belonging to SG70 for the Wyckoff's position $16e$ and spanned by the primitive lattice vectors $\boldsymbol{a}_i$. (b) BZ for SG70 with primitive reciprocal lattice vectors $\boldsymbol{b}_i$.\cite{Bilbao} $\Gamma$, $Y$, $T$ and $Z$ are high-symmetry points of the first BZ. $\Gamma_2$, $Y_2$, $T_2$ and $Z_2$ are the equivalent high-symmetry points of the next BZ. 
}
\end{figure}

The hyperhoneycomb lattice, shown in Fig.~\ref{fig_HHL}(a), belongs to the nonsymmorphic space group no.~$70$ $Fddd$ (SG70), i.e.~ it is an orthorhombic face-centered Bravais lattice spanned by the primitive lattice vectors $\{\boldsymbol{a}_1,\boldsymbol{a}_2,\boldsymbol{a}_3\}$. It corresponds to the Wyckoff's position $16e$ with four inequivalent lattice sites per primitive unit cell (i.e.~the ``sub-lattice" degrees of freedom) that we label $i=\{1,2,3,4\}$ and color green, red, yellow and blue, respectively in Fig.~\ref{fig_HHL}(a).\cite{ITC} The point group is $D_{2h}$, with three $C_2$ rotations with respect to each of the cartesian directions $\{\hat{x},\hat{y},\hat{z}\}$, inversion and three glide reflections with respect to the three mirrors perpendicular to the Cartesian directions. It is useful to think of the hyperhoneycomb lattice as consisting of two kinds of bonds: the horizontal bonds, blue in Fig.~\ref{fig_HHL}(a), and the bonds of the zigzag chains, green and red in Fig.~\ref{fig_HHL}(a). In the following we refer to the six inequivalent nearest-neighbor (NN) bonds as $\nu=\{a,b,c,d,e,f\}$. 

In this work we concentrate on the simplest possible but still unconventional superconducting phases supported by the bare sub-lattice degrees of freedom. As such we consider only spin-singlet superconductivity. Moving beyond the trivial on-site and isotropic $s$-wave state, we thus consider all stable pairing states found within a tight-binding model with up to NN hopping and pairing terms. Nevertheless, we argue in the end that many of the qualitative results discussed in this work must hold even when longer ranged hopping and pairing terms are included, as long as the space group and topological classes are conserved. 
Physically, if we ignore the on-site pairing, this model corresponds exactly to the renormalized mean-field theory of the $t-J$ model obtained for strongly correlated materials within the limit of strong on-site Coulomb repulsion.\cite{ZhangGrosandRiceShiba, AndersonRice_04, EdeggerGros_07, LeHurRice_09} In this model superconductivity arises only in the spin-singlet pairing channel on NN bonds and is a consequence of the anti-ferromagnetic Heisenberg interaction. Since already discovered hyperhoneycomb materials within the iridate family are both strongly correlated and with a magnetic ground state,\cite{Takagi15, Kim_HHL_magnetic_pd, Veiga_HHL} this model is also directly applicable to these materials.

The tight-binding Bogoliubov-de Gennes (BdG) with NN interactions and spin-singlet pairing and allowed by the symmetries of the bare hyperhoneycomb lattice takes the form
\begin{eqnarray}
	\mathcal{H}^{\mathrm{BdG}} &=& \sum\limits_{\boldsymbol{k}} \left(\begin{array}{c} \hat{\boldsymbol{C}}^{\dagger}_{\boldsymbol{k},\uparrow} \nonumber \\  \hat{\boldsymbol{C}}_{-\boldsymbol{k},\downarrow} \end{array}\right)^T 
	H(\boldsymbol{k})
	 \left(\begin{array}{c} \hat{\boldsymbol{C}}_{\boldsymbol{k},\uparrow} \\  \hat{\boldsymbol{C}}^{\dagger}_{-\boldsymbol{k},\downarrow} \end{array}\right)  \nonumber \\
	 &+& \left(\begin{array}{c} \hat{\boldsymbol{C}}^{\dagger}_{\boldsymbol{k},\downarrow} \\  \hat{\boldsymbol{C}}_{-\boldsymbol{k},\uparrow} \end{array}\right)^T 
	\tau_z H(\boldsymbol{k})\tau_z
	 \left(\begin{array}{c} \hat{\boldsymbol{C}}_{\boldsymbol{k},\downarrow} \\  \hat{\boldsymbol{C}}^{\dagger}_{-\boldsymbol{k},\uparrow} \end{array}\right)		\;,\\
\label{H_BdG}
	 H(\boldsymbol{k}) &=& \left(\begin{array}{cc} H_0(\boldsymbol{k}) & H_{\Delta}(\boldsymbol{k}) \\ 
	H^{\dagger}_{\Delta}(\boldsymbol{k}) & -H_0^T(-\boldsymbol{k}) \end{array}\right) \;,
\end{eqnarray}
with $\tau_z = \sigma_z \otimes \mathbb{I}_{4\times 4}$, where $\sigma_z$ acts in particle-hole space and $\mathbb{I}_{4\times 4}$ in sub-lattice space. Here $\hat{\boldsymbol{C}}^{\dagger}_{\boldsymbol{k},\sigma} = \left( \hat{c}^{\dagger}_{1,\boldsymbol{k},\sigma},\hat{c}^{\dagger}_{2,\boldsymbol{k},\sigma},\hat{c}^{\dagger}_{3,\boldsymbol{k},\sigma},\hat{c}^{\dagger}_{4,\boldsymbol{k},\sigma} \right)$ are defined in terms of the tight-binding sub-lattice basis set
\begin{equation}
	\hat{c}^{\dagger}_{i,\boldsymbol{k},\sigma} = \dfrac{1}{\sqrt{N }}	\sum\limits_{\boldsymbol{R}_{n}} \mathrm{e}^{ i \boldsymbol{k}\cdot (\boldsymbol{R}_{n}+\boldsymbol{r}_{i})}  \hat{c}^{\dagger}_{i,\boldsymbol{R}_{n},\sigma}  \;,
\end{equation}
where $\boldsymbol{R}_{n}$ is a vector of the Bravais lattice, $\{\boldsymbol{r}_{i}\}$ locate the four sub-lattice sites within each primitive unit cell, and $\boldsymbol{k}$ is a point of the Brillouin zone (BZ) for SG70, shown in Fig.~\ref{fig_HHL}(b). We omit here any constant terms that are not relevant to our discussion. 

Up to NN hopping, the normal part of the Hamiltonian is given by 
\begin{equation}
\label{H0}
	H_{0} = \left[\begin{array}{cccc}
		 - \mu & t f_a & 0 & t f_d^* + t f_e^* \\
		t f_a^* &  - \mu   &  t f_b +   t f_c & 0 \\
		 0 &  t f_b^* + t f_c^*  &  - \mu & t f_a \\
		 t f_d +   t f_e & 0 & t f_a^* &  - \mu
		\end{array}\right],
\end{equation}
where we have introduced $f_{\nu} = \ex^{i \boldsymbol{k} \cdot \boldsymbol{\delta}_{\nu}}$ for each NN sub-lattice bond vector $\{\boldsymbol{\delta}_{\nu}\}=\{ \boldsymbol{r}_j -\boldsymbol{r}_i\}_{ij=12,23,23',41,41',34}$. Note here that, while there is only one way to connect sites $1$ and $2$ and similarly $3$ and $4$, through a horizontal bond, there are two ways to connect sites $1$ and $4$ and similarly $2$ and $3$, through zigzag bonds. 

\begin{figure}[t!]
\centering
\begin{tabular}{c}
\includegraphics[width=0.5\linewidth]{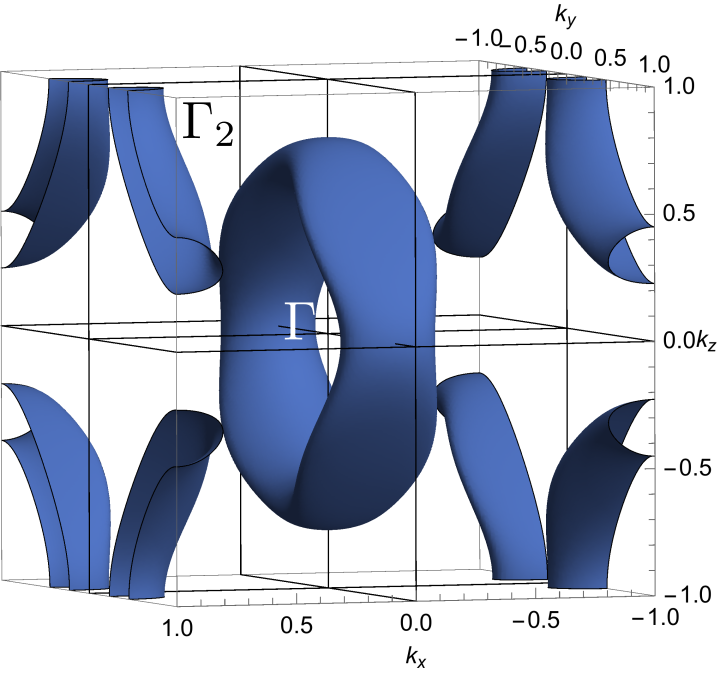}  
\end{tabular}
\caption{\label{fig_FS} Toroidal Fermi surface of the normal state band structure for a finite doping ($\mu=0.06$) away from half-filling. The plotted domain spans two BZ such that two copies of the Fermi surface are visible (second copy is split into eight eighths shifted by reciprocal lattice vectors, e.g.~$ \boldsymbol{k}(\Gamma_2)-\boldsymbol{k}(\Gamma)= \boldsymbol{b}_1+\boldsymbol{b}_2+\boldsymbol{b}_3$.}
\end{figure}

The symmetries of the hyperhoneycomb lattice leads to a global band topology that imposes the presence of a nodal line between two valence bands and two conduction bands, independently of the details of the Hamiltonian considered.\cite{ABABS_HHL_2} In the case of Eq.~(\ref{H0}) at half-filling ($\mu=0$) an extra chiral symmetry is also satisfied (symmetry under sub-lattice sites exchange) leading to a line nodal Fermi surface, i.e.~the whole nodal line appears necessarily at zero energy.\cite{Ezawa_HHL} Under doping the line node inflates into a toroidal Fermi surface, see Fig.~\ref{fig_FS}. 
We note that the four-dimensional sub-lattice space of the hyperhoneycomb lattice is intrinsically related to the global band topology and the whole sub-lattice space must be included in any tight-binding Hamiltonian in order to comply with the symmetry requirements of SG70.\footnote{The nonsymmorphicity of SG70 leads to double degeneracies at the boundaries of the BZ, such that each band crossing the Fermi level is connected to an other band that does not cross the Fermi level. Since two bands are involved in the line nodal Fermi surface, a minimum of four bands must be taken into account.}    

The superconducting off-diagonal part in Eq.~(\ref{H_BdG}) can up to NN interactions be described by one on-site gap parameter on each sub-lattice site $\{\Delta_{0,i}\}$ and one gap parameter on each sub-lattice bond $\{\Delta_{\nu}\}$, leading to
\begin{equation}
\label{Hoff}
	H_{\Delta} =
	 \left[\begin{array}{cccc}
		\Delta_{0,1} & \Delta_a f_a & 0 &  \begin{array}{l} (\Delta_d f_d^* \\ + \Delta_e f_e^*) \end{array}\\
		\Delta_a f_a^* & \Delta_{0,2}   &     \begin{array}{l} (\Delta_b f_b \\ + \Delta_c f_c) \end{array} & 0 \\
		 0 &\begin{array}{l} (\Delta_b f_b^* \\ + \Delta_c f_c^*) \end{array}  &  \Delta_{0,3} & \Delta_f f_a \\
		\begin{array}{l} (\Delta_d f_d \\ + \Delta_e f_e) \end{array}   & 0 & \Delta_f f_a^* & \Delta_{0,4}
		\end{array}\right] .
\end{equation}
Every spin-singlet pairing state must correspond to one of the even irreducible representations of $D_{2h}$, i.e.~$\{\Gamma_1^+,\Gamma_2^+,\Gamma_3^+,\Gamma_4^+\}$ in the Koster \textit{et al.} notations,\cite{Koster} each of which characterizes a different set of constraints over the gap parameters. Moreover, $D_{2h}$ splits the gap parameters into three groups, such that we always find $\vert \Delta_{0,1}\vert=\vert \Delta_{0,2}\vert=\vert \Delta_{0,3}\vert=\vert \Delta_{0,4}\vert$ (on-site), $\vert\Delta_a \vert= \vert\Delta_f\vert$ (NN horizontal), and $\vert \Delta_b \vert = \vert \Delta_c \vert=\vert \Delta_d \vert=\vert \Delta_e \vert$ (NN zigzag). Using the vector notation $\boldsymbol{\Delta}_{ 0} = \left( \Delta_{0,1},\Delta_{0,2}, \Delta_{0,3}, \Delta_{0,4}\right)$, $\boldsymbol{\Delta}_{h} = \left( \Delta_{a},\Delta_{f}\right)$ and $\boldsymbol{\Delta}_{z} = \left(\Delta_{b},\Delta_{c},\Delta_{d},\Delta_{e}\right)$, we thus find that every pairing state is given by 
\begin{align}
	\boldsymbol{\Delta}_{\Gamma_j} &= \boldsymbol{\Delta}_{ 0}^{\Gamma_j} \oplus \boldsymbol{\Delta}_{h}^{\Gamma_j} \oplus \boldsymbol{\Delta}_{z}^{\Gamma_j} \nonumber \\
\label{sym_states}
	&= \Delta_0 \boldsymbol{v}_{0}^{\Gamma_j} \oplus \Delta_{h} \boldsymbol{v}_{h}^{\Gamma_j} \oplus \Delta_{z} \boldsymbol{v}_{z}^{\Gamma_j}\;,
\end{align}
where the basis vectors $\{\boldsymbol{v}^{\Gamma_j}_{0},\boldsymbol{v}^{\Gamma_j}_{h},\boldsymbol{v}^{\Gamma_j}_{z}\}$ are defined for each irreducible representation $\Gamma_j$ according to Table \ref{table_PS}.
\begin{table}
\begin{equation*}
\begin{tabular}{c|ccc}
\hline 
\hline 
	& $ \boldsymbol{v}_{0}$ & $\boldsymbol{v}_{h}$ & $\boldsymbol{v}_{z}$ \\
	\hline
	$\Gamma_{1}^+$ & $ (1,1,1,1)$  & $(1,1)$ & $ (1,1,1,1)$ \\
	$\Gamma_{2}^+$ & $(0,0,0,0,0)$ & $(0,0)$ & $(1,-1,-1,1)$ \\
	$\Gamma_{3}^+$ & $(0,0,0,0,0)$ & $(0,0)$ & $(1,-1,1,-1)$ \\
	$\Gamma_{4}^+$ & $(1,-1,-1,1)$ & $(0,0)$ & $(1,1,-1,-1)$ \\
\hline 
\hline 
\end{tabular}\;.
\end{equation*}
\caption{\label{table_PS} Basis vectors $\boldsymbol{v}$ for each irreducible representation $\Gamma_i^+$.}
\end{table}
Therefore, due to the point symmetries, only three independent pairing parameters remain: $\{\Delta_0,\Delta_{h},\Delta_{v}\}$. Moreover, because of SU(2)-spin symmetry, the time-reversal operator can simply be taken as the complex conjugation, such that TRS holds when $\left(\Delta_0, \Delta_{h},\Delta_{z}\right) = \mathrm{e}^{i \theta} \left( \vert\Delta_{0}\vert , s \vert\Delta_{h}\vert, s' \vert\Delta_{z} \vert\right)$, still keeping the relative signs $s,s'=\pm1$ free. For the TRS phases, we can always choose the gauge in which $\Delta_0, \Delta_{h},\Delta_{z} \in \mathbb{R}$. In Appendix \ref{app_lattice} we give an alternative form of the tight-binding BdG Hamiltonian that is explicitly based on the lattice symmetries and needed for the analytical derivation of the topological phase transition discussed in Section \ref{Top_PT}.

Within the $t-J$ model, the most stable superconducting pairing states at zero temperature are obtained by solving the self-consistent gap equations
\begin{equation}
\label{self_eq}
	\Delta_{\nu} = - J \sum_{\boldsymbol{k}} e^{-i \boldsymbol{k}\cdot \boldsymbol{\delta}_{\nu}} \left\langle \hat{c}_{i,\boldsymbol{k},\downarrow}\hat{c}_{j,-\boldsymbol{k},\uparrow} - \hat{c}_{i,\boldsymbol{k},\uparrow}\hat{c}_{j,-\boldsymbol{k},\downarrow} \right\rangle_{\rm mf},
\end{equation}
where the expectation value is taken with respect to the ground state of the mean-field BdG Hamiltonian in Eq.~(\ref{H_BdG}). 
Depending on $J$ and $\mu$, four distinct stable superconducting phases have previously already been established for this interaction.\cite{SBBS_16} 
For completeness we here briefly describe that phase diagram. At high-doping a fully gapped pairing state satisfying TRS dominates. It belongs to the trivial representation of $D_{2h}$ and we call it $\Gamma_{1,a}^+$ in the following. At lower doping, there is a stable nodal state also conserving TRS. It also belongs to the trivial representation and we call it $\Gamma_{1,b}^+$. The direct phase transition region between these two phases, i.e.~at intermediary doping values, is hindered by an intervening sliver of a fully gapped state that breaks TRS. Also this state belongs to the trivial representation and we call it $\Gamma_{1,c}^+$. Finally, there is also a stable nodal state breaking TRS at very low doping. This state mixes different representations, breaking point group symmetries spontaneously, and we call it $\Gamma_{d}^+$. 

In the following we start the discussion with the pairing states that conserve TRS, $\Gamma_{1,a}^+$ and $\Gamma_{1,b}^+$, and then consider the states with broken time-reversal symmetry (BTRS), $\Gamma_{1,c}^+$ and $\Gamma_{d}^+$. In these discussions we set $\Delta_0 = 0 $, as found in the $t-J$-model due to strong on-site repulsion. Finally, since much of the overall character of the phase diagram is set by the phase transition between the two TRS states, we study the details of this phase transition. In particular, we consider a much more generic phase transition, where we keep all the gap parameters, including the on-site pairing, in order to achieve the most general analytical expressions. We also there relax the condition $\vert\Delta_{z}\vert = \vert\Delta_{h}\vert$, which exist in the $t-J$ model when all bonds are equivalent but is not a necessary condition in a generic hyperhoneycomb lattice.

\section{Fully gapped and line-nodal states with time-reversal symmetry}\label{TRS_states}
We first study the properties of the TRS states $\Gamma_{1,a}^+$ and $\Gamma_{1,b}^+$, found in the $t-J$ model at high and low doping, respectively. They both belong to the $\Gamma_1^+$, i.e.~the trivial irreducible representation, of $D_{2h}$ and are of the form $\boldsymbol{\Delta}_{\Gamma_1^+} = \Delta \left[(1,1) \oplus s  (1,1,1,1)\right]$.

\subsection{Bulk topology}\label{bulk_top_TRS}
\begin{figure}[t!]
\centering
\begin{tabular}{cc} 
	\begin{overpic} [width=0.5\linewidth]{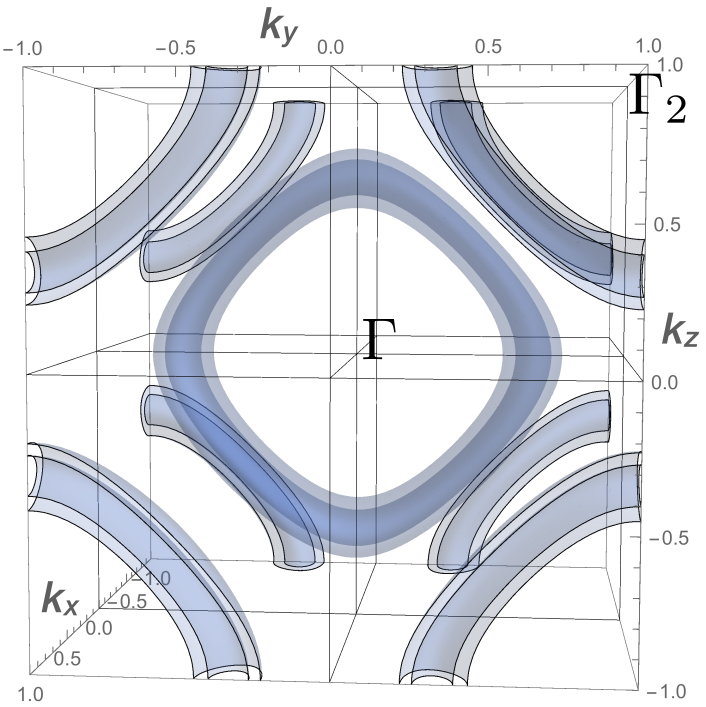}
 		\put (10,-3) {(a)}
	\end{overpic} & 
	\begin{overpic} [width=0.5\linewidth]{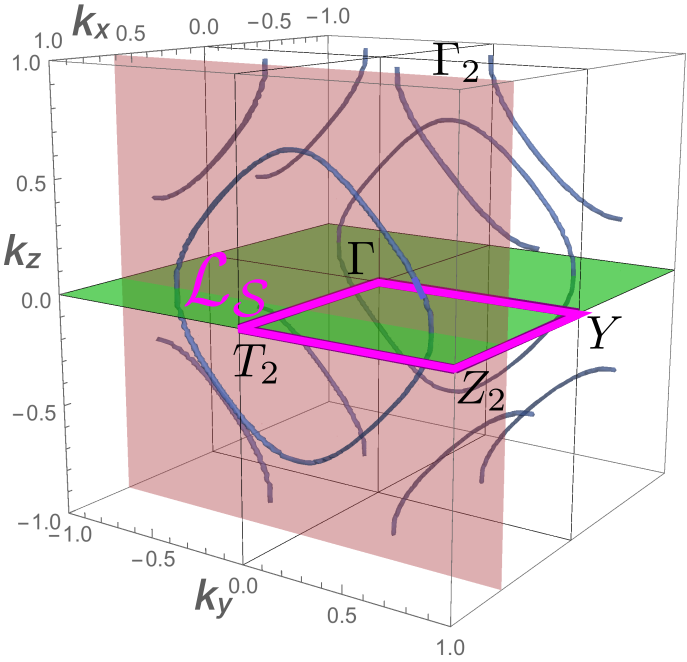}
 		\put (10,-3) {(b)}
	\end{overpic}
\end{tabular}
\caption{\label{fig_nodal} (a) BdG energy gap for the state $\Gamma_{1,a}^+$ represented through the energy iso-surfaces at $E  \approx -\Delta_{\mathrm{bulk}} = -0.5 t $. (b) Nodal lines (zero-energy iso-lines) of state $\Gamma_{1,b}^+$. Two BZs are shown, and thus two copies of the spectrum are visible in (a) and (b).}
\end{figure}

The state $\Gamma_{1,a}^+$ is obtained for $s=+1$ and is fully gapped. We show in Fig.~\ref{fig_nodal}(a) the highest occupied iso-energy surfaces of the BdG spectrum for this state over 2 BZs for clarity. Since the pairing order parameter has the same sign on every NN bond it corresponds to an ``extended $s$-wave'' state and it gaps out every point of the toroidal Fermi surface  shown in Fig.~\ref{fig_FS}.
In this case, the BdG Hamiltonian satisfies particle-hole symmetry by construction, which leads, combined with TRS, to an effective chiral symmetry. Together with the SU(2)-spin symmetry the system belongs to the three-dimensional Altland-Zirnbauer class CI of topological superconductors.\cite{Schnyder08} In the basis that makes the chiral symmetry operator diagonal the BdG Hamiltonian takes a block off-diagonal form,\cite{Schnyder08} i.e.~we find (see Appendix \ref{CI})
\begin{align}
	\tilde{H}(\boldsymbol{k}) &= \left(\begin{array}{cc} 0 & D(\boldsymbol{k})\\
	D^{\dagger}(\boldsymbol{k}) & 0 \end{array}\right)\;, \nonumber\\
\label{chiral_blockoff}
	D(\boldsymbol{k}) &= H_{\Delta}(\vk) + i H_{0}(\vk) \;.
\end{align}
This form is useful because the topological invariants for class CI can be expressed through the smaller matrix $D(\boldsymbol{k})$. The fully gapped phase in three dimensions is characterized by a winding number written in terms of the $D$ matrix,\cite{Schnyder08} which we directly find to be zero for the $\Gamma_{1,a}^+$-phase.
We further note that due to the block off-diagonal form of Eq.~(\ref{chiral_blockoff}), the eigenvalues are given through $\mathrm{det}[D(\boldsymbol{k}) \cdot D^{\dagger}(\boldsymbol{k}) - E^2 \mathbb{I}_{4\times 4}] = 0 $. The problem of finding the BdG spectrum is then reduced to an eigenvalue problem of a $4\times4$ matrix, which can be solved analytically. However, the expressions are cumbersome and add little to the discussion so we do not discuss them here, although in Section \ref{pheno} we use the qualitative features of the analytical BdG spectrum. 

Turning to the nodal state $\Gamma_{1,b}^+$, it is obtained by changing the relative sign between the zigzag- and horizontal-bond pairing, i.e.~$\boldsymbol{\Delta}_{\Gamma_{1,b}^+} = \Delta [ (1,1) \oplus - (1,1,1,1)]$. This phase exhibits two inequivalent nodal lines at zero energy within the first BZ, see Fig.~\ref{fig_nodal}(b) which shows four nodal lines as it covers two BZs. This can be understood intuitively by noting that the order parameter changes sign between the horizontal bonds (in the $x$-direction) and the zigzag bonds (mainly in the $y,z$-plane). Therefore, up to leading order in an expansion in spherical harmonics, this state corresponds to a $d_{3x^2-r^2}$-wave state with a double cone of zeros centered on the $x$-axis.  
Nodal lines in the three-dimensional CI class are characterized by a winding number that is inherited from the Altland-Zirnbauer class AIII, since TRS does not trivialize the topology.\cite{Sato11,Zhao13,Class_sym_review} The winding number is given in terms of the $D$ matrix (\ref{chiral_blockoff}) as\cite{WenZee_nodal02, Volovik_lect_notes07, Beri_nodal, Schnyder_nodal_0}
\begin{equation}
\label{WN_FS}
	\nu[\mathcal{L}] = \dfrac{i}{2\pi} \oint_{P\mathcal{L}}	dq~  \mathrm{tr} D^{-1}(q) \partial_q D(q) \in \mathbb{Z} \;, 
\end{equation}
where $\mathcal{L}$ is a closed loop in momentum space and $P$ means that the integral is path-ordered. The path ordering implicitly defines an orientation of the loop $\mathcal{L}$. We can interpret $\nu$ as a signed chirality attached to every nodal line in class CI after the orientation of the base loops has been fixed once and for all. Whenever $\mathcal{L}$ encircles nodal lines, the winding number $\nu[\mathcal{L}]$ counts the signed number of these. Using this we find that the two nodal lines of the $\Gamma_{1,b}^+$-state are topologically non-trivial with $\vert \nu \vert = 1$ and with opposite chiralities. Taking into account both spin species, we actually get $\nu(\uparrow) +\nu(\downarrow) = 2 \nu \in 2 \mathbb{Z}$ for each nodal line, but we here choose to only use the spin polarized quantities since with the full SU(2)-spin symmetry the spin plays no role. 

\subsubsection{Berry phase approach}\label{Berrys}
While we were able to calculate the topology of the line nodes in the $\Gamma_{1,b}^+$ above, let us here introduce an alternative and numerically much simpler approach to the bulk topology of line nodes in class CI based on the Berry phase. The Berry phase and its underlying Wilson loop approach has already proven to be an extremely useful tool to characterize topological insulators and topological semimetals.\cite{Bernevig1, Bernevig_point_groups, Bernevig3, AlexBernevig_berryphase,Alex1,Alex2,ABABS_points,ABABS_lines} One of the advantages of the Wilson loop approach as developed in\cite{Bernevig1, Bernevig_point_groups, Bernevig3, AlexBernevig_berryphase, Alex1, Alex2, ABABS_points, ABABS_lines} lies in its high efficiency for numerical computations, and here show how to extend it also to topological superconductors in class CI. 

Discretizing a closed loop in momentum space, i.e.~$\mathcal{L}=\{\vk_1,\dots,\vk_{N_k}=\vk_1\}$, the total Berry phase of the occupied bands over $\mathcal{L}$ can be efficiently\footnote{The phases of the eigenvalues of the Wilson loops are very stable and only a few number of points are necessary in the discretization of the base loop.} computed through \cite{Bernevig1, Bernevig_point_groups, Bernevig3, AlexBernevig_berryphase,Alex1,Alex2}
\begin{eqnarray}
\label{Berry_phase}
	\gamma_{B}\left[\mathcal{L}\right] &=&  \mathrm{Arg}\left\{\det \mathcal{W}[\mathcal{L}]   \right\}	\;,\\
\label{Wilson_loop_matrix}
	\mathcal{W}[\mathcal{L}] &=&  M_{1,N_k} \cdot \left(\prod\limits_{i=1}^{N_k-1}   M_{i+1,i}   \right)  \;, \\
M_{i+1,i} &=&  U^{\dagger}_{\mathrm{occ}}(\vk_{i+1}) \cdot  U_{\mathrm{occ}}(\vk_{i}) 	\nonumber\;, 
\end{eqnarray}
where $\gamma_B$ is the Berry phase, $\mathcal{W}[\mathcal{L}]$ is the Wilson loop matrix, and the column vectors of the matrices $U_{\mathrm{occ}}(\vk)$ are composed of the occupied BdG eigenstates, i.e.~we have $H(\boldsymbol{k}) \vert u_n,\boldsymbol{k} \rangle = E_{n}(\boldsymbol{k})  \vert u_n,\boldsymbol{k} \rangle$ with $E_{n}(\boldsymbol{k})<0$ and $[U_{\mathrm{occ}}(\vk_{i})]_n = \vert u_n,\boldsymbol{k} \rangle$. 
Previously it has been shown that chiral symmetry leads to a $\mathbb{Z}_2$ quantization of the Berry phase.\cite{Hatsugai_graphene_chiral, RyuSchnyder_10ways, Hatsugai_nonabelian} Fixing explicitly the global gauge that satisfies the chiral symmetry it can be shown that\cite{RyuSchnyder_10ways}
\begin{equation}
\label{Berry_quantized}
	\mathrm{exp}\{ i \gamma_B [\mathcal{L}] \} = \mathrm{exp}\{ i \pi \nu [\mathcal{L}] \} \in \{+1,-1\}\;,
\end{equation} 
where $\nu [\mathcal{L}] \in \mathbb{Z}$ is exactly the winding number Eq.~(\ref{WN_FS}). Choosing the sector $[0,2\pi)$ it follows that $\gamma_B[\mathcal{L}] \in\{0,\pi\} \cong \mathbb{Z}_2$. 
Due to its numerical efficiency, we use this Berry phase approach below in our discussion of the bulk-boundary correspondence.

\subsection{Bulk-boundary correspondence and surface Majorana flat bands}\label{BBC}
A very useful bulk-boundary correspondence exists that relates the winding number\cite{Sato11} in Eq.~(\ref{WN_FS}) or the quantized Berry phase\cite{RyuHatsugai_BulkEC,Hatsugai_graphene_chiral} in Eq.~(\ref{Berry_quantized}), both evaluated in the bulk, to the existence of surface Majorana states. Here we illustrate the bulk-boundary correspondence  in the hyperhoneycomb lattice by relating the bulk topological number with the existence of surface states and numerically calculating the surface spectrum. This shows that the nodal line $\Gamma^+_{1,b}$ state has surface Majorana flat bands.

Let us define a surface-cut orientation through the normal direction $\boldsymbol{r}_{\perp} = x_{\perp} \boldsymbol{n}$, where $\boldsymbol{n}$ is the unit vector perpendicular to the surface. We can then always numerically solve the lattice BdG equations for a slab geometry with two infinitive parallel surfaces, say at $x_{\perp}=x_0$ and $x_0-L$. Hence only $\boldsymbol{k}_{\parallel}=(k_{1,\parallel},k_{2,\parallel})$, with $\boldsymbol{k}_{\parallel}$ in the surface BZ, are good quantum numbers of the surface Hamiltonian $H(x_{\perp},\boldsymbol{k}_{\parallel} )$. Let us now consider a momentum path perpendicular to the surface BZ $\mathcal{L}_{\boldsymbol{k}_{\parallel}} = \{ \boldsymbol{k} = (\boldsymbol{k}_{\parallel}, k_{\perp}) \vert k_{\perp} \in [-G_{\perp}/2,G_{\perp}/2) \}$ at fixed $\boldsymbol{k}_{\parallel}$ and with $\boldsymbol{G}_{\perp} $ a reciprocal lattice vector in the $k_{\perp}$-direction. It is a non-contractible loop by periodicity of the Bloch states under a translation by a reciprocal lattice vector. We then write the winding number in Eq.~(\ref{WN_FS}) computed along such path as $\nu[\mathcal{L}_{\boldsymbol{k}_{\parallel}}]$, and equivalently the Berry phase $\gamma_{B}[\mathcal{L}_{\boldsymbol{k}_{\parallel}}]$. The bulk-boundary correspondence tells that whenever $\nu[\mathcal{L}_{\boldsymbol{k}_{\parallel}}] \propto \gamma_{B}[\mathcal{L}_{\boldsymbol{k}_{\parallel}}]\neq 0$ the surface spectrum has a zero-energy surface state at $\boldsymbol{k}_{\parallel}$ protected by chiral symmetry.\cite{Sato11} Since topological numbers are invariant under gap-preserving adiabatic transformations, the bulk numbers, $\nu$ and $\gamma_B$, remain unchanged under parallel shifts of the path $\mathcal{L}_{\boldsymbol{k}_{\parallel}}$, as long as we avoid any bulk-gap closing points:~in our case, as long as we do not cross a bulk nodal line. Therefore, the projection of bulk nodal lines on the surface BZ defines two-dimensional domains $\boldsymbol{k}_{\parallel}\in \Omega_{\parallel}$, with or without surface states, that together forms Majorana flat bands.\cite{Sato11,Schnyder_nodal_2}  The surface Majorana states are however not robust if the surface breaks chiral symmetry, e.g.~through a spontaneous breaking of TRS.\cite{Majorana_PLee} 

We note in passing that the quantized Berry phase $\gamma_{B}[\mathcal{L}_{\boldsymbol{k}_{\parallel}}]$ for the non-contractible loop $\mathcal{L}_{\boldsymbol{k}_{\parallel}}$ is the analogue of the Zak phase of one-dimensional systems with TRS and inversion symmetry.\cite{Zak_1D_phase} Moreover, King-Smith and Vanderbilt\cite{KingVanderbilt} have shown the equivalence of the Zak phase with the quantized electronic polarization. This lead to a bulk-boundary correspondence in terms of the Zak phase that has been widely used as an indicator of accumulated surface charges.\cite{VanderbiltKingSmith_surface,KariyadoHatsugai} However, it is not as robust\cite{Intercellular_Zak} as in the case of chiral symmetry since surfaces always break inversion symmetry.  

\begin{figure}[t]
\centering
\begin{tabular}{cc} 
	\begin{overpic} [width=0.5\linewidth]{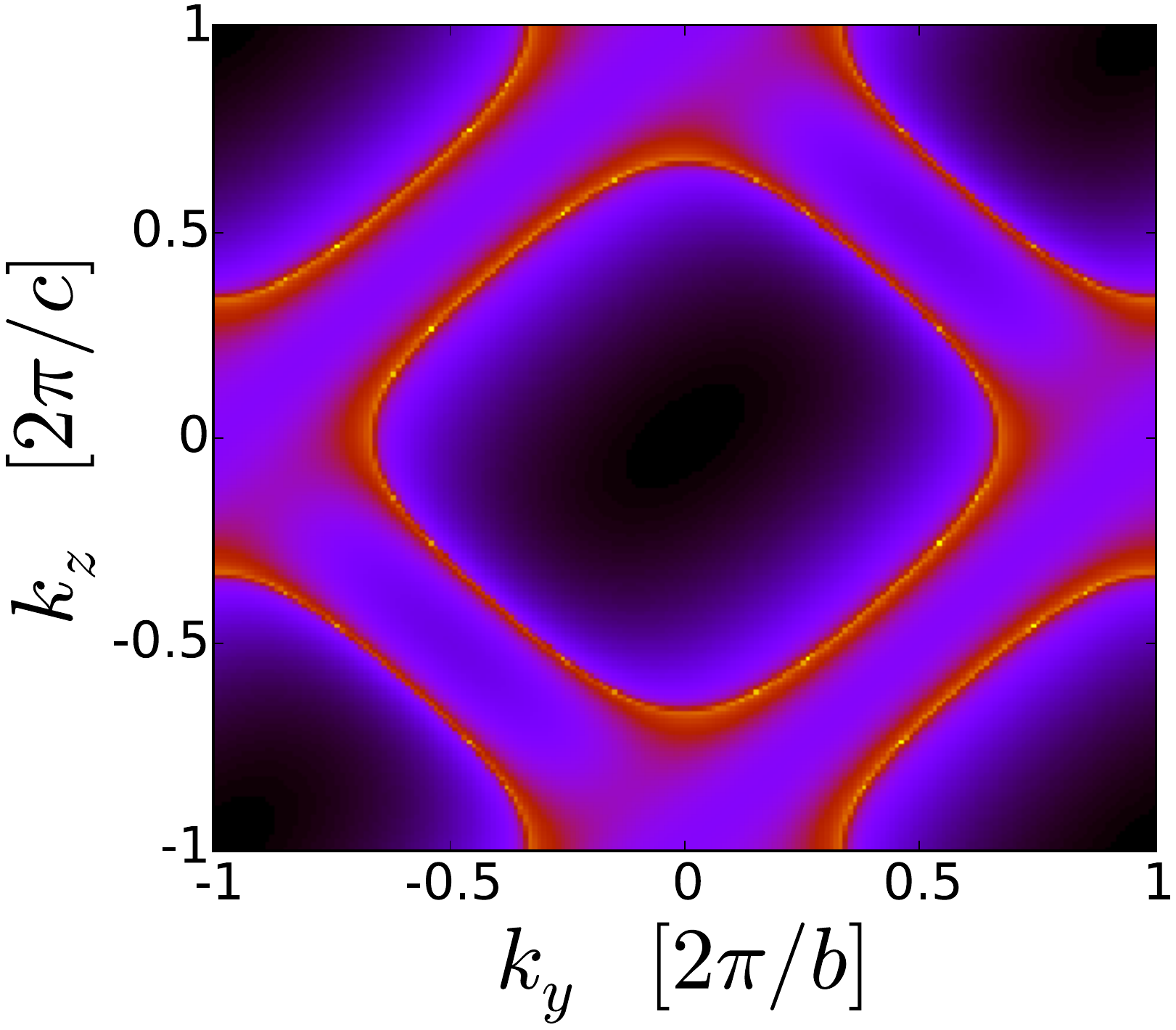}
 		\put (0,0) {(a)}
	\end{overpic} &
	\begin{overpic} [width=0.5\linewidth]{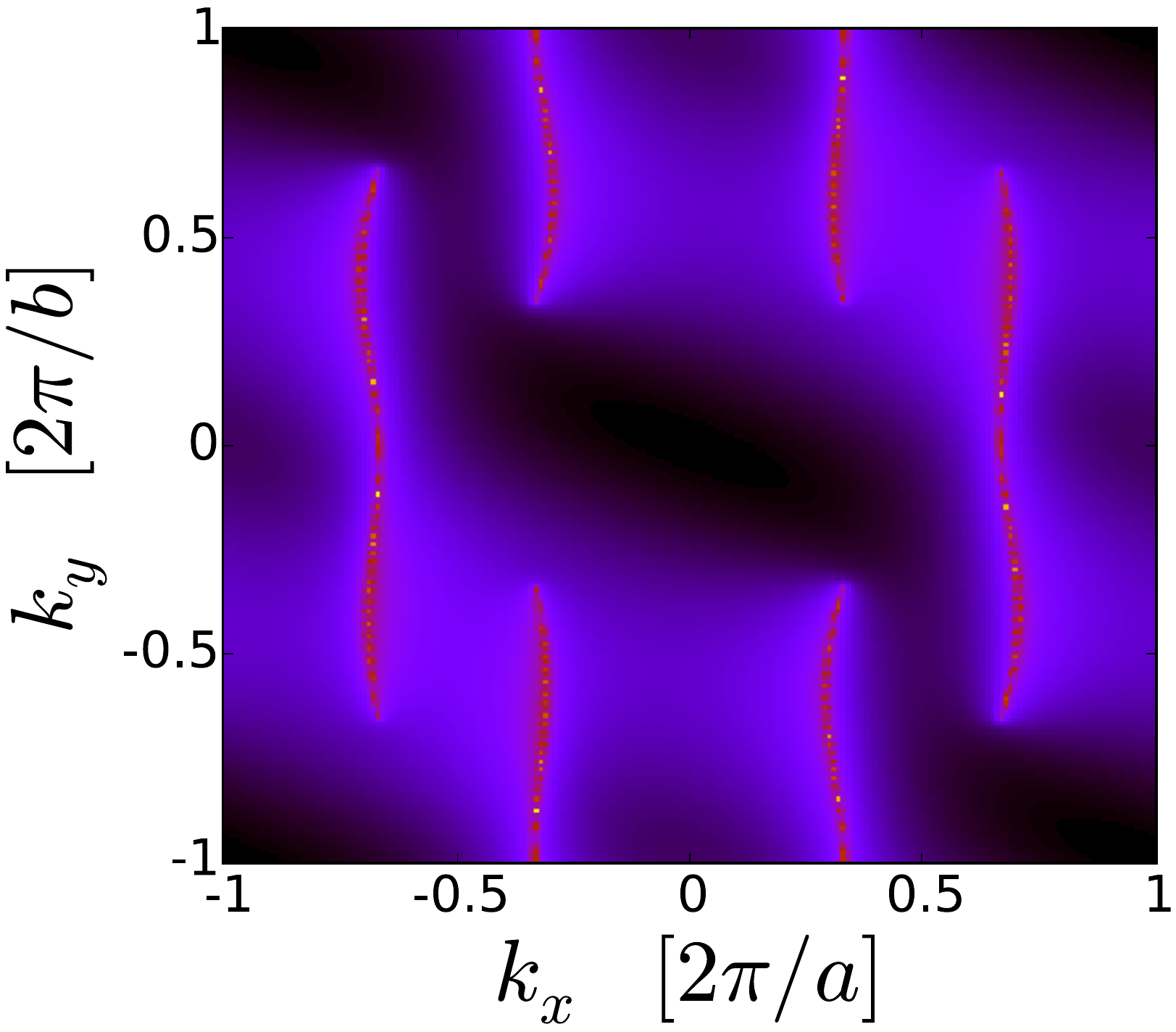}
 		\put (0,0) {(b)}
	\end{overpic}
\end{tabular}
\caption{\label{surface_spec} Surface BdG spectral function at zero energy for the line-nodal state $\Gamma_{1,b}^+$ at a $(100)$-surface (a) and $(001)$-surface (b). Yellow indicates high spectral weight (not seen), black indicates exponentially suppressed spectral weight.}
\end{figure}
To explicitly demonstrate the bulk-boundary correspondence, we compute the surface BdG spectral function at zero energy, $-1/\pi \Im G^{\mathrm{BdG}}(x_{\perp}=x_0,\boldsymbol{k}_{\parallel},E=0)$, in the line-nodal state $\Gamma_{1,b}^+$ for different slab geometries and also compare with the bulk Berry phase. We choose here the Berry phase as the bulk indicator since it can be computed numerically very easily. 
Figure~\ref{surface_spec}(a) shows the spectral weight at a $(100)$-surface, i.e.~with a normal vector $\boldsymbol{n} = (1,0,0)\propto  \boldsymbol{a}_1 + \boldsymbol{a}_2- \boldsymbol{a}_3 \propto \boldsymbol{b}_1+\boldsymbol{b}_2$.~Fig.~\ref{surface_spec}(b) shows the spectral weight at a $(001)$-surface, i.e.~with the normal vector $\boldsymbol{n} = (0,0,1)\propto \boldsymbol{a}_1- \boldsymbol{a}_2+ \boldsymbol{a}_3 \propto \boldsymbol{b}_1+\boldsymbol{b}_3$. None of these surfaces show surface Majorana flat bands. At the (100)-surface the two bulk nodal lines, see Fig.~\ref{fig_nodal}(b), project exactly on top of each other on the surface BZ leading to a cancelation of the Berry phase $\gamma_B[\mathcal{L}_{\boldsymbol{k}_{\parallel}}]=+1-1=0$ for every $\boldsymbol{k}_{\parallel}$. At the (001)-surface the bulk nodal lines instead project into one-dimensional segments on the surface BZ (Fig.~\ref{surface_spec}(b), again resulting in vanishing surface Majorana flat bands.

\begin{figure}[t]
\centering
\begin{tabular}{cc} 
	\begin{overpic} [width=0.5\linewidth]{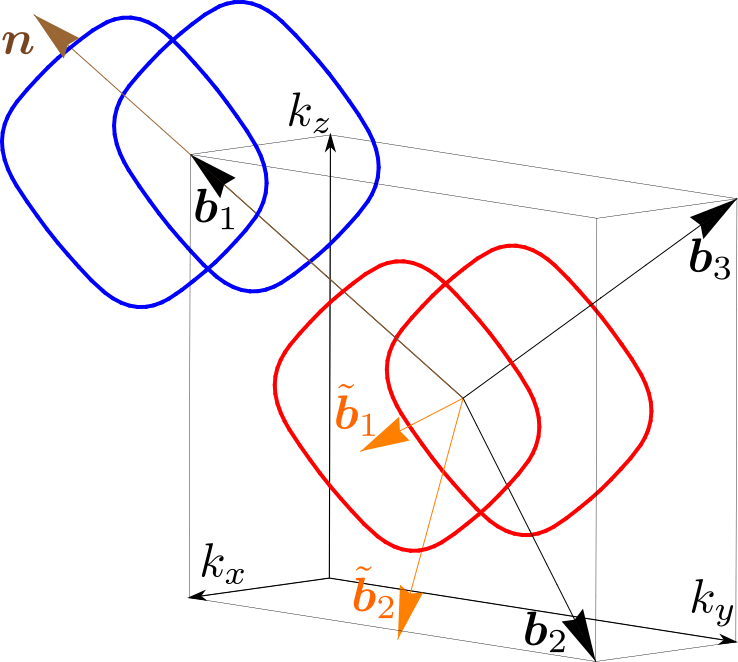}
 		\put (0,5) {(a)}
	\end{overpic}&
	\begin{overpic} [width=0.5\linewidth]{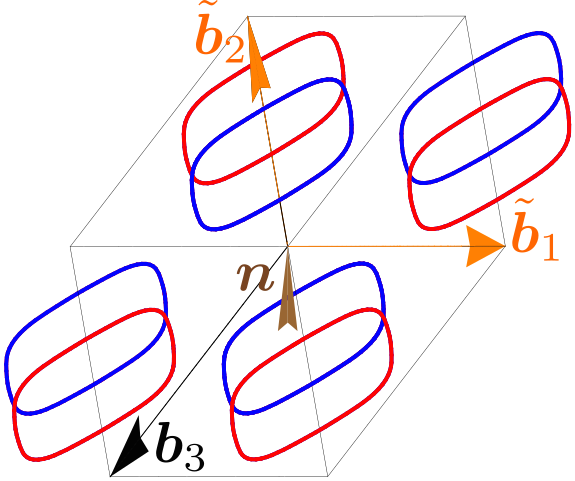}
 		\put (70,5) {(b)}
	\end{overpic}\\
		\begin{overpic} [width=0.5\linewidth]{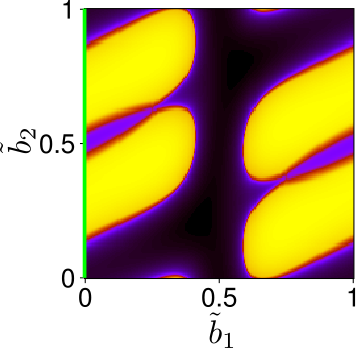}
 		\put (0,5) {(c)}
	\end{overpic} &
	\begin{overpic} [width=0.47\linewidth]{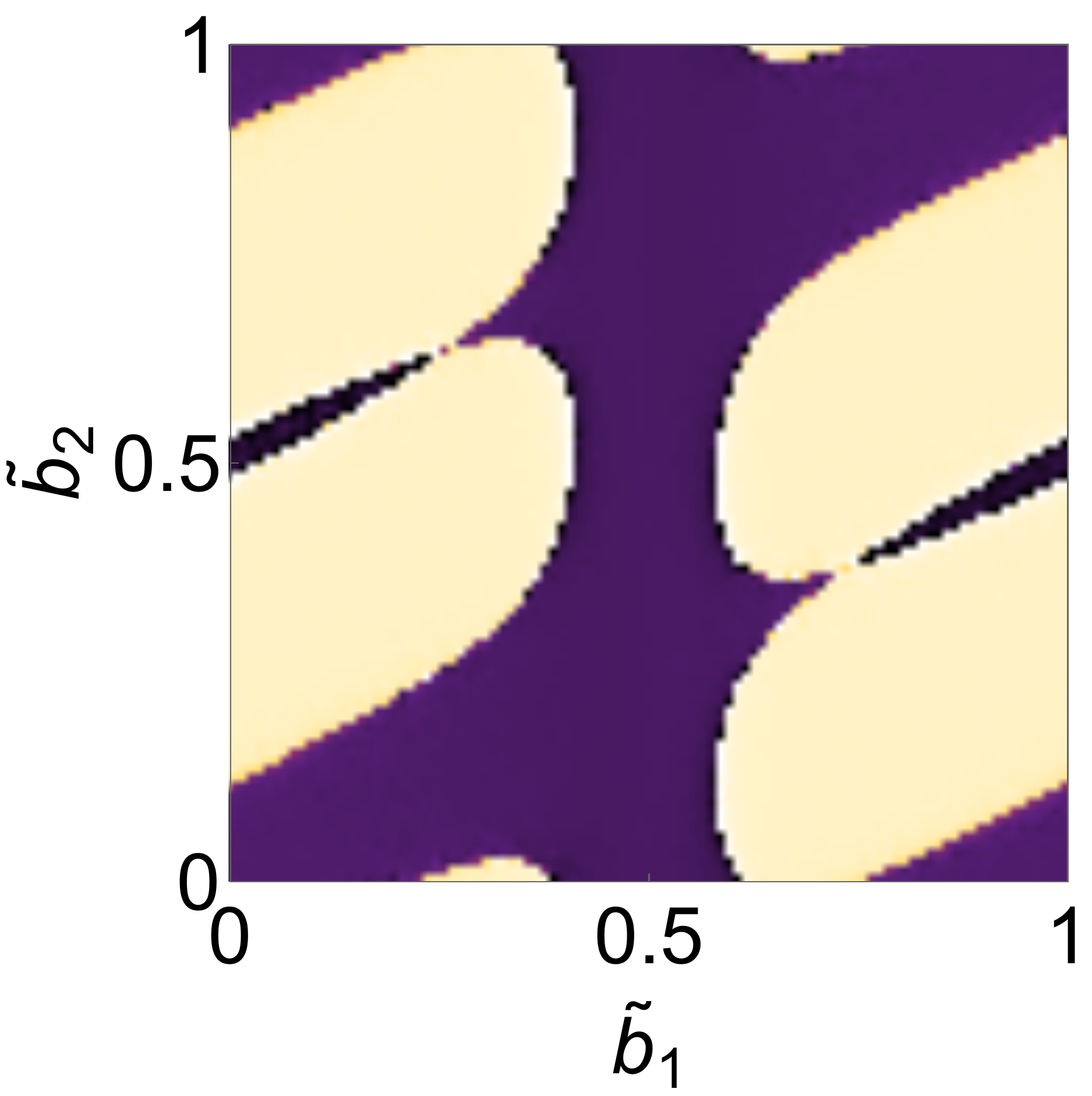}
 		\put (0,5) {(d)}
	\end{overpic} 
\end{tabular}
\begin{tabular}{c} 
	\begin{overpic} [width=0.7\linewidth]{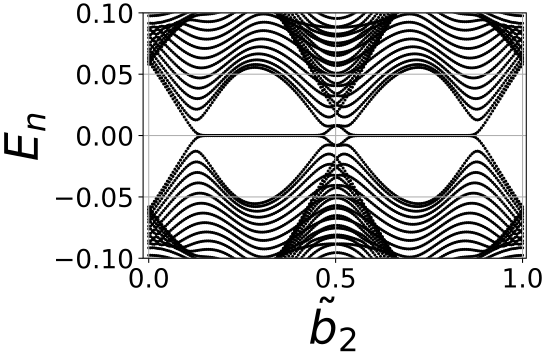}
 		\put (0,5) {(e)}
	\end{overpic}
\end{tabular}
\caption{\label{surface_spec_B} (a) Schematic bulk nodal lines of the state $\Gamma_{1,b}^+$ and surface normal vector $\boldsymbol{n} \propto \boldsymbol{b}_1$ (brown) for surface spanned by $\tilde{\boldsymbol{a}}_1=\boldsymbol{a}_3$ and $\tilde{\boldsymbol{a}}_2=-\boldsymbol{a}_2+\boldsymbol{a}_3$. (b) Projection in parallel to $\boldsymbol{n} \propto \boldsymbol{b}_1$ of the bulk nodal lines onto the surface BZ with $\{\tilde{\boldsymbol{b}}_1,\tilde{\boldsymbol{b}}_2\}$ reciprocal primitive vectors (orange). (c) Surface BdG spectral function at zero energy over the surface BZ. (d) Bulk Berry phase over a non-contractible loop $\gamma_B[\mathcal{L}_{\boldsymbol{k}_{\parallel}}]$, with $\gamma_B = 0$ (dark) and $\gamma_B = \pi$ (light). (e) BdG spectrum for the full slab geometry as a function of $\tilde{b}_2$ and at fixed $\tilde{b}_1=0$, i.e.~along a slice of the surface BZ marked by a light green line in (d).}
\end{figure}

We next consider the diagonal surface spanned by the lattice vectors $\tilde{\boldsymbol{a}}_1= \boldsymbol{a}_3$ and $\tilde{\boldsymbol{a}}_2=- \boldsymbol{a}_2+ \boldsymbol{a}_3$.~It has the normal vector $\boldsymbol{n} \propto \tilde{\boldsymbol{a}}_1\times \tilde{\boldsymbol{a}}_2 \propto \boldsymbol{b}_1$, shown in Fig.~\ref{surface_spec_B}(a) (brown) together with the schematic bulk nodal lines and the primitive reciprocal vectors $\{\tilde{\boldsymbol{b}}_1,\tilde{\boldsymbol{b}}_2\}$ (orange) of the surface. Figure~\ref{surface_spec_B}(b) shows the projection in parallel to $\boldsymbol{n} \propto \boldsymbol{b}_1$ of the schematic bulk nodal lines onto the surface BZ. In Fig.~\ref{surface_spec_B}(c) we display the surface BdG spectral weight at zero-energy across the whole surface BZ, using $\boldsymbol{k}_{\parallel} = \tilde{b}_1 \tilde{\boldsymbol{b}}_1+ \tilde{b}_2 \tilde{\boldsymbol{b}}_2$. Alternatively, in Fig.~\ref{surface_spec_B}(e) we display the BdG spectrum for the whole slab geometry as a function of $\tilde{b}_2$, but keeping a fixed $\tilde{b}_1=0$. Based on these figures we can conclude that the yellow regions of Fig.~\ref{surface_spec_B}(c) represent surface Majorana flat bands.
Finally, we show that the existence of Majorana flat bands are directly related ot the bulk Berry phase in Fig.~\ref{surface_spec_B}(d). We here compute the bulk Berry phase over the non-contractible loops $\mathcal{L}_{\boldsymbol{k}_{\parallel}}$ for every point $\boldsymbol{k}_{\parallel}$ of the surface BZ.
The dark regions indicate $\gamma_B[\mathcal{L}_{\boldsymbol{k}_{\parallel}}] =  0\mod2\pi$, while the light regions indicate $\gamma_B[\mathcal{L}_{\boldsymbol{k}_{\parallel}}]  =\pi\mod2\pi\neq 0$. Comparing with the BdG spectral weight in Fig.~\ref{surface_spec_B}(c), this becomes a perfect substantiation of the bulk-boundary correspondence as introduced above. Further comparing with the projected schematic bulk nodal lines, see Fig.~\ref{surface_spec_B}(b), we find that the surface Majorana states are gapped in the regions where projected line-nodal contours overlap (corresponding to the purple regions of Fig.~\ref{surface_spec_B}(c) and dark in Fig.~\ref{surface_spec_B}(d)).

\section{Fully gapped and point-nodal states with broken time-reversal symmetry}\label{BTRS_states}
Having classified the TRS states appearing as mean-field solutions to the $t-J$ model, we now turn to the states found that breaks TRS: $\Gamma_{1,c}^+$ and $\Gamma_d^+$. Here, $\Gamma_{1,c}^+$ is found at an intermediary filling between the states $\Gamma_{1,a}^+$ and $\Gamma_{1,b}^+$ in the phase diagram, while the $\Gamma_d^+$ states lies at the lowest filling levels. $\Gamma_{1,c}^+$ is fully gapped and belongs to the trivial representation. $\Gamma_d^+$ is nodal and mixes representations. Both of these BTRS states conserve the translational symmetry of the lattice. While the fully gapped phase is topologically trivial, we show that the other state has nodal points and shares several main features with Weyl semimetals.

\subsection{Bulk topology}
\begin{figure}[t]
\centering
\begin{tabular}{c} 
	\begin{overpic} [width=0.5\linewidth]{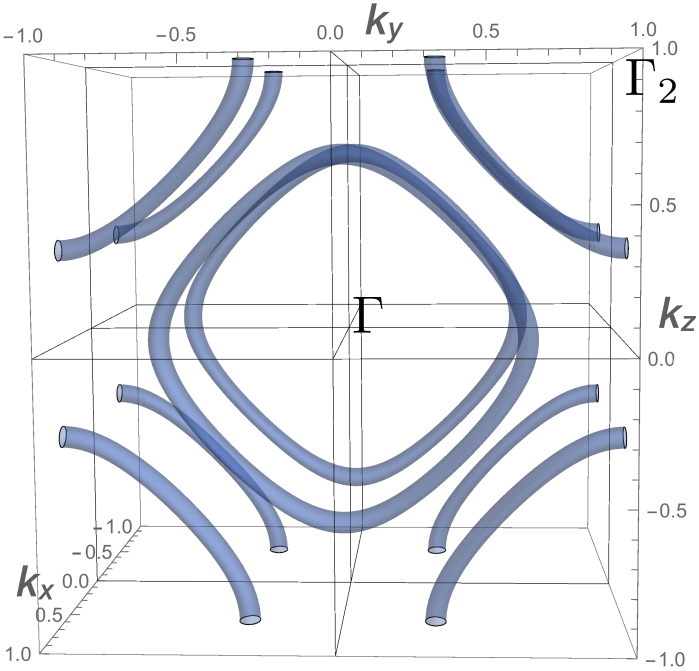}
 		\put (-10,0) {(a)}
	\end{overpic}
\end{tabular}
\begin{tabular}{cc} 
	\begin{overpic} [width=0.45\linewidth]{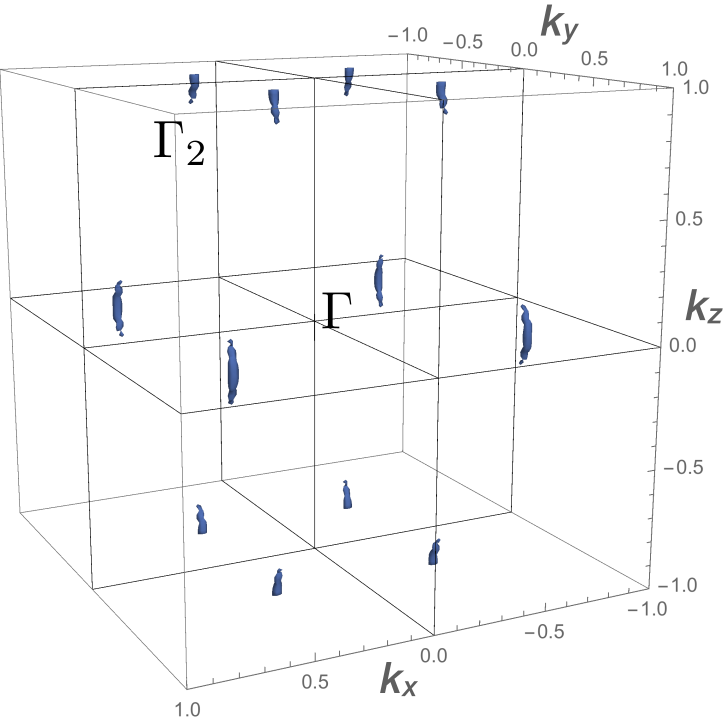}
 		\put (0,0) {(b)}
	\end{overpic} &
	\begin{overpic} [width=0.55\linewidth]{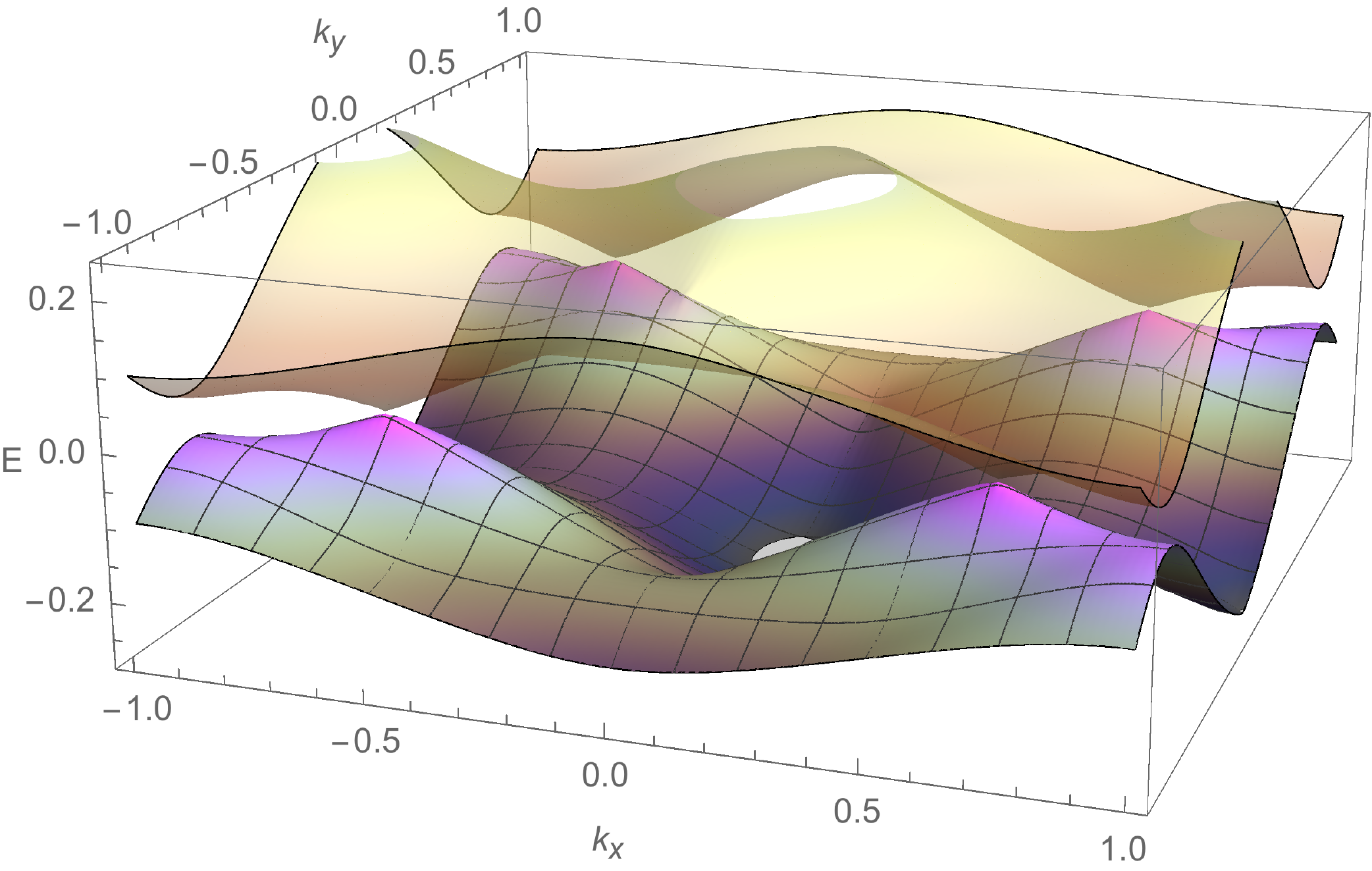}
 		\put (0,0) {(c)}
	\end{overpic}
\end{tabular}
\caption{\label{BTRS_ab} (a) BdG energy gap for the state $\Gamma_{1,c}^+$ represented through the maximum occupied iso-energy surface, here for $\Delta_{h} =\Delta_{z}  = 0.1t$ and $\phi=0.9\pi$. (b) Positions in the BZ of the nodal points of the state $\Gamma_{d}^+$ (numerical accuracy makes them look like cigars). (c) BdG dispersion in the vicinity of the nodal points over the plane $k_z=0$ containing four nodal points, here for $\vert \Delta_{h}\vert =\vert \Delta_{z} \vert  = 0.5 t$ and $\phi = 0.2\pi$.}
\end{figure}

The $\Gamma^+_{1,c}$ state is a fully gapped state and takes the form $\boldsymbol{\Delta}_{\Gamma_{1,c}^+} = \vert \Delta_{h} \vert (1,1) \oplus \mathrm{e}^{i \phi} \vert\Delta_{z} \vert (1,1,1,1)$. The complex phase factor leads to the complete gapping of the nodal lines of the state $\boldsymbol{\Delta}_{\Gamma_{1,b}^+}$, see Fig.~\ref{BTRS_ab}(a) where the BdG energy gap is displayed by plotting the highest iso-energy momentum surface of the occupied bands. By breaking TRS the complex phase factor $e^{i\phi}$ can be seen as interpolating between the fully gapped state $\Gamma_{1,a}^+$ at $\phi = 0 $ and the line-nodal state $\Gamma_{1,b}^+$ at $\phi = \pi $. When TRS is broken, chiral symmetry is also absent such that the system now belongs to the three-dimensional Altland-Zirnbauer class C (we still assume a full SU(2)-spin symmetry).\cite{Schnyder08} Fully gapped phases of the class C always have a trivial topology in three dimensions.\cite{Schnyder08} 

At the lowest doping levels we find the BTRS state $\Gamma_{d}^+$, which is realized with the form $\boldsymbol{\Delta}_{\Gamma_{d}^+} =  \vert\Delta_{h}\vert (1,1) \oplus  -\vert\Delta_{z}\vert  \left(  e^{i\phi},e^{-i\phi},e^{-i\phi},e^{i\phi}\right)$ where $\phi \gtrsim 0$. This state gaps partially the nodal lines of the state $\Gamma_{1,b}^+$ (recovered for $\phi=0$), leaving four nodal points within the BZ, see Fig.~\ref{BTRS_ab}(b) and (c). We note that on top of breaking TRS $\Gamma_{d}^+$ also breaks spontaneously the $D_{2h}$ symmetry of the lattice since it mixes different representations, in this case~$\Gamma_1^+$ and $\Gamma_2^+$ of Table \ref{table_PS}. This lowers the point group to $C^{(y)}_{2h}$ with the principal axis chosen in the $k_y$-direction. 
Disregarding particle-hole symmetry, we can consider the $\Gamma^+_{d}$ state as being in the three-dimensional Altland-Zirnbauer class A. This class is well known to realize Weyl semimetals with nodal (Weyl) points characterized through Chern numbers.\cite{Schnyder08,Ishikawa_86,ShiozakiSato_I} 
Since particle-hole symmetry does not trivialize the topology of the Weyl points, the nodal points in the correct class C inherit the Chern number of class A and we can thus speak of a Weyl superconducting state.\cite{Class_sym_review} 

\begin{figure}[t]
\centering
\begin{tabular}{cc} 
	\begin{overpic} [width=0.5\linewidth]{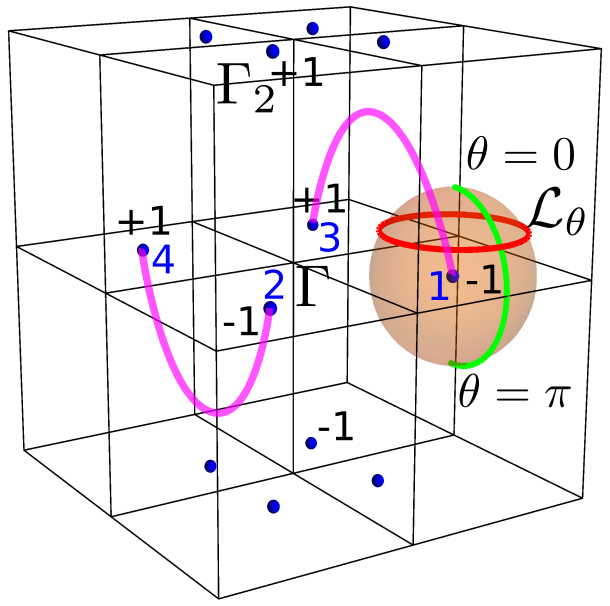}
 		\put (0,0) {(a)}
	\end{overpic} &
	\begin{overpic} [width=0.5\linewidth]{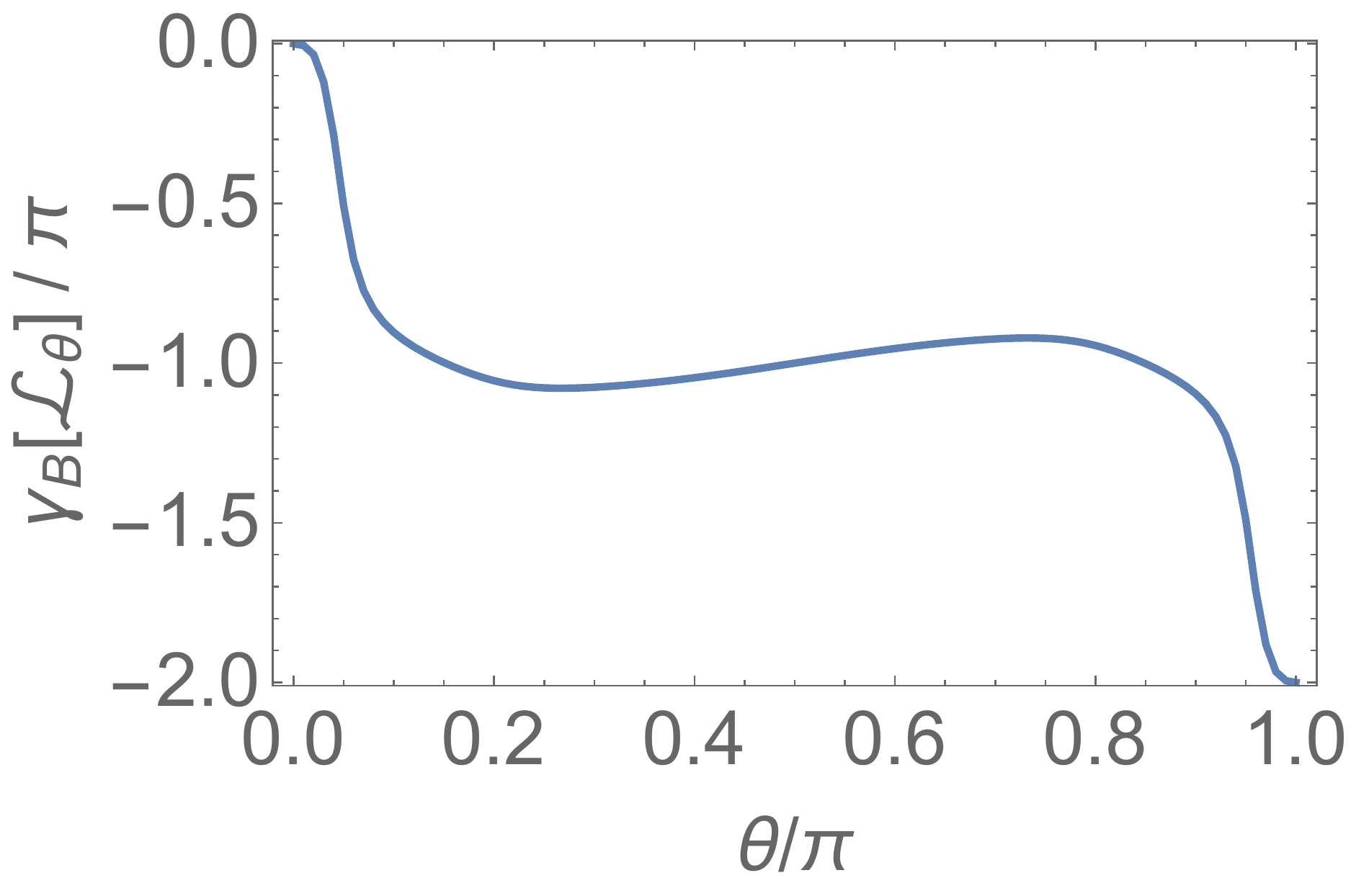}
 		\put (0,0) {(b)}
	\end{overpic}
\end{tabular}
\caption{\label{BTRS_points} (a) Schematic nodal points (blue, labeled from 1 to 4) for $\Gamma_d^+$ state with Chern numbers ($\pm1$) and Berry flux lines joining them (pink). Orange sphere surrounding one nodal point illustrates the base loop $\mathcal{L}_{\theta}$ (red) at a fixed polar angle $\theta\in [0,\pi]$. (b) Flow of Berry phase as base loop covers the full sphere leading to a Chern number $C_1=-1$.}
\end{figure}

The Berry phase approach of Section \ref{Berrys} turns out to be very practical for the computation also of the Chern number for the nodal points in the $\Gamma^+_{d}$ state. However, since TRS is broken, chiral symmetry is absent and the Berry phase is not quantized. Therefore, it can not be taken as a topological invariant as such. Instead, we track the continuous flow of the Berry phase as we sweep a base loop over a closed manifold, such as the orange sphere in Fig.~\ref{BTRS_points}(a), that encircles a nodal point.\cite{Stone_76, ABABS_points, ABABS_lines} The Chern number is then given as the total flow of Berry phase from one pole (NP) of the surrounding sphere to the opposite pole (SP), i.e.~$C_1 =  (\gamma_B[\mathrm{NP}]-\gamma_B[\mathrm{SP}])/2\pi \in \mathbb{Z}$. Let us first take the sphere in Fig.~\ref{BTRS_points}(a) surrounding the nodal point at $p_1=(-p_x,p_y,0)$ and parametrize it with the base loops $\mathcal{L}_{\theta}$ (red) at constant polar angle $\theta \in [0,\pi]$. We show in Fig.~\ref{BTRS_points}(b) the flow of Berry phase $\gamma_B[\mathcal{L}_{\theta}]$ as we sweep the base loop from the polar angle $\theta=0$ to $\theta=\pi$. Since the Berry phase winds by $-2\pi$, we find Chern number $C_1 =-1$, i.e.~this nodal point is a sink of Berry curvature. Since the Berry curvature transforms as a vector under rotational symmetries and as a pseudo-vector under mirror symmetries, we easily conclude from the lower point group $C^{(y)}_{2h}$ that the nodal point at $p_2=(p_x,p_y,0)=C_{2y} p_1$ (i.e.~image under the rotation $C_{2y}$) must have the same charge, while those at $p_3=(-p_x,-p_y,0)=m_yp_1$ (image under the reflection $m_y$) and $p_4=(p_x,-p_y,0)=I p_1$ (image under the inversion $I$) must have the opposite charge, i.e.~the two latter are sources of Berry curvature. We illustrate these sinks and sources in Fig.~\ref{BTRS_points}(a) by drawing lines of Berry flux (pink) connecting them. 

\subsection{Surface Fermi arcs}\label{arcs}
In complete analogy with Weyl semimetals, there is a bulk-boundary correspondence also for Weyl superconductors, according to which the projection of the bulk Berry flux lines on a surface BZ traces out surface Fermi arcs that connect the projected nodal points.\cite{Vishwanath_weyl0,Bernevig_Weyl0}
\begin{figure}[t]
\centering
\begin{tabular}{cc} 
	\begin{overpic} [width=0.5\linewidth]{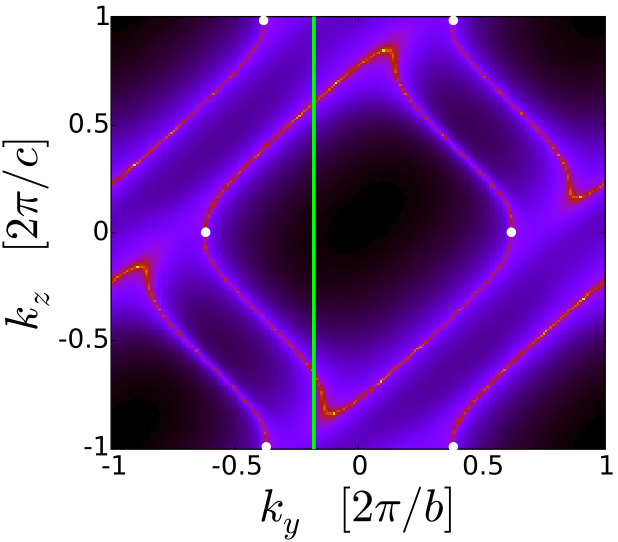}
 		\put (0,0) {(a)}
	\end{overpic} &
	\begin{overpic} [width=0.5\linewidth]{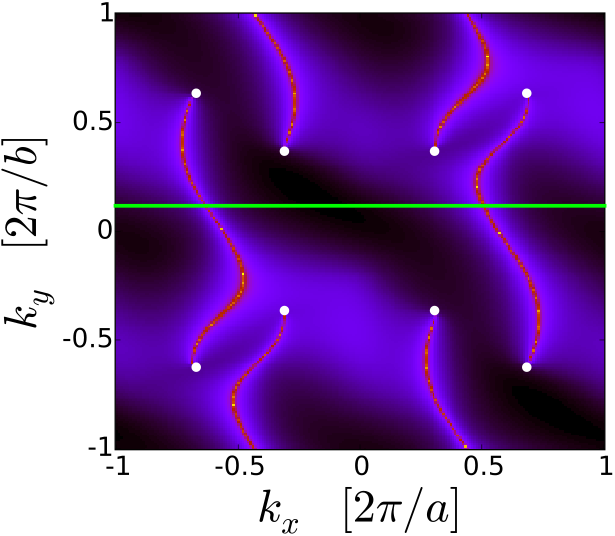}
 		\put (0,0) {(b)}
	\end{overpic} \\
	\begin{overpic} [width=0.5\linewidth]{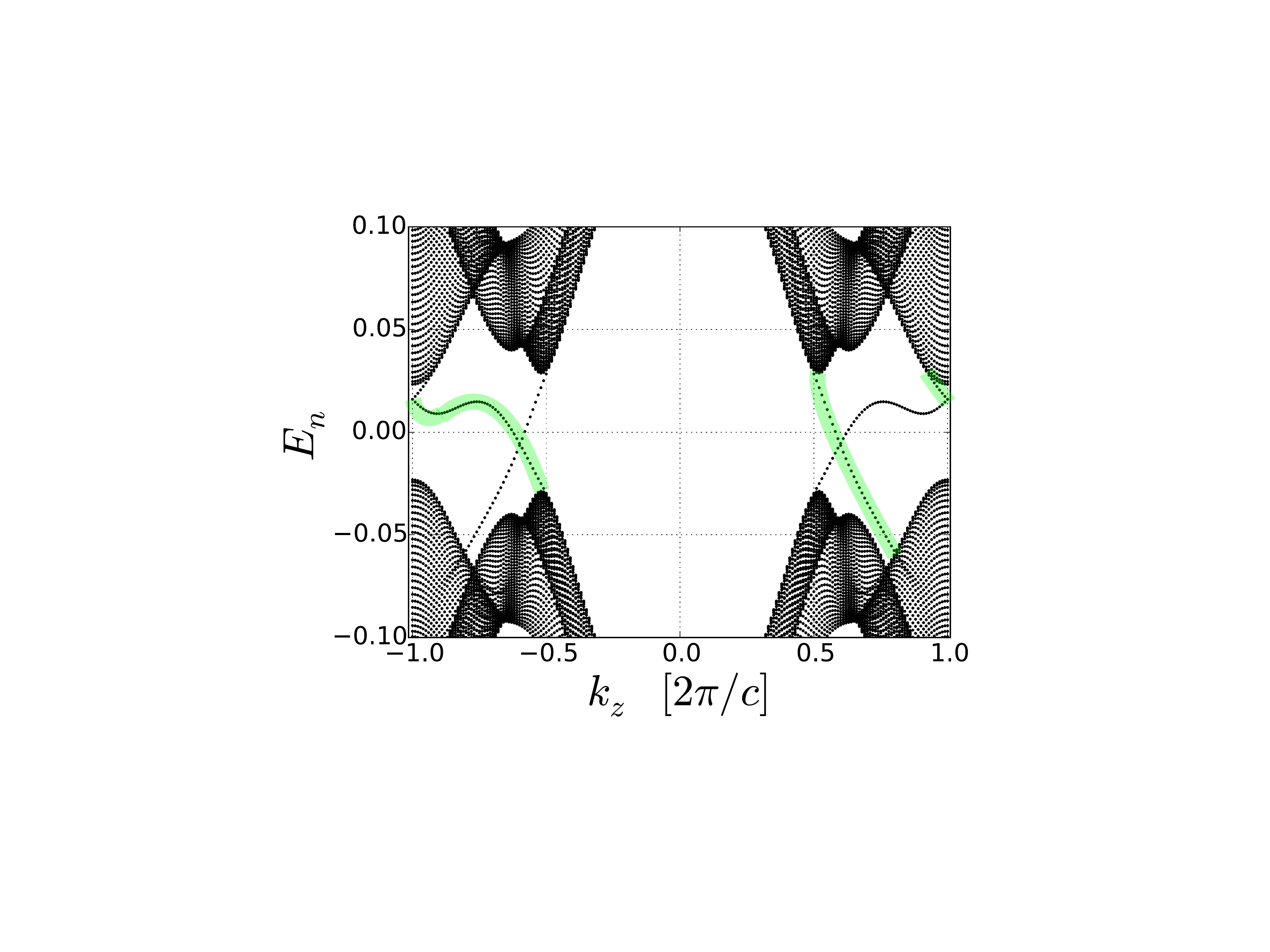}
 		\put (0,0) {(c)}
	\end{overpic} &
	\begin{overpic} [width=0.5\linewidth]{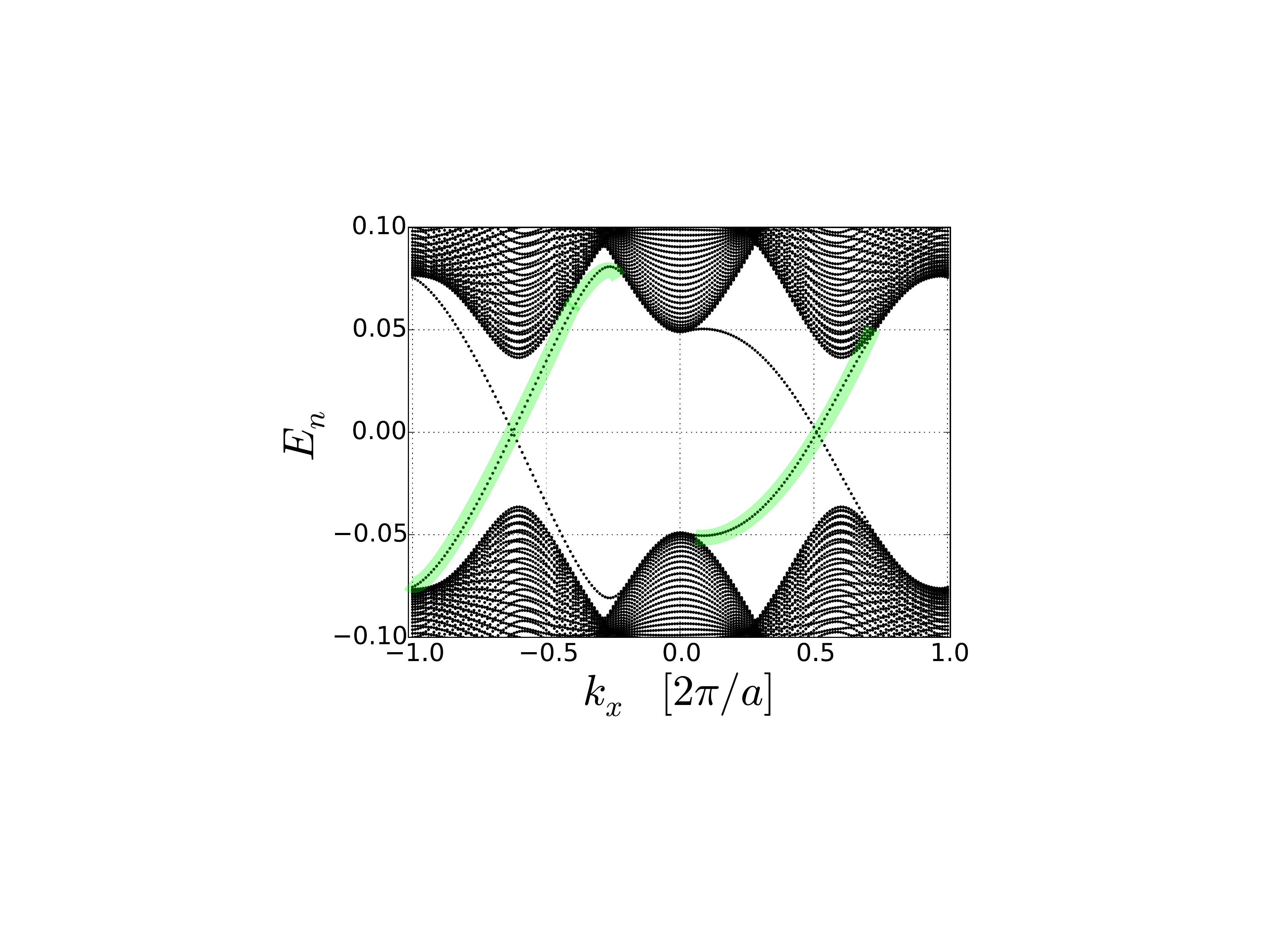}
 		\put (0,0) {(d)}
	\end{overpic}
\end{tabular}
\caption{\label{BTRS_surface_spec} Surface properties for the point nodal state $\Gamma^+_d$. (a) Surface BdG spectral function at zero energy for a $(100)$ surface. (b) Surface BdG spectral function at zero energy for a $(001)$ surface. (c) Surface BdG spectrum as a function of $k_z$ for fixed $k_y=-0.2~[2\pi/b]$ for the $(100)$ surface, i.e.~along the light green line of (a). (d) Surface BdG spectrum as a function of $k_x$ for fixed $k_y=0.1~[2\pi/b]$ for the $(001)$ surface, i.e.~along the light green line of (b).}
\end{figure}
Here we illustrate this by computing the surface spectral weight of the point-nodal state $\Gamma_{d}^+$ by solving the BdG equations in a slab geometry. The spectral weight at zero energy is shown in Fig.~\ref{BTRS_surface_spec}(a) for a $(100)$-surface and in Fig.~\ref{BTRS_surface_spec}(b) for a $(001)$-surface. Both reveal surface Fermi arcs (red lines). We also show the BdG spectrum along one-dimensional cuts of the surface BZ: in Fig.~\ref{BTRS_surface_spec}(c) for the $(100)$-surface at fixed $k_y=-0.2~[2\pi/b]$ and in Fig.~\ref{BTRS_surface_spec}(d) for the $(001)$-surface  at fixed $k_y=0.1~[2\pi/b]$. Contrary to the case with TRS and chiral symmetry, for each surface momentum $\boldsymbol{k}_{\parallel}$ there is now a single surface branch crossing the gap.\footnote{Since chiral symmetry is absent, the BdG spectrum is not symmetric under $E_n \rightarrow -E_n$ at a given $\boldsymbol{k}_{\parallel}$. However, particle-hole symmetry still imposes the symmetry of the spectrum under $E_n(\boldsymbol{k}_{\parallel}) \rightarrow -E_n(-\boldsymbol{k}_{\parallel}) $.} 
We highlight in green the branches corresponding to the surface spectral weight of Figs.~\ref{BTRS_surface_spec}(a) and (b), while the other sub-gap branches are states localized at the opposite surface of the slab. It it is clear for Fig.~\ref{BTRS_surface_spec}(c,d) that the surface Fermi arcs come from sub-gap branches that cross the gap non-trivially, such that they cannot be removed without closing the bulk gap. 

Projecting the bulk nodal points, see Fig.~\ref{BTRS_ab}(b), and the schematic Berry flux lines, see Fig.~\ref{BTRS_points}(a), onto the surface BZ for the $(100)$- and $(001)$-surfaces, we find an exact agreement with the positions of the surface Fermi arcs found in Figs.~\ref{BTRS_surface_spec}(a) and (b), respectively. In particular, the projected nodal points, depicted by white dots in Fig.~\ref{BTRS_surface_spec}), act as start/end points for the Fermi arcs. Note that for the $(100)$-surface, pairs of nodal points with identical charge are projected on top of each other such that each projected point is the origin of two different Fermi arcs: this leads to the apparent closed Fermi loops of Fig.~\ref{BTRS_surface_spec}(a). The situation is clearer for the $(001)$-surface, which has disconnected surface Fermi arcs.

\section{Topological phase transition}\label{Top_PT}
Having analyzed in detail all the TRS and BTRS states appearing in the self-consistently calculated phase diagram of the $t-J$ model,\cite{SBBS_16} we turn our attention to the topological phase transition between the TRS fully gapped $\Gamma^+_{1,a}$ and the line-nodal $\Gamma^+_{1,b}$ states. This phase transition is conceptually interesting because it is a transition between a fully gapped and a nodal state and, as such, a transition that changes the topology of the system. But even more interestingly, these two phases dominates the phase diagram and thus this phase transition to a large degree determines the overall phase diagram.
In this section we are able to show that this topological phase transition between the  $\Gamma^+_{1,a}$ and $\Gamma^+_{1,b}$ can actually be fully determined analytically by using only topological arguments. We then show how simple energy arguments further support this result and can also be extended to explain why the topological phase transition in a self-consistent phase diagram often will be hidden behind an intermediary phase of the fully gapped $\Gamma^+_{1,c}$ with BTRS.

\subsection{Analytical phase diagram from topological arguments}
We start by only considering the fully gapped $\Gamma^+_{1,a}$ and the line-nodal $\Gamma^+_{1,b}$ phases with TRS. 
Since they are both realized within the trivial irreducible representation, the nodal lines are necessarily accidental, i.e.~they are not imposed by symmetry, and appear at general positions of the BZ. As a consequence, the presence of the nodal lines cannot be directly deduced from the normal band structure. Nevertheless, combining group theory and topology we here derive an analytical condition for the existence of the nodal lines. Based on this the phase diagram can be analytically constructed.
As pointed out earlier, assuming $J$ equal on every NN bond in the self-consistent equation Eq.~(\ref{self_eq}) leads to stable pairing states that all satisfy $\vert \Delta_{h}\vert = \vert \Delta_{z}\vert$. This is relevant when the horizontal NN bonds and the zigzag NN bonds have the same length. However, SG70 is compatible with horizontal and zigzag bonds of different lengths, leading to different NN coupling constants, i.e.~$J_h \neq J_z$. 
To allow for general results we therefore expand the parameter space and consider both on-site pairing and different NN pairing strengths on the zigzag and horizontal bonds, thus modeling all possibilities within a NN model. 
This results in seeking the topological phase transition between fully gapped and line-nodal phase in terms of all the parameters $\{\mu,t,\Delta_0, \Delta_{h}, \Delta_{z}\}$. 

Since the topological phase transition is marked by the appearance/disappearance of nodal lines, the winding number in  Eq.~(\ref{WN_FS}) should be a convenient indicator of the phase transition.
Choosing the base loop $\mathcal{L}_{S} = \Gamma YZ_2T_2\Gamma$, see purple loop in Fig.~\ref{fig_nodal}(b), the winding number $\nu[\mathcal{L}_{S}]$ counts the total number of signed nodal lines crossing the area encircled by the loop, $\nu[\mathcal{L}_{S}] = n_+ - n_-$, where $n_{\pm}$ is the number of nodal lines crossing the area with the chirality $\pm1$. Since all the lattice symmetry operators commute with the chiral symmetry operator, the matrix $D$ in Eq.~(\ref{chiral_blockoff}) takes a block-diagonal form over high-symmetry points of the BZ. Accordingly, $\mathcal{L}_{S}$ has been chosen along high-symmetry lines of the BZ, hence simplifying the computation of the winding number and allowing for an analytical solution. While the calculation is straightforward, it is not particularly enlightening and we refer to Appendix \ref{PT_app} for the details in deriving the analytical expression for the winding number. Here, let us report the results and we first consider the situation 
\begin{equation}
\label{alpha}
	(\Delta_{z},\Delta_{h}) = \Delta (\alpha,1)\;,
\end{equation}
i.e.~we fix $\Delta_{h}=\Delta>0$ and focus on the phase transition between the fully gapped and the line nodal states as a function of $\alpha = \Delta_{z}/\Delta \in [-1,1]$. The range $1>\alpha>0$ can be phenomenologically interpreted as a result of compressing the zigzag bonds while keeping the horizontal bonds unchanged. 

\begin{figure}[t]
\centering
\begin{tabular}{c} 
	\begin{overpic} [width=0.68\linewidth]{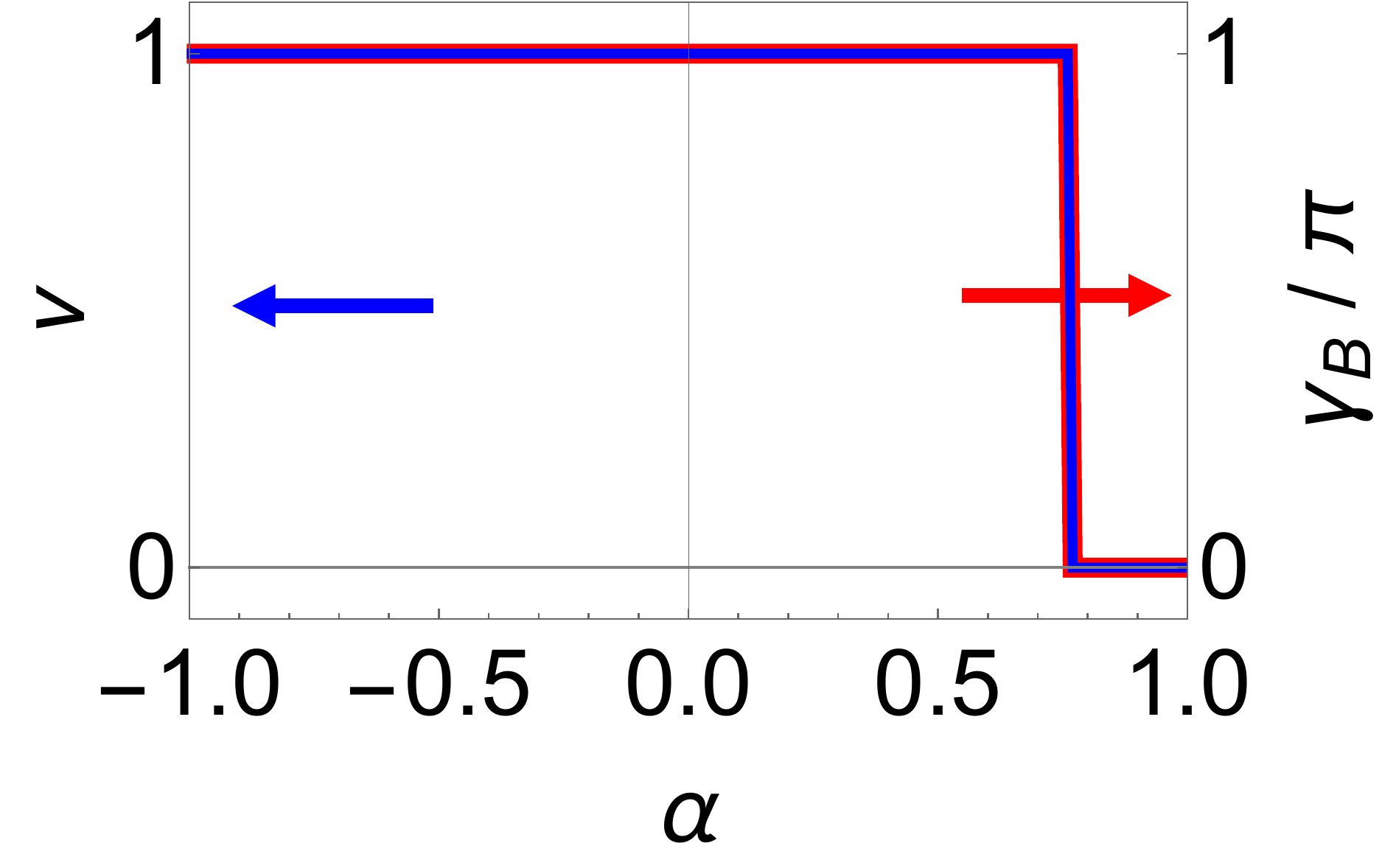}
 		\put (0,0) {(a)}
	\end{overpic} \\
	\begin{overpic} [width=0.6\linewidth]{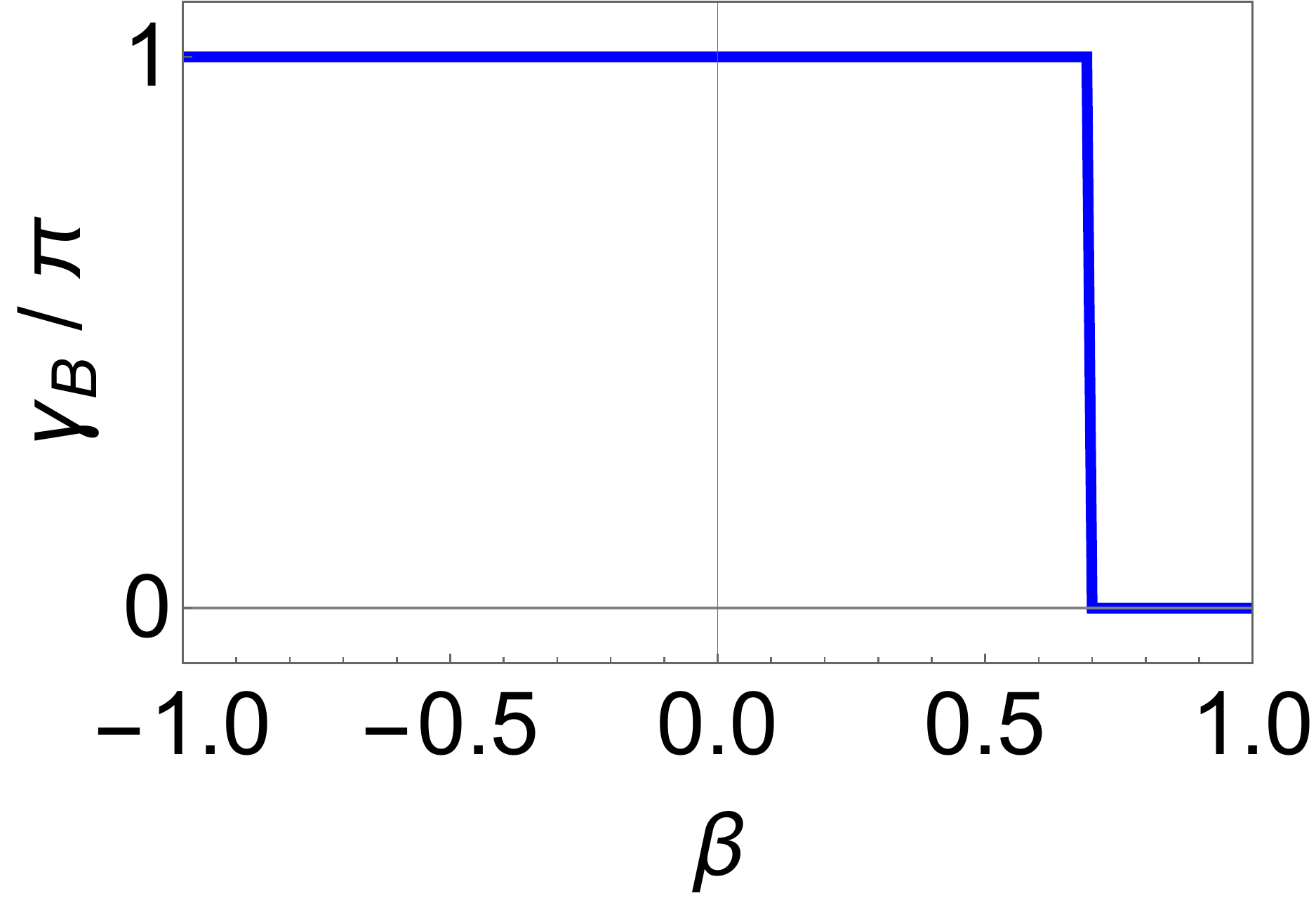}
 		\put (0,0) {(b)}
	\end{overpic} 
\end{tabular}
\caption{\label{phase_transition_G1p} (a) Winding number (blue) and Berry phase (red) computed over the base loop $\mathcal{L}_{S}=\Gamma Y Z_2 T_2 \Gamma$ [see Fig.~\ref{fig_nodal}(b)] as a function of $\alpha=\Delta_{z}/\Delta$ for $(\mu,t,\Delta,\Delta_0) = (0.06,0.2,0.1,0)$. Both capture the topological phase transition between the fully gapped state $\Gamma_{1,a}^+$ ($\alpha=+1$) and the line nodal state $\Gamma_{1,b}^+$ ($\alpha=-1$) at $\alpha_c \approx 0.77$. (b) Berry phase computed over $\mathcal{L}_{S}$ as a function of $\beta=\Delta_{h}/\Delta$ capturing the topological phase transition at $\beta_c = 0.7$ for the same choice of remaining parameters as in (a). 
}
\end{figure}

We plot in Fig.~\ref{phase_transition_G1p}(a) the analytically derived winding number (blue line) over the loop $\mathcal{L}_{S}$ as function of $\alpha$ with the other parameters are chosen as $(\mu,t,\Delta,\Delta_0) = (0.06,0.2,0.1,0)$. There is a clear jump in the winding number, marking the transition between fully gapped and line-nodal phases. From this we can directly define a critical value $\alpha_c$. For $\alpha > \alpha_c$, we find $\nu[\mathcal{L}_{S}] = n_+ - n_- = 0$ such that only zero or pairs of nodal lines of opposite chirality are allowed to cross the area bounded by $\mathcal{L}_{S}$. Since the extended $s$-wave phase is realized with a fully gapped BdG spectrum for $\alpha =1$, we conclude that the phase must have zero nodal lines and be fully gapped for $\alpha > \alpha_c$. For $\alpha < \alpha_c$, we find $ \nu[\mathcal{L}_{S}]  = n_+ - n_- = 1$, such that an odd number of nodal lines must cross the area bounded by $\mathcal{L}_{S}$. Since the simultaneous creation of three distinct nodal lines is impossible without exceptional conditions (e.g.~lattice symmetries), we conclude that a single nodal line is present inside the loop $\mathcal{L}_{S}$ for all values of $\alpha < \alpha_c$. 
Figure~\ref{phase_transition_G1p}(a) also shows the Berry phase (red) computed numerically over the same closed loop $\mathcal{L}_{S}=\Gamma Y Z_2 T_2 \Gamma$ and as a function of $\alpha$. We see a perfect match with the analytical winding number (blue line) as expected. 
We also derive in Appendix \ref{PT_app} the analytical expression of $\alpha_c$.~Choosing $\Delta_0$, $\Delta_{h}$ and $\Delta_{z} \in \mathbb{R} $, done by fixing the free U(1)-gauge phase of TRS states, and assuming $t \geq \mu\geq0$, $\Delta \geq \Delta_0 \geq 0 $, we get
\begin{equation}
\label{alpha_critical}
	\alpha_c = \dfrac{ t(\Delta + \Delta_0) }{ \Delta (t+\mu)} \;. 
\end{equation}
Taking $(\mu,t,\Delta,\Delta_0) = (0.06,0.2,0.1,0)$ as in Fig~.\ref{phase_transition_G1p}(a), we obtain $\alpha_c \approx 0.77$, which perfectly matches the jump of the winding number and the Berry phase.

Let us next consider the complementary region to Eq.~(\ref{alpha}) by setting
\begin{equation}
\label{beta}
	(\Delta_{z},\Delta_{h}) = \Delta (1,\beta) \;,
\end{equation}
i.e.~we fix $\Delta_{z}=\Delta>0$ and focus on the phase transition between the fully gapped and the line nodal states as a function of $\beta = \Delta_{h}/\Delta \in [-1,1]$. Similarly as to before, $1>\beta>0$ can be interpreted as resulting from the compression of the horizontal bonds while keeping the zigzag bonds unchanged. We show in Fig.~\ref{phase_transition_G1p}(b) the Berry phase computed over $\mathcal{L}_{S}$ as a function of $\beta$ for the same choice of remaining parameters as in Fig.~\ref{phase_transition_G1p}(a). We now find a topological phase transition at $\beta_c=0.7$ between the fully gapped and the line-nodal phases. Assuming again $t \geq \mu\geq0$, $\Delta \geq \Delta_0 \geq 0 $, we find an analytical critical point between the fully gapped phase and the line nodal phase (see Appendix \ref{PT_app})
\begin{equation}
\label{beta_critical}
	\beta_c = \dfrac{ t \Delta_0 + t \Delta- \mu \Delta }{t \Delta} \;, 
\end{equation}
which matches perfectly the $\beta_c =0.7$ in Fig.~\ref{phase_transition_G1p}(b) for that set of parameters.
Note especially here that $\beta_c$ is not identical to $1/\alpha_c$ as they capture two distinct phase transitions. 

%

\begin{figure}[t]
\centering
\begin{tabular}{c} 
	\begin{overpic} [width=0.5\linewidth]{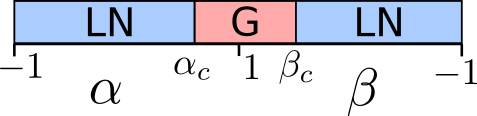}
 		\put (-10,0) {(a)}
	\end{overpic} \\
	\begin{overpic} [width=0.7\linewidth]{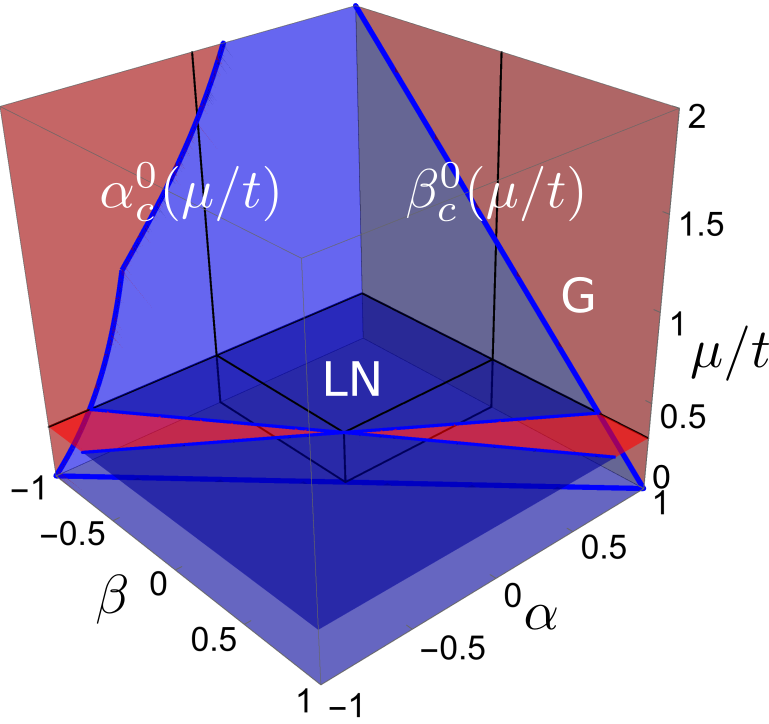}
 		\put (-10,5) {(b)}
	\end{overpic}
\end{tabular}
\caption{\label{phase_transition} Non-self-consistent phase diagram for the fully gapped (G, blue) and line-nodal (LN, red) phases with TRS as a function of $\alpha=\Delta_{z}/\Delta$ and $\beta=\Delta_{h}/\Delta$ with $\alpha,\beta\in[-1,1]$, and $\mu/t$. (a) Phase diagram when $\alpha$ and $\beta$ are changed separately for fixed $\mu/t=0.3$, $\Delta_0 = 0$ and $\Delta=0.5t$. (b) Complete phase diagram for $\alpha$, $\beta$ and $\mu/t$. Imbedded horizontal plane (darker colors) gives phase diagram at the fixed doping $\mu/t=0.3$. Remaining parameters as in (a).
}
\end{figure}

The results as a function of $\alpha$ and $\beta$ can be combined into the schematic phase diagram in Fig.~\ref{phase_transition}(a) where LN marks the line-nodal phase (blue) and G marks the fully gapped phase (red) found when $\alpha$ and $\beta$ are varied independently. This shows that symmetry allowed deformation which change the length between zigzag and horizontal bonds (in either way) can trigger a gapped to line-nodal topological phase transition. 
Focusing on the case relevant for the $t-J$ model with $\Delta_0=0$, the separate critical points simplify to
\begin{equation}
\label{criticals}
\begin{aligned}
	\alpha_c[\Delta_0=0] = \alpha^0_c(\mu/t)&= \dfrac{ 1   }{  1+(\mu/t)} \\
	 \beta_c[\Delta_0=0] = \beta^0_c(\mu/t)&= 1 - (\mu/t) \;,
\end{aligned}
\end{equation}
which gives $\alpha^0_c(-\mu/t) = 1/\beta^0_c(\mu/t)$, and hence recovering the particle-hole symmetry. 

From this we can directly conclude that increasing the doping $\mu$ leads to the decreasing of the value of the critical points, $\alpha_c$ and $\beta_c$. Thus the nodal phase is suppressed as we increase the doping, and we start to favor the fully gapped state at higher doping levels. This is in fact exactly what has previously been found in the self-consistent phase diagram; the fully gapped state was found at high doping level, while the line-nodal state resides at lower doping levels instead.\cite{SBBS_16} By simply analyzing the topology of non-self-consistent but generic solutions, we have thus been able to derive the overall structure of the phase diagram of the fully self-consistent solution. Moreover, the fact that the gapped region is always limited to a region around the symmetric lattice (zigzag and horizontal bonds equal), makes the hyperhoneycomb lattice very prone to topological phase transitions driven by lattice deformations.

The full phase diagram for when $\alpha$, $\beta$, and $\mu/t$ are all allowed to vary is presented in Fig.~\ref{phase_transition}(b).
The critical line $\alpha^0_c(\mu/t)$ (and $\beta^0_c(\mu/t)$) is now a section of the critical surface of the three-dimensional phase diagram between the two states G and LN. The imbedded horizontal plane shows the phase diagram at the fixed doping $\mu/t=0.3$. While Eqs.~(\ref{alpha_critical}) and (\ref{beta_critical}) were derived under the assumption that $t\geq \mu \geq 0$, we show in Appendix \ref{PT_app} that for $2t\geq \mu > t $, on the one hand, $\beta_c$ is unchanged (Eq.~(\ref{beta_critical})), and, on the other hand, $\alpha_c$ is now obtained as the root of a cubic polynomial (see Appendix \ref{PT_app} for more details). This phase diagram confirms that (i) the symmetry allowed lattice deformation of the hyperhoneycomb lattice triggers a topological G-LN phase transition, and (ii) the non-self-consistent phase diagram predicts the trend in the self-consistent phase diagram,\cite{SBBS_16} where the fully gapped phase is favored as the doping increases. 
We finally remark that the phase diagram in Fig.~\ref{phase_transition}(b) is symmetric under the inversion' $(\alpha,\beta) \rightarrow (-\alpha, -\beta)$, which corresponds to a U(1) gauge transformation leaving the BdG spectrum unchanged. We also note that the method presented here, entirely based on topological invariants, is straightforwardly applicable to more complicated models where arbitrary hoping and pairing terms are taken into account and can correctly predict the phase diagram without cumbersome self-consistent calculations.

\subsection{Phase diagram from energy arguments}\label{pheno}
Above we  established using only topology how the gapped $\Gamma^+_{1,a}$ state is favored at higher doping level compared to the line-nodal state $\Gamma^+_{1,b}$. In fact, it is possible to even further understand why this is the case using simple energy arguments. And also why there is an intervening BTRS state overshadowing the topological phase transition between them in a fully self-consistent solution. 

As we have pointed out in Section \ref{model} the normal Hamiltonian Eq.~(\ref{H0}) exhibits a Fermi nodal line at half-filling ($\mu=0$) located on the $k_x=0$ plane.\cite{Ezawa_HHL} We have also noted that due to chiral symmetry the BdG spectrum can be obtained from the equation $\mathrm{det}[D(\boldsymbol{k})\cdot D(\boldsymbol{k})^{\dagger} - E^2 \mathbb{I}_{4\times4}]=0$, where $D(\boldsymbol{k}) = H_{\Delta}(\boldsymbol{k})+i H_{0}(\boldsymbol{k})$. 
Let us first consider the state $\Gamma_{1,a}^+$ where we set every NN bond pairing parameters identical, i.e.~$\Delta_h=\Delta_z=\Delta$, hence the $\boldsymbol{k}$-dependence of the hopping terms and the pairing terms are the same. This we can formally write $H_{\Delta}[\Delta_0,\Delta](\boldsymbol{k}) = H_{0}[-\mu\rightarrow\Delta_0;t\rightarrow\Delta](\boldsymbol{k})$, by which we mean, take Eq.~(\ref{H0}) for the normal state and simply substitute the parameters. 
As a consequence, $D_{\Gamma_{1,a}^+}(\boldsymbol{k}) = H_{0}[-\mu\rightarrow\Delta_0-i\mu;t\rightarrow\Delta+it](\boldsymbol{k})$. Now setting $\Delta_0=\mu=0$, we find 
\begin{align*}
D_{\Gamma_{1,a}^+}(\boldsymbol{k}) D^{\dagger}_{\Gamma_{1,a}^+}(\boldsymbol{k}) &= (\Delta^2+t^2)H_{0}[t\rightarrow1](\boldsymbol{k}) H^{\dagger}_{0}[t\rightarrow1](\boldsymbol{k}) \\
&= (\Delta^2+t^2)H^2_{0}[t\rightarrow1](\boldsymbol{k})\;.
\end{align*} 
The BdG spectrum is thus given by the solution of  $\mathrm{det}\{\pm \sqrt{\Delta^2+t^2} H_{0}[t\rightarrow1](\boldsymbol{k}) - E \mathbb{I}_{4\times4}\}=0$, which in turn is readily given by the band structure of the normal state with a global scaling in energy by $\sqrt{1+(\Delta/t)^2}$. 
This result both verifies that the  $\Gamma_{1,a}^+$ states has ``extended $s$-wave'' symmetry and that the superconducting order becomes a hidden order, i.e.~not visible in the energy spectrum, at half-filling. As a direct consequence, the condensation energy gained when entering the superconducting phase is heavily reduced for this state. 
On the contrary, the line-nodal state $\Gamma_{1,b}^+$ fully gaps out the Fermi nodal line at half-filling. Hence, even though the $\Gamma_{1,b}^+$ state has two nodal lines at general positions of the BZ at half-filling, it features a strong gain in condensation energy by gapping out the Fermi nodal line in the normal state. These simple energy arguments explain why the line-nodal state is favored at low doping. At higher doping the fully gapped $\Gamma_{1,a}^+$ state has a superconducting energy gap throughout the BZ and thus the $\Gamma_{1,b}^+$ state with its line nodes is having the relatively smaller condensation energy in this doping regime. In summary, this shows that by knowing both the normal state and superconducting nodal lines and points, a qualitative phase diagram can be constructed, fully consistent with both the topological derivation presented in the previous subsection and earlier self-consistent calculations. 

We can also understand the BTRS solution $\Gamma^+_{1,c}$ by a very similar energy argument. 
In Section \ref{BTRS_states} we showed that the fully gapped state $\Gamma^+_{1,c}$ with BTRS is the simplest extrapolation in terms of real space order parameters between the line-nodal and fully gapped with TRS. It is then very natural that this state can be allowed to appear as an intermediary phase in-between the two TRS states. 
Indeed, the $\Gamma^+_{1,c}$ state allows a energy compromise at intermediary values of the doping between the line-nodal phase, where the line nodes cost condensation energy, and the extended $s$-wave state with a gap that is still borderline too small to be stabilized. By breaking TRS the $\Gamma^+_{1,c}$ allows a fully gapped phase that becomes the energetically most favorable state in topological phase transition region.

\section{Conclusion and outlook}\label{conclusions}
We here first conclude our results and then discuss their stability under generalizations to more long-range tight-binding models and including spin-orbit coupling.
In this work we study superconducting pairing in the bare hyperhoneycomb lattice, focusing on spin-singlet pairing. An earlier numerical mean-field study has found a very rich spin-singlet phase diagram with multiple stable states\cite{SBBS_16} and we here analyze in detail the topological properties of these states and the topological phase transitions in-between them. We reveal in this work that the line nodal phase with TRS dominating low doping, $\Gamma^+_{1,b}$ belongs to the topological class CI and exhibits surface Majorana flat bands. We show the bulk-boundary correspondence in terms of the $\mathbb{Z}_2$ quantized Berry phase by finding a perfect match between the bulk prediction of the Majorana states and the actual surface spectrum computed numerically for several slab geometries. 
We also find that the point nodal phase breaking TRS at very low doping levels, $\Gamma^+_d$ belongs to the Altland-Zirnbauer topological class D and, in analogy with Weyl semimetals, exhibits surface Fermi arcs. We compute the bulk Chern numbers and determine the qualitative geometry of the Berry flux lines as constrained by the symmetries of the system. From these, we show a perfect match between the prediction of Fermi arcs from the bulk-boundary correspondence and the surface spectra obtained numerically. We also establish that the two fully gapped states, $\Gamma^+_{1,a}$ at high doping levels with TRS and $\Gamma^+_{1,c}$ with BTRS find in a sliver at intermediary doping levels, has only trivial topology.

Having established the topology of all stable superconducting states we study the topological phase transition between the two dominating states: the fully gapped $\Gamma^+_{1,a}$ phase with TRS and the line-nodal $\Gamma^+_{1,b}$ phase with TRS. Using only symmetry and topology we extract completely analytically the whole phase diagram as a function of the paring parameters and doping. From this we predict not only the phase transition with increasing doping from line-nodal to fully gapped state as also found previously,\cite{SBBS_16} but also that the same transition occurs readily by symmetry-allowed lattice deformations compressing either the horizontal or zigzag bonds. Thus, the fully gapped state is only found at higher doping levels and around a point where the lattice is as most symmetric.
Extending the argument to also include energy considerations in conjunction with the global band topology of the normal state, we are able to predict that the fully gapped $\Gamma^+_{1,c}$ state with BTRS appears at intermediary doping levels, usually covering the topological phase transition between the low-doping line-nodal $\Gamma^+_{1,b}$ and high-doping fully gapped $\Gamma^+_{1,a}$ states. By only using general symmetry, topology, and energy arguments to derive the qualitative phase diagram we automatically establish that its overall properties are very general and therefore remarkably insensitive to the details of the materials and models. 

While we establish that superconducting phase diagram is very general, the underlying symmetry and topology arguments we use technically fail if the effective model changes symmetry or topological class. As a final discussion we provide a brief account to show that our results are likely largely unchanged despite this.
First consider more general models including an arbitrary number of neighbors in the tight-binding model. In order to address this question properly, the global band topology of the normal state must first be considered. An exhaustive discussion is beyond the scope of this work, however, as long as the number of degrees of freedom are conserved (four sub-lattice sites and no spin-orbit-coupling), the three non-commuting glide symmetries of SG70 impose the existence of a nodal line at half-filling since the nodal line connects two unoccupied bands and two occupied bands.\cite{ABABS_HHL_2} The difference with the model considered here, Eq.~(\ref{H0}), is that allowing terms beyond the NN terms can break the artificial sub-lattice symmetry at half-filling such that the nodal line does not appear at constant energy.\cite{Ezawa_HHL} This leads at exactly half-filling to a toroidal Fermi surface with point nodal bottlenecks\cite{Thomas_line,ABABS_lines} or Fermi cyclides.\cite{Ahn_cyclides} Still, the toroidal Fermi surface is recovered at a finite amount of doping. Thus, while the topology of the Fermi surface is changed from half-filling up to this threshold value of the doping, the phase diagram beyond this threshold is unchanged. Considering that at low doping levels there is already the interfering $\Gamma^+_d$ disrupting the general competition between the nodal-line and fully gapped TRS states, this at most produces only small alteration in a small part of the phase diagram.

An other generalization comes from including spin-orbit-coupling. Since SG70 has inversion symmetry, only Kane-Mele type spin-orbit-coupling is allowed. Since strong spin-orbit-coupling changes the normal band topology and induces strong spin-triplet pairing, it is beyond the scope of this work to give a detailed account.\footnote{Contrary to the two-dimensional honeycomb lattice, the SU(2)-spin symmetry is fully broken in the hyperhoneycomb lattice since the second neighbor hopping processes cannot be coplanar. Also, since the spinless Fermi surface lies at general positions of the BZ, spin-orbit-coupling gaps out the Fermi surface at half-filling possibly leading to a topological insulator.\cite{Kim_HHL_topins} } 
However, we can still determine the effect of an adiabatic switching on of spin-orbit-coupling and thus assume that the topology of the normal Fermi surface is intact (note the Kramers degeneracy of the normal bands due to TRS and inversion symmetry) and the pairing remains within the spin-singlet channel.

On one hand, the line-nodal paring state, originally in class CI, conserves the Kramers degeneracy of the normal state and thus introducing spin-orbit-coupling, resulting in class DIII, does not gap out the nodal lines. We have confirmed this prediction numerically: a rather large value of spin-orbit-coupling must be used in order to reshape the normal band structure and remove the nodal lines in the pairing state. We thus conclude that the nodal lines of a centro-symmetric spin-singlet pairing state are robust under the change of class CI$\rightarrow$DIII. In fact, it has been shown that class DIII supports robust line nodes in arbitrary pairing channels, making this result likely more general.\cite{Sato11} As a consequence, the overall competition between the line-nodal and gapped states with TRS which are dominating the phase diagram are stable for small spin-orbit coupling.

On the other hand, the point-nodal state with BTRS in class C at very low doping levels has no Kramers degeneracy. Therefore, by introducing spin-orbit-coupling, resulting in class D, we expect the point nodes to not be robust. Actually, it has been shown recently that with inversion symmetry the class D supports monopole nodal surfaces.\cite{Agterberg_BdGsurface, Bzdusek_mulitnodes} We find numerically that by introducing spin-orbit coupling the point nodes of the state with BTRS are inflated into nodal surfaces. This will change the phase diagram, but again only at very low doping levels. Based on these considerations we argue that the overall structure of the phase diagram that we establish in this work based on symmetry and topology considerations is remarkably stable even beyond the formal requirements. This opens the door for a generic superconducting phase diagram for hyperhoneycomb materials.

\begin{acknowledgments}
We thank thank A.~Furusaki, M.~Sato, and A.~P.~Schnyder for insightful discussions at the early stage of this project, and E.~Sj{\"o}qvist for advice on the computation of the Berry phase. A. B.~would also like to thank J.~Goryo and M.~Takafumi for their very kind and generous hospitality at Hirosaki University where part of this work was performed.
\end{acknowledgments}

\appendix

\section{Tight-binding Bogoliubov-de Gennes Hamiltonian from symmetry}\label{app_lattice}
In this appendix we provide an alternative form of the BdG Hamiltonian in Eqs.~(\ref{H0}) and (\ref{Hoff}) based on the lattice symmetries. This is used in the analytical derivation of the winding number and critical points in Appendix \ref{PT_app}. The hyperhoneycomb lattice corresponds to the nonsymmorphic space group $Fddd$ (SG70). Choosing the center of point symmetry at the middle of a horizontal bond, SG70 can be decomposed into cosets with respect to the representatives of $D_{2h}$ as
\begin{align}
	Fddd &= \left\{ E \vert \boldsymbol{0}\right\}\mathbf{T} \oplus
	 \left\{ C_{2z} \vert \boldsymbol{0}\right\}\mathbf{T} \oplus
	  \left\{ C_{2y} \vert \boldsymbol{0}\right\}  \mathbf{T}\nonumber\\
	   &\oplus\left\{ C_{2x} \vert \boldsymbol{0}\right\} \mathbf{T}\oplus 
	    \left\{ I \vert \boldsymbol{\tau}\right\} \mathbf{T}\oplus
	     \left\{ m_z \vert \boldsymbol{\tau}\right\}\mathbf{T}  \\
	      &\oplus\left\{ m_y \vert \boldsymbol{\tau}\right\}\mathbf{T} \oplus
	       \left\{ m_x \vert \boldsymbol{\tau}\right\}\mathbf{T}   \;,\nonumber
\end{align}
with $\mathbf{T}$ the Abelian normal subgroup of Bravais translations, i.e.~$ \cup_{\forall n,m,l \in \mathbb{N}} ~n \boldsymbol{a}_1+m \boldsymbol{a}_2+l \boldsymbol{a}_3$, and the fractional translation $\boldsymbol{\tau} = (\boldsymbol{a}_1+ \boldsymbol{a}_2+\boldsymbol{a}_3)/4=(a/4,b/4,c/4) $. 
We then introduce a symmetrized basis set according to the irreducible representation of $D_{2h}$, given in Table~\ref{chartab_D2h}, which gives
\begin{eqnarray}
\label{sym_basis}
	\hat{\boldsymbol{\Psi}}^{\dagger T}_{\boldsymbol{k}} &= &   \left( \hat{\psi}^{\dagger}_{\Gamma_1^+,\boldsymbol{k}} ,
	\hat{\psi}^{\dagger}_{\Gamma_1^-,\boldsymbol{k}} ,
	\hat{\psi}^{\dagger}_{\Gamma_4^+,\boldsymbol{k}} ,
	\hat{\psi}^{\dagger}_{\Gamma_4^-,\boldsymbol{k}} \right)^T  =  \hat{\boldsymbol{C}}^{\dagger T}_{\boldsymbol{k},\uparrow}
	 U_{0,S}\;,\\
	U_{0,S} &=& \dfrac{1}{2}\left(\begin{array}{rrrr} 
		1 & 1 & 1 & 1 \\
		1 & 1 & -1 & -1 \\
		1 & -1 & -1 & 1 \\
		1 & -1 & 1 & -1 
	\end{array}\right) \;,
\end{eqnarray}
in which the normal part of the tight-binding Hamiltonian takes the form,
\begin{equation}
\label{H0S}
	H_{0,S}(\boldsymbol{k}) = \left(\begin{array}{rrrr}
		h^{(1)}_{11} & h^{(\bar{1})}_{1\bar{1}} & h^{(4)}_{14} & h^{(\bar{4})}_{1\bar{4}} \\
		-h^{(\bar{1})}_{1\bar{1}} & h^{(1)}_{\bar{1}\bar{1}} & h^{(\bar{4})}_{\bar{1}4} & h^{(4)}_{\bar{1}\bar{4}} \\
		h^{(4)}_{14} & -h^{(\bar{4})}_{\bar{1}4} & h^{(1)}_{44} & h^{(\bar{1})}_{4\bar{4}} \\
		-h^{(\bar{4})}_{1\bar{4}} & h^{(4)}_{\bar{1}\bar{4}} & -h^{(\bar{1})}_{4\bar{4}} & h^{(1)}_{\bar{4}\bar{4}}
	\end{array}\right)(\boldsymbol{k}) \;,
\end{equation}
with the elements
\begin{align*}
	h^{(1)}_{11}  &=h^{(1)}_{\bar{1}\bar{1}}  = h^{(1)}_{44} = h^{(1)}_{\bar{4}\bar{4}}  =   - \mu +  t_h  f^{(1)}_{h} + t_z f^{(1)}_{z}  \;,\\
	h^{(\bar{1})}_{1\bar{1}} &= h^{(\bar{1})}_{4\bar{4}}=  i t_z  f^{(\bar{1})}_{z} \;,\\
	h^{(4)}_{14} &= h^{(4)}_{\bar{1}\bar{4}}= t_z  f^{(4)}_{z} \;,\\
	h^{(\bar{4})}_{1\bar{4}} &= h^{(\bar{4})}_{\bar{1}4} = i t_h f^{(\bar{4})}_h + i t_z  f^{(\bar{4})}_{z}  \;,
\end{align*}
and the functions 
\begin{align*}
 	f^{(1)}_{h}(\boldsymbol{k}) &= \cos \boldsymbol{\delta}_a\cdot \boldsymbol{k} \;,\\
 	f^{(\bar{4})}_{h}(\boldsymbol{k}) &= \sin \boldsymbol{\delta}_a\cdot \boldsymbol{k} \;,\\
	f^{(1)}_{z}(\boldsymbol{k}) &=  \cos \boldsymbol{\delta}_b\cdot \boldsymbol{k}+  \cos \boldsymbol{\delta}_c\cdot \boldsymbol{k} +  \cos \boldsymbol{\delta}_d\cdot \boldsymbol{k} +  \cos \boldsymbol{\delta}_e\cdot \boldsymbol{k}\;,\\
	f^{(\bar{1})}_{z}(\boldsymbol{k}) &= \sin \boldsymbol{\delta}_b\cdot \boldsymbol{k}+  \sin \boldsymbol{\delta}_c\cdot \boldsymbol{k} -  \sin \boldsymbol{\delta}_d\cdot \boldsymbol{k} -  \sin \boldsymbol{\delta}_e\cdot \boldsymbol{k}\;,\\
	f^{(4)}_{z}(\boldsymbol{k}) &=  \cos \boldsymbol{\delta}_b\cdot \boldsymbol{k}+  \cos \boldsymbol{\delta}_c\cdot \boldsymbol{k} -  \cos \boldsymbol{\delta}_d\cdot \boldsymbol{k} -  \cos \boldsymbol{\delta}_e\cdot \boldsymbol{k} \;,\\
	f^{(\bar{4})}_{z}(\boldsymbol{k}) &= \sin \boldsymbol{\delta}_b\cdot \boldsymbol{k}+  \sin \boldsymbol{\delta}_c\cdot \boldsymbol{k} +  \sin \boldsymbol{\delta}_d\cdot \boldsymbol{k} + \sin \boldsymbol{\delta}_e\cdot \boldsymbol{k}\;,
\end{align*}
with the NN bond vectors defined in Section \ref{model}. We have here used the shortened notation $\Gamma_j^+ = j$ and $\Gamma_j^- = \bar{j}$, where the lower indices of e.g.~$h^{(\bar{1})}_{1\bar{4}}(\boldsymbol{k}) =h^{(\Gamma_{1}^-)}_{\Gamma_{1}^+,\Gamma_{4}^-}(\boldsymbol{k})$ means that it connects the fields $\hat{\psi}^{\dagger}_{\Gamma_1^+,\boldsymbol{k}}\hat{\psi}_{\Gamma_4^-,\boldsymbol{k}}$, while the upper index means that it is a basis function of the irreducible representation $\Gamma_{1}^-$. In the main text we assume $t_h=t_z=t$ and arrive at Eq.~(\ref{H0}) in the  main text.

The BdG Hamiltonian in the symmetrized basis is given through 
\begin{align}
\label{HBdG_sym}
H_{S}(\boldsymbol{k}) &= U^{\dagger}_{S} H(\boldsymbol{k})U_{S}\;,\\
\label{symmetrized}
U_{S} &= (\sigma_0\otimes U_{0,S})\;.
\end{align}
Assuming that the pairing state belongs to the trivial representation of $D_{2h}$, the off-diagonal part of the BdG Hamiltonian takes a similar form as the normal part Eq.~(\ref{H0S}), i.e.~$H_{\Delta ,S}[\Delta_0;\Delta_h;\Delta_z] (\boldsymbol{k}) = H_{0,S}[-\mu\rightarrow \Delta_0; t_h\rightarrow \Delta_h; t_z\rightarrow \Delta_z](\boldsymbol{k})$.
\begin{table}[t]
\caption{\label{chartab_D2h} Character table of $D_{2h}$.}
\begin{equation*}
\begin{array}{c|rrrrrrrr}
	D_{2h} & E & C_{2z} & C_{2y} & C_{2x} & I & m_z & m_y & m_x \\
	\hline
	\Gamma_{1}^+ & 1 & 1 & 1 & 1 & 1 & 1 & 1 & 1 \\
	\Gamma_{2}^+ & 1 & -1 & 1 & -1 & 1 & -1 & 1 & -1 \\
	\Gamma_{3}^+ & 1 & 1 & -1 & -1 & 1 & 1 & -1 & -1 \\
	\Gamma_{4}^+ & 1 & -1 & -1 & 1 & 1 & -1 & -1 & 1 \\
	\Gamma_{1}^- & 1 & 1 & 1 & 1 & -1 & -1 & -1 & -1 \\
	\Gamma_{2}^- & 1 & -1 & 1 & -1 & -1 & 1 & -1 & 1 \\
	\Gamma_{3}^- & 1 & 1 & -1 & -1 & -1 & -1 & 1 & 1 \\
	\Gamma_{4}^- & 1 & -1 & -1 & 1 & -1 & 1 & 1 & -1 
\end{array}
\end{equation*}
\end{table}
There is an advantage in using the Hamiltonian in the symmetrized form of Eq.~(\ref{HBdG_sym}) with $H_{0,S}$ in Eq.~(\ref{H0S}) and similarly for $H_{\Delta,S}$. Indeed, it takes a block-diagonal form over the high-symmetry regions of the BZ, hence simplifying calculations.~For instance, over the $k_z=0$ plane of the BZ that is invariant under $m_z$, all the terms of $H_S$ that are odd under $m_z$ must vanish, i.e.~$\left.h^{(\bar{1})}_{\mu\nu}(\boldsymbol{k})\right|_{k_z=0}=\left.h^{(4)}_{\mu\nu}(\boldsymbol{k})\right|_{k_z=0} = 0$ for all $\mu,\nu\in \{1,\bar{1},4,\bar{4}\}$.

\section{Class CI}\label{CI}
In this appendix we derive the important properties of class CI which are used in the main text and in the next Section.
Without spin-orbit coupling and as we only consider $s$-wave type electronic orbitals, TRS can be chosen as $\mathcal{T} =  \mathcal{K}$ (complex conjugation) such that $\mathcal{T}^2 = +1$. TRS is then given by $\mathcal{T} H(\boldsymbol{k}) \mathcal{T}^{-1} \cong H(-\boldsymbol{k}) $ which leads to $H^*(-\boldsymbol{k}) \cong H(\boldsymbol{k})$.
\footnote{With the BdG spectrum invariant under global U(1) gauge transformations, i.e.~$\hat{\psi} \mapsto \mathrm{e}^{i\theta}\hat{\psi}$ for every fermion field, we call two Hamiltonians equivalent if $\tilde{H}(\boldsymbol{k}) = U_{\theta}^{\dagger} H(\boldsymbol{k}) U_{\theta} $ with $U_{\theta} = \mathrm{diag}[\mathrm{e}^{-i \theta},\mathrm{e}^{i \theta}] \otimes \mathbb{I}_{4\times 4}$, which we write $\tilde{H} \cong H$. As we define symmetries only up to such a gauge transformation, we use the equivalence relation `$\cong $' instead of the strict equality `$=$'.} 
Moreover, particle-hole symmetry is given by $\mathcal{C} H(\boldsymbol{k}) \mathcal{C}^{-1} \cong - H(-\boldsymbol{k})$ with $\mathcal{C} = C \mathcal{K}$, where $C = -i\sigma_y \otimes \mathbb{I}_{4\times 4}$ and $\mathcal{C}^2 = - \mathbb{I}_{8\times 8} $.
Combining TRS and particle-hole symmetry we readily obtain the chiral symmetry as $\mathcal{J} H(\boldsymbol{k}) \mathcal{J}^{-1} \cong - H(\boldsymbol{k})$ with $\mathcal{J} = i \mathcal{C}\mathcal{T} = i C$ and $\mathcal{J}^2 = \mathbb{I}_{8\times 8} $.

In the basis that makes the chiral symmetry operator diagonal the BdG Hamiltonian takes a block off-diagonal form. This is achieved through
\begin{eqnarray}
\label{chiral_blockoff_B}
	\tilde{H}(\boldsymbol{k}) = U_{\mathcal{J}}^{\dagger} H(\boldsymbol{k}) U_{\mathcal{J}} &=& \left(\begin{array}{cc} 0 & D(\boldsymbol{k})\\
	D^{\dagger}(\boldsymbol{k}) & 0 \end{array}\right)\;, \\
\label{chiral_op}
	U_{\mathcal{J}} &=& (\sigma_0 + i \sigma_x)/\sqrt{2} \otimes \mathbb{I}_{4\times 4} \;,
\end{eqnarray}
where $D(\boldsymbol{k}) = H_{\Delta}(\vk) + i H_{0}(\vk)$. The topological invariants for class CI are then expressed through the matrix $D(\boldsymbol{k})$. For instance, the fully gapped phase in three dimensions is characterized by the winding number
\begin{multline}
	\nu = \int \dfrac{d^3k}{24 \pi^2} \epsilon^{\mu\nu\rho} \mathrm{tr} \left(D^{-1} \partial_{\mu} D \right) \left(D^{-1} \partial_{\nu} D \right)   \left(D^{-1} \partial_{\rho} D \right) \\ \in 2\mathbb{Z} 	\;,
\end{multline}
where $\mu,\nu,\rho=k_x,k_y,k_z$. We obtain that this is aways zero for the TRS fully gapped $\Gamma_{1,a}^+$state on the hyperhoneycomb lattice.

Due to the chiral symmetry and the block off-diagonal form of the BdG Hamiltonian in Eq.~(\ref{chiral_blockoff_B}), the eigenvalue problem 
\begin{equation*}
\mathrm{det}[\tilde{H}(\boldsymbol{k}) - E(\boldsymbol{k}) \mathbb{I}_{8\times8}]=0\;,
\end{equation*} 
readily simplifies to 
\begin{equation*}
\mathrm{det}[D(\boldsymbol{k})\cdot D^{\dagger}(\boldsymbol{k}) - E^2(\boldsymbol{k}) \mathbb{I}_{4\times4}]=0\;.
\end{equation*} 
Since the chiral symmetry operator Eq.~(\ref{chiral_op}) commutes with all the lattice symmetry operators, $D$ in the symmetrized basis given  in Eq.~(\ref{sym_basis}) takes the same form as $H_{0,S}$ in Eq.~(\ref{H0S}) and similarly for $H_{\Delta,S}$,
\begin{equation}
\label{D_sym}
 D_{S}(\boldsymbol{k}) = H_{\Delta,S}(\boldsymbol{k}) + i H_{0,S}(\boldsymbol{k}) \;.
 \end{equation}
Starting from Eq.~(\ref{H_BdG}), we obtain this form through the basis transformation
\begin{equation}
\label{lattice_chiral_basis}
	\left(\begin{array}{c} \hat{\boldsymbol{\Theta}}^{\dagger}_{\boldsymbol{k}} \\ \hat{\boldsymbol{\Theta}}_{-\boldsymbol{k}} \end{array}\right)^T =  \left(\begin{array}{c} \hat{\boldsymbol{\Psi}}^{\dagger}_{\boldsymbol{k}} \\ \hat{\boldsymbol{\Psi}}_{-\boldsymbol{k}} \end{array}\right)^T U_{\mathcal{J}} =  \left(\begin{array}{c} \hat{\boldsymbol{C}}^{\dagger}_{\boldsymbol{k}} \\ \hat{\boldsymbol{C}}_{-\boldsymbol{k}} \end{array}\right)^T U_{S} U_{\mathcal{J}} \;,
\end{equation}
with $U_{S}$ defined in Eq.~(\ref{symmetrized}) and $U_{\mathcal{J}}$ in Eq.~(\ref{chiral_op}).

\section{Analytical derivation of winding number and critical points}\label{PT_app}
In Section \ref{Top_PT} we use the winding number in Eq.~(\ref{WN_FS}) as the indicator of the topological phase transition between the fully gapped and the line-nodal phases with TRS. We derive here the analytical expression of the winding number computed over the base loop $\mathcal{L}_{S} = \Gamma Y Z_2 T_2 \Gamma$ (magenta loop in Fig.~\ref{fig_nodal}(b)). 

In the preceding appendix we have argued that in the basis Eq.~(\ref{lattice_chiral_basis}) the matrix $D_S$ takes the same form as $H_{0,S}$ in Eq.~(\ref{H0S}). Therefore, by choosing a base loop that follows the high-symmetry lines of the BZ, we bring $D_S$ to a block-diagonal form. Since the chosen base loop belongs to a $m_z$-invariant plane ($k_z=0$), we have already noted that the terms odd under $m_z$ must vanish, such that we find
\begin{equation}
	\left.D_{S}\right|_{k_z=0} = 
	\left[\begin{array}{cccc}
		D^{(1)}_{11} & 0 & 0 & D^{(\bar{4})}_{1\bar{4}} \\
		0 & D^{(1)}_{\bar{1}\bar{1}} & D^{(\bar{4})}_{\bar{1}4} & 0 \\
		0 & -D^{(\bar{4})}_{\bar{1}4} & D^{(1)}_{44} & 0 \\
		-D^{(\bar{4})}_{1\bar{4}} & 0 & 0 & D^{(1)}_{\bar{4}\bar{4}}
	\end{array}\right]\;.
\end{equation}
Let us now decompose the base loop as $\mathcal{L}_S = l_1+l_2+l_3+l_4$ with the high-symmetry segments of the BZ: $l_1 = \overline{\Gamma Y}$, $l_2 = \overline{YZ_2}$, $l_3 =\overline{Z_2T_2}$, and $l_4 =\overline{T_2\Gamma}$, see Fig.~\ref{fig_HHL}(b). Since the lines $l_1$ and $l_3$ are invariant under $C_{2y}$, $D_{S}$ takes a diagonal form on these high-symmetry lines (indeed, the terms $D^{(\bar{4})}_{\mu\nu}(\boldsymbol{k})$ are odd under $C_{2y}$ and must then vanish on $l_1$ and $l_3$). 

After some straightforward but tedious algebra we find the winding number
\begin{align}
	\nu[\mathcal{L}_s] &= \dfrac{1}{i 2\pi} \int_{l_1+l_2+l_3+l_4} dq~ \mathrm{tr} D^{-1}_{\mathcal{S}}(q)  \partial_q D_{\mathcal{S}}(q) \nonumber\\
	&= \dfrac{1}{i 2\pi} \left(\Delta I_1+\Delta I_2+ \Delta I_3+ \Delta I_4\right)\;,
\end{align}
where
\begin{align*}
	\Delta I_{1,3} &= I_{1,3}(k_y=1)- I_{1,3}(k_y=0)\;,\\
	\Delta I_{2,4} &= I_{2,4}(k_x=1)- I_{2,4}(k_x=0)\;,
\end{align*}
with
\begin{equation}
\begin{aligned}
	I_1(k_y)  &= \mathrm{Log}[z_{1,a}(k_y) ] + \mathrm{Log}[z_{1,b}(k_y) ] \\
	&+\mathrm{Log}[z_{1,c}(k_y) ] +\mathrm{Log}[z_{1,d}(k_y) ] \;,\\
	I_2(k_x)  &= 0\;, \\
	  I_3(k_y)   &= 2 \mathrm{Log}\left[z_{3}(k_y)\right]  \;, \\
   I_4(k_x) &=  \mathrm{Log}\left[z_{4}(k_x)\right]\;.
\end{aligned}
\end{equation}
\begin{widetext}
and with the complex functions
\begin{equation}
\begin{aligned}
	z_{1,a}(k_y)  &= t + i (\Delta_0 + \Delta_h) + \mu - 2(t+i \Delta_z) \cos (k_y \pi/2) \;,\\
	 z_{1,b}(k_y)&= t - i \Delta_0 +i \Delta_h - \mu + 2(t+i \Delta_z) \cos (k_y \pi/2) \;,\\
	z_{1,c}(k_y)&= -t + i \Delta_0 - i \Delta_h + \mu + 2(t+i \Delta_z) \cos (k_y \pi/2) \;,\\
	z_{1,d}(k_y)&= t + i \Delta_0 +i \Delta_h + \mu + 2(t+i \Delta_z) \cos (k_y \pi/2) \;, \\
	  z_{3}(k_y)   &= 3 t^2 - \Delta_h^2 - 2 \Delta_z^2 + 2 i t (\Delta_h + 2 \Delta_z) + ( \Delta_0 -i \mu)^2 + 2 (t + i \Delta_z)^2 \cos (k_y \pi)  \;, \\
   z_4(k_x) &=  1+ \dfrac{16 (t+i\Delta_h)^2 (t+i\Delta_z)^2 \cos(k_x \pi)}
  {16  (t+i\Delta_h)^2 (t+i\Delta_z)^2  -2 \left(5t^2-\Delta_h^2 -4 \Delta_z^2 + 2 i t (\Delta_h +4 \Delta_z )+ (\Delta_0 - i \mu)^2 \right)}  \;.
\end{aligned}
\end{equation}
\end{widetext}
Here we also recall the definition of the logarithm of a complex number $z$, $\mathrm{Log}~z = \ln r + i \theta$, where $r$ is the norm and $\theta$ is the argument of $z$.

By construction the winding number must be an integer, therefore we can track the topological phase transition through the jumps by $i 2\pi$ of the imaginary parts of the functions $\mathrm{Log}[z_i(k)]$, where $k=k_x,k_y$. Let us consider the smooth graphs in the complex plane of the complex functions $\{z_i(k) = (\Re z_i(k),\Im z_i(k)) \vert  k\in [0,1]\}$ for a fixed set of parameters $\{t,\mu,\Delta_0,\Delta_h,\Delta_z\}$. Whenever one graph crosses the origin of the complex plan under the variation of one parameter, the phase difference $\Delta \theta_i = \mathrm{Arg}~ z_i(k=1)-\mathrm{Arg}~ z_i(k=0)$ must jump by $2\pi$. Therefore, the critical points of the topological phase transition in the parameter space are tracked by the zeros of the complex functions $z_i(k)$. Without loss of generality we take $t , \mu,\Delta,\Delta_0 \geq 0 $, and we also restrict to $\Delta \geq \Delta_0$ valid if electron repulsion is taken into account. We first derive the critical points in the case of $t \geq \mu$, and then we take $2t\geq \mu>t$. 

\subsection{$t \geq \mu$}
First we set $(\Delta_h,\Delta_z) = \Delta(1,\alpha)$ with $\alpha \in [-1,1]$, and concentrate on the critical $\alpha$ at which the topological phase transition takes place. By inspection, we find that only the complex function $z_{1,a}(k_y)$ supports a zero for this choice of parameters. Since the real part does not depend on the parameter $\alpha$, we can set $\Re z_{1,a}(\bar{k}_y)=0$, from which we find $\bar{k}_y = (2/\pi) \mathrm{arccos} \dfrac{t+\mu}{2t}$. Then, substituting $\bar{k}_y$ for $k_y$ in the equation for the imaginary part, i.e.~$\Im z_{1,a}(\bar{k}_y)=0$, we find 
\begin{equation}
\label{alpha_critical_B}
	\alpha_c = \dfrac{ t(\Delta + \Delta_0) }{ \Delta (t+\mu)} \;. 
\end{equation}

We plot in Fig.~\ref{phase_transition_G1p}(a) the real (dashed lines) and imaginary (solid lines) parts of $\mathrm{Log}[z_{1,a}(k_y)]$ as a function of $k_y$ for $\alpha \gtrsim \alpha_c$ (red) and $\alpha \lesssim \alpha_c$ (blue). We see the jump by $2\pi$ in the phase difference of the graphs, i.e.~$\left\vert \Delta \theta_{1,a}^{\alpha \gtrsim \alpha_c} -  \Delta \theta_{1,a}^{\alpha \lesssim \alpha_c}\right\vert =2\pi$ with $\alpha_c \approx 0.769$ in Fig.~\ref{phase_transition_G1p}(a). We note that the divergence of $\Re \mathrm{Log} [z_{1,a}(k_y)]$, i.e.~the norm of $z_{1,a}(k_y)$, at $\bar{k}_y$ also marks the graph of the complex function crossing the origin of the complex plane at $\alpha_c$.
\begin{figure}[t]
\centering
\begin{tabular}{c} 
	\includegraphics[width=0.7\linewidth]{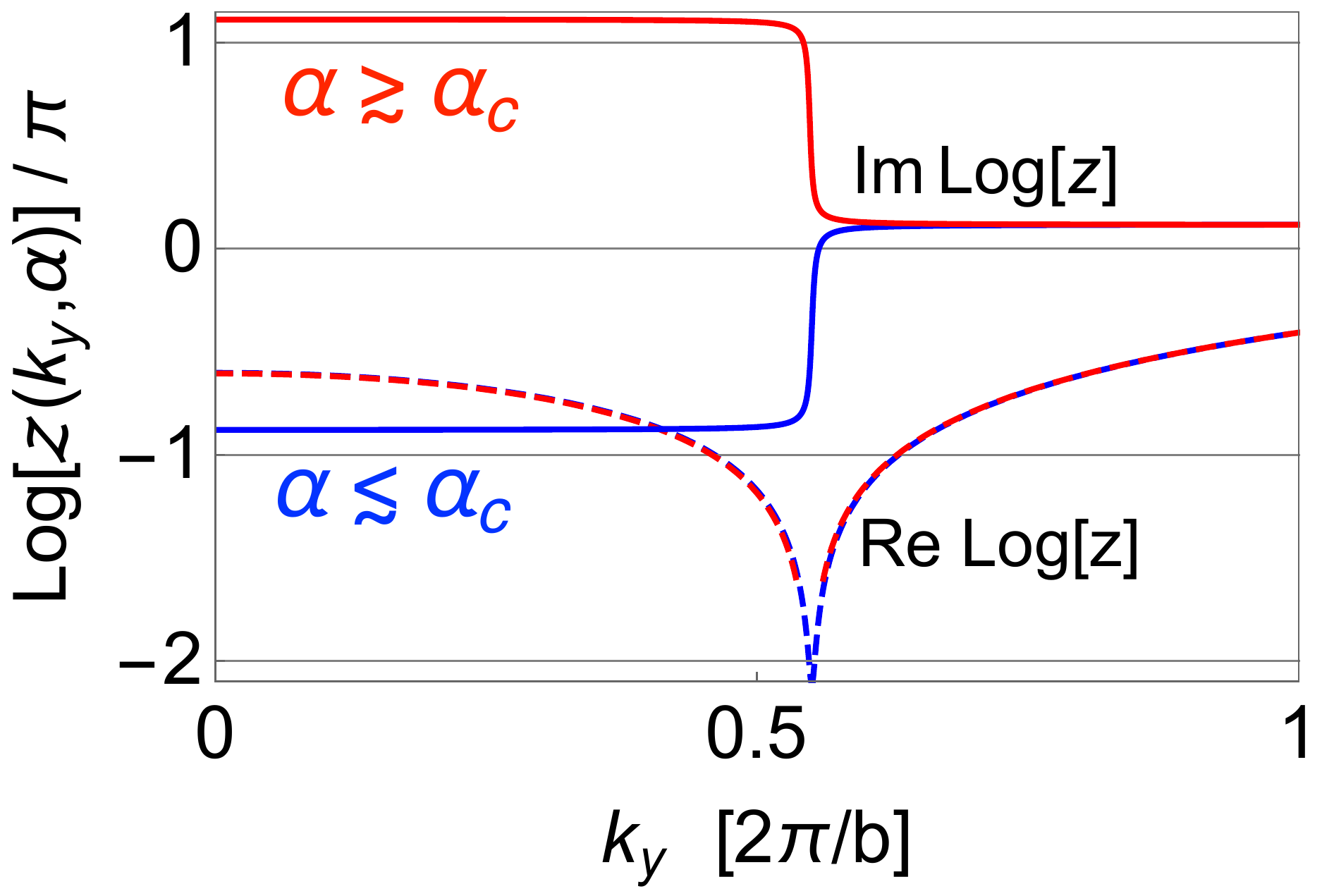}  \\
	(a)
\end{tabular}
\caption{\label{phase_transition_G1p_app} Norm $\Re \mathrm{Log} [z_{1,a}(k_y)]$ (dashed lines) and argument $\Im \mathrm{Log} [z_{1,a}(k_y)]$) (solid lines) of the complex function $z_{1,a}(k_y)$ as a function of $k_y$ for two different values of the parameter $\alpha$ (red and blue). Remaining parameters are $(\mu, t, \Delta, \Delta_0 ) = (0.06, 0.2, 0.1, 0)$ from which we find $\alpha_c \approx 0.769$.
}
\end{figure}

Next we set $(\Delta_h,\Delta_z) = \Delta(\beta,1)$ with $\beta \in [-1,1]$, and concentrate on the critical $\beta$ at which the topological phase transition takes place. Again by inspection, we find that only the complex function $z_{1,c}(k_y)$ supports a zero for this choice of parameters. Similarly as above, we find $\bar{k}_y = (2/\pi) \mathrm{arccos} \dfrac{t-\mu}{2t}$ from the zero of the real part of $z_{1,c}(k_y)$, and, substituting $\bar{k}_y$, the zero of the imaginary part finally gives
\begin{equation}
\label{beta_critical_B}
	\beta_c = \dfrac{ t\Delta_0 + t \Delta- \mu \Delta }{t \Delta} \;. 
\end{equation}

\subsection{$2t\geq \mu>t$} 
Again, first setting $(\Delta_h,\Delta_z) = \Delta(1,\alpha)$ with $\alpha \in [-1,1]$, we find by inspection that the complex function $z_{4}(k_x)$ supports a zero. As previously, we determine the conditions for the simultaneous vanishing of the real and the imaginary parts of $z_{4}(k_x)$.~For simplicity, we set $\Delta_0=0$ in the following.~After some algebra, we find that the term $\cos(k_x \pi)$ can be factorized in the imaginary part, i.e. it takes the form $\Im z_{4}(k_x) \propto \cos(k_x \pi) N_1(\alpha) $, with 
\begin{align*}
	N_1(\alpha) &= 4 \Delta^2 (\alpha^3 - \alpha^2) + (3t^2-\Delta^2+\mu^2) \alpha \\
	&\quad -3t^2 + \Delta^2 + \mu^2\;,
\end{align*}
where we have kept hidden factors that are not relevant for our choice of parameters. Hence the zeros of $\Im z_{4}(k_x)$ are readily given by the zeros of $N_1(\alpha)$.~The unique real zero of $N_1(\alpha)$ then gives $\alpha^0_c$ (the zero indicates that $\Delta_0=0$, see Section \ref{Top_PT}).~The real part can then be made to vanish through the appropriate choice of $k_x$, i.e.~$\Re z_{4}(\alpha_c, \bar{k}_x)=0$.~The plot of $\alpha^0_c(\mu/t)$ for $2\geq \mu/t >1 $ is shown in Fig.~\ref{phase_transition}(b) of Section \ref{Top_PT}.
       
Then setting $(\Delta_h,\Delta_z) = \Delta(\beta,1)$ with $\beta \in [-1,1]$, we find by inspection that the complex function $z_{1,b}(k_y)$ supports a zero. From the zero of the real part, we have $\bar{k}_y =  (2/\pi) \mathrm{arccos} \dfrac{\mu-t}{2t}$, and from the zero of the imaginary part, we find the same expression of the critical point $\beta_c$ as Eq.~(\ref{beta_critical_B}).


\bibliography{mybib}

\begin{thebibliography}{62}%
\makeatletter
\providecommand \@ifxundefined [1]{%
 \@ifx{#1\undefined}
}%
\providecommand \@ifnum [1]{%
 \ifnum #1\expandafter \@firstoftwo
 \else \expandafter \@secondoftwo
 \fi
}%
\providecommand \@ifx [1]{%
 \ifx #1\expandafter \@firstoftwo
 \else \expandafter \@secondoftwo
 \fi
}%
\providecommand \natexlab [1]{#1}%
\providecommand \enquote  [1]{``#1''}%
\providecommand \bibnamefont  [1]{#1}%
\providecommand \bibfnamefont [1]{#1}%
\providecommand \citenamefont [1]{#1}%
\providecommand \href@noop [0]{\@secondoftwo}%
\providecommand \href [0]{\begingroup \@sanitize@url \@href}%
\providecommand \@href[1]{\@@startlink{#1}\@@href}%
\providecommand \@@href[1]{\endgroup#1\@@endlink}%
\providecommand \@sanitize@url [0]{\catcode `\\12\catcode `\$12\catcode
  `\&12\catcode `\#12\catcode `\^12\catcode `\_12\catcode `\%12\relax}%
\providecommand \@@startlink[1]{}%
\providecommand \@@endlink[0]{}%
\providecommand \url  [0]{\begingroup\@sanitize@url \@url }%
\providecommand \@url [1]{\endgroup\@href {#1}{\urlprefix }}%
\providecommand \urlprefix  [0]{URL }%
\providecommand \Eprint [0]{\href }%
\providecommand \doibase [0]{http://dx.doi.org/}%
\providecommand \selectlanguage [0]{\@gobble}%
\providecommand \bibinfo  [0]{\@secondoftwo}%
\providecommand \bibfield  [0]{\@secondoftwo}%
\providecommand \translation [1]{[#1]}%
\providecommand \BibitemOpen [0]{}%
\providecommand \bibitemStop [0]{}%
\providecommand \bibitemNoStop [0]{.\EOS\space}%
\providecommand \EOS [0]{\spacefactor3000\relax}%
\providecommand \BibitemShut  [1]{\csname bibitem#1\endcsname}%
\let\auto@bib@innerbib\@empty
\bibitem [{\citenamefont {Schnyder}\ and\ \citenamefont
  {Ryu}(2011)}]{Schnyder_nodal_0}%
  \BibitemOpen
  \bibfield  {author} {\bibinfo {author} {\bibfnamefont {A.~P.}\ \bibnamefont
  {Schnyder}}\ and\ \bibinfo {author} {\bibfnamefont {S.}~\bibnamefont {Ryu}},\
  }\href@noop {} {\bibfield  {journal} {\bibinfo  {journal} {Phys. Rev. B}\
  }\textbf {\bibinfo {volume} {84}},\ \bibinfo {pages} {060504(R)} (\bibinfo
  {year} {2011})}\BibitemShut {NoStop}%
\bibitem [{\citenamefont {Brydon}\ \emph {et~al.}(2011)\citenamefont {Brydon},
  \citenamefont {Schnyder},\ and\ \citenamefont {Timm}}]{Schnyder_nodal_0b}%
  \BibitemOpen
  \bibfield  {author} {\bibinfo {author} {\bibfnamefont {P.~M.~R.}\
  \bibnamefont {Brydon}}, \bibinfo {author} {\bibfnamefont {A.~P.}\
  \bibnamefont {Schnyder}}, \ and\ \bibinfo {author} {\bibfnamefont
  {C.}~\bibnamefont {Timm}},\ }\href@noop {} {\bibfield  {journal} {\bibinfo
  {journal} {Phys. Rev. B}\ }\textbf {\bibinfo {volume} {84}},\ \bibinfo
  {pages} {020501} (\bibinfo {year} {2011})}\BibitemShut {NoStop}%
\bibitem [{\citenamefont {Schnyder}\ \emph {et~al.}(2012)\citenamefont
  {Schnyder}, \citenamefont {Brydon},\ and\ \citenamefont
  {Timm}}]{Schnyder_nodal_1}%
  \BibitemOpen
  \bibfield  {author} {\bibinfo {author} {\bibfnamefont {A.~P.}\ \bibnamefont
  {Schnyder}}, \bibinfo {author} {\bibfnamefont {P.~M.~R.}\ \bibnamefont
  {Brydon}}, \ and\ \bibinfo {author} {\bibfnamefont {C.}~\bibnamefont
  {Timm}},\ }\href@noop {} {\bibfield  {journal} {\bibinfo  {journal} {Phys.
  Rev. B}\ }\textbf {\bibinfo {volume} {85}},\ \bibinfo {pages} {024522}
  (\bibinfo {year} {2012})}\BibitemShut {NoStop}%
\bibitem [{\citenamefont {Schnyder}\ and\ \citenamefont
  {Brydon}(2015)}]{Schnyder_nodal_2}%
  \BibitemOpen
  \bibfield  {author} {\bibinfo {author} {\bibfnamefont {A.~P.}\ \bibnamefont
  {Schnyder}}\ and\ \bibinfo {author} {\bibfnamefont {P.~M.~R.}\ \bibnamefont
  {Brydon}},\ }\href@noop {} {\bibfield  {journal} {\bibinfo  {journal} {J.
  Phys.: Condens. Matter}\ }\textbf {\bibinfo {volume} {27}},\ \bibinfo {pages}
  {243201} (\bibinfo {year} {2015})}\BibitemShut {NoStop}%
\bibitem [{\citenamefont {Sato}(2006)}]{Sato_nodal_2006}%
  \BibitemOpen
  \bibfield  {author} {\bibinfo {author} {\bibfnamefont {M.}~\bibnamefont
  {Sato}},\ }\href@noop {} {\bibfield  {journal} {\bibinfo  {journal} {Phys.
  Rev. B}\ }\textbf {\bibinfo {volume} {73}},\ \bibinfo {pages} {214502}
  (\bibinfo {year} {2006})}\BibitemShut {NoStop}%
\bibitem [{\citenamefont {B{\'e}ri}(2010)}]{Beri_nodal}%
  \BibitemOpen
  \bibfield  {author} {\bibinfo {author} {\bibfnamefont {B.}~\bibnamefont
  {B{\'e}ri}},\ }\href@noop {} {\bibfield  {journal} {\bibinfo  {journal}
  {Phys. Rev. B}\ }\textbf {\bibinfo {volume} {81}},\ \bibinfo {pages} {134515}
  (\bibinfo {year} {2010})}\BibitemShut {NoStop}%
\bibitem [{\citenamefont {Sato}\ \emph {et~al.}(2011)\citenamefont {Sato},
  \citenamefont {Tanaka}, \citenamefont {Yada},\ and\ \citenamefont
  {Yokoyama}}]{Sato11}%
  \BibitemOpen
  \bibfield  {author} {\bibinfo {author} {\bibfnamefont {M.}~\bibnamefont
  {Sato}}, \bibinfo {author} {\bibfnamefont {Y.}~\bibnamefont {Tanaka}},
  \bibinfo {author} {\bibfnamefont {K.}~\bibnamefont {Yada}}, \ and\ \bibinfo
  {author} {\bibfnamefont {T.}~\bibnamefont {Yokoyama}},\ }\href@noop {}
  {\bibfield  {journal} {\bibinfo  {journal} {Phys. Rev. B}\ }\textbf {\bibinfo
  {volume} {83}},\ \bibinfo {pages} {224511} (\bibinfo {year}
  {2011})}\BibitemShut {NoStop}%
\bibitem [{\citenamefont {Goswami}\ and\ \citenamefont
  {Balicas}()}]{GoswamiBalicas_URu2Si2}%
  \BibitemOpen
  \bibfield  {author} {\bibinfo {author} {\bibfnamefont {P.}~\bibnamefont
  {Goswami}}\ and\ \bibinfo {author} {\bibfnamefont {L.}~\bibnamefont
  {Balicas}},\ }\href@noop {} {\bibinfo  {journal} {arXiv:1312.3632v1}\
  }\BibitemShut {NoStop}%
\bibitem [{\citenamefont {Volovik}(1999)}]{Volovik_nodal_99}%
  \BibitemOpen
\bibfield  {journal} {  }\bibfield  {author} {\bibinfo {author} {\bibfnamefont
  {G.~E.}\ \bibnamefont {Volovik}},\ }\href@noop {} {\bibfield  {journal}
  {\bibinfo  {journal} {Proc. Natl. Acad. Sci. USA}\ }\textbf {\bibinfo
  {volume} {96}},\ \bibinfo {pages} {6042} (\bibinfo {year}
  {1999})}\BibitemShut {NoStop}%
\bibitem [{\citenamefont {Volovik}(2003)}]{Volovik}%
  \BibitemOpen
  \bibfield  {author} {\bibinfo {author} {\bibfnamefont {G.~E.}\ \bibnamefont
  {Volovik}},\ }\href@noop {} {\emph {\bibinfo {title} {The Universe in a
  Helium Droplet}}}\ (\bibinfo  {publisher} {Oxford University Press},\
  \bibinfo {year} {2003})\BibitemShut {NoStop}%
\bibitem [{\citenamefont {Takayama}\ \emph {et~al.}(2015)\citenamefont
  {Takayama}, \citenamefont {Kato}, \citenamefont {Dinnebier}, \citenamefont
  {Nuss}, \citenamefont {Kono}, \citenamefont {Veiga}, \citenamefont {Fabbris},
  \citenamefont {Haskel},\ and\ \citenamefont {Takagi}}]{Takagi15}%
  \BibitemOpen
  \bibfield  {author} {\bibinfo {author} {\bibfnamefont {T.}~\bibnamefont
  {Takayama}}, \bibinfo {author} {\bibfnamefont {A.}~\bibnamefont {Kato}},
  \bibinfo {author} {\bibfnamefont {R.}~\bibnamefont {Dinnebier}}, \bibinfo
  {author} {\bibfnamefont {J.}~\bibnamefont {Nuss}}, \bibinfo {author}
  {\bibfnamefont {H.}~\bibnamefont {Kono}}, \bibinfo {author} {\bibfnamefont
  {L.~S.~I.}\ \bibnamefont {Veiga}}, \bibinfo {author} {\bibfnamefont
  {G.}~\bibnamefont {Fabbris}}, \bibinfo {author} {\bibfnamefont
  {D.}~\bibnamefont {Haskel}}, \ and\ \bibinfo {author} {\bibfnamefont
  {H.}~\bibnamefont {Takagi}},\ }\href@noop {} {\bibfield  {journal} {\bibinfo
  {journal} {Phys. Rev. Lett.}\ }\textbf {\bibinfo {volume} {114}},\ \bibinfo
  {pages} {077202} (\bibinfo {year} {2015})}\BibitemShut {NoStop}%
\bibitem [{\citenamefont {Kim}\ \emph {et~al.}(2016)\citenamefont {Kim},
  \citenamefont {Lee},\ and\ \citenamefont {Kim}}]{Kim_HHL_kitaev}%
  \BibitemOpen
  \bibfield  {author} {\bibinfo {author} {\bibfnamefont {H.-S.}\ \bibnamefont
  {Kim}}, \bibinfo {author} {\bibfnamefont {E.~K.-H.}\ \bibnamefont {Lee}}, \
  and\ \bibinfo {author} {\bibfnamefont {Y.~B.}\ \bibnamefont {Kim}},\
  }\href@noop {} {\bibfield  {journal} {\bibinfo  {journal} {EPL}\ }\textbf
  {\bibinfo {volume} {112}},\ \bibinfo {pages} {6} (\bibinfo {year}
  {2016})}\BibitemShut {NoStop}%
\bibitem [{\citenamefont {Lee}\ and\ \citenamefont
  {Kim}(2015)}]{Kim_HHL_magnetic_pd}%
  \BibitemOpen
  \bibfield  {author} {\bibinfo {author} {\bibfnamefont {E.~K.-H.}\
  \bibnamefont {Lee}}\ and\ \bibinfo {author} {\bibfnamefont {Y.~B.}\
  \bibnamefont {Kim}},\ }\href@noop {} {\bibfield  {journal} {\bibinfo
  {journal} {Phys. Rev. B}\ }\textbf {\bibinfo {volume} {91}},\ \bibinfo
  {pages} {064407} (\bibinfo {year} {2015})}\BibitemShut {NoStop}%
\bibitem [{\citenamefont {Veiga}\ \emph {et~al.}(2017)\citenamefont {Veiga},
  \citenamefont {Etter}, \citenamefont {Glazyrin}, \citenamefont {Sun},
  \citenamefont {C.~A.~Escanhoela}, \citenamefont {Fabbris}, \citenamefont
  {Mardegan}, \citenamefont {Malavi}, \citenamefont {Deng}, \citenamefont
  {Stavropoulos}, \citenamefont {Kee}, \citenamefont {Yang}, \citenamefont {van
  Veenendaal}, \citenamefont {Schilling}, \citenamefont {Takayama},
  \citenamefont {Takagi},\ and\ \citenamefont {Haskel}}]{Veiga_HHL}%
  \BibitemOpen
  \bibfield  {author} {\bibinfo {author} {\bibfnamefont {L.~S.~I.}\
  \bibnamefont {Veiga}}, \bibinfo {author} {\bibfnamefont {M.}~\bibnamefont
  {Etter}}, \bibinfo {author} {\bibfnamefont {K.}~\bibnamefont {Glazyrin}},
  \bibinfo {author} {\bibfnamefont {F.}~\bibnamefont {Sun}}, \bibinfo {author}
  {\bibfnamefont {J.}~\bibnamefont {C.~A.~Escanhoela}}, \bibinfo {author}
  {\bibfnamefont {G.}~\bibnamefont {Fabbris}}, \bibinfo {author} {\bibfnamefont
  {J.~R.~L.}\ \bibnamefont {Mardegan}}, \bibinfo {author} {\bibfnamefont
  {P.~S.}\ \bibnamefont {Malavi}}, \bibinfo {author} {\bibfnamefont
  {Y.}~\bibnamefont {Deng}}, \bibinfo {author} {\bibfnamefont {P.~P.}\
  \bibnamefont {Stavropoulos}}, \bibinfo {author} {\bibfnamefont {H.-Y.}\
  \bibnamefont {Kee}}, \bibinfo {author} {\bibfnamefont {W.~G.}\ \bibnamefont
  {Yang}}, \bibinfo {author} {\bibfnamefont {M.}~\bibnamefont {van
  Veenendaal}}, \bibinfo {author} {\bibfnamefont {J.~S.}\ \bibnamefont
  {Schilling}}, \bibinfo {author} {\bibfnamefont {T.}~\bibnamefont {Takayama}},
  \bibinfo {author} {\bibfnamefont {H.}~\bibnamefont {Takagi}}, \ and\ \bibinfo
  {author} {\bibfnamefont {D.}~\bibnamefont {Haskel}},\ }\href@noop {}
  {\bibfield  {journal} {\bibinfo  {journal} {Phys. Rev. B}\ }\textbf {\bibinfo
  {volume} {96}},\ \bibinfo {pages} {140402(R)} (\bibinfo {year}
  {2017})}\BibitemShut {NoStop}%
\bibitem [{\citenamefont {Ezawa}(2016)}]{Ezawa_HHL}%
  \BibitemOpen
  \bibfield  {author} {\bibinfo {author} {\bibfnamefont {M.}~\bibnamefont
  {Ezawa}},\ }\href@noop {} {\bibfield  {journal} {\bibinfo  {journal} {Phys.
  Rev. Lett.}\ }\textbf {\bibinfo {volume} {116}},\ \bibinfo {pages} {127202}
  (\bibinfo {year} {2016})}\BibitemShut {NoStop}%
\bibitem [{\citenamefont {Schmidt}\ \emph {et~al.}(2016)\citenamefont
  {Schmidt}, \citenamefont {Bouhon},\ and\ \citenamefont
  {Black-Schaffer}}]{SBBS_16}%
  \BibitemOpen
  \bibfield  {author} {\bibinfo {author} {\bibfnamefont {J.}~\bibnamefont
  {Schmidt}}, \bibinfo {author} {\bibfnamefont {A.}~\bibnamefont {Bouhon}}, \
  and\ \bibinfo {author} {\bibfnamefont {A.~M.}\ \bibnamefont
  {Black-Schaffer}},\ }\href@noop {} {\bibfield  {journal} {\bibinfo  {journal}
  {Phys. Rev. B}\ }\textbf {\bibinfo {volume} {94}},\ \bibinfo {pages} {104513}
  (\bibinfo {year} {2016})}\BibitemShut {NoStop}%
\bibitem [{\citenamefont {Hahn}(2006)}]{ITC}%
  \BibitemOpen
  \bibfield  {author} {\bibinfo {author} {\bibfnamefont {T.}~\bibnamefont
  {Hahn}},\ }\href {http://it.iucr.org/Ac/} {\emph {\bibinfo {title}
  {International Tables for Crystallography. Volume A, Space-group symmetry}}}\
  (\bibinfo  {publisher} {online},\ \bibinfo {year} {2006})\BibitemShut
  {NoStop}%
\bibitem [{\citenamefont {Aroyo}\ \emph {et~al.}(2006)\citenamefont {Aroyo},
  \citenamefont {Kirov}, \citenamefont {Capillas}, \citenamefont {Perez-Mato},\
  and\ \citenamefont {Wondratschek}}]{Bilbao}%
  \BibitemOpen
  \bibfield  {author} {\bibinfo {author} {\bibfnamefont {M.~I.}\ \bibnamefont
  {Aroyo}}, \bibinfo {author} {\bibfnamefont {A.}~\bibnamefont {Kirov}},
  \bibinfo {author} {\bibfnamefont {C.}~\bibnamefont {Capillas}}, \bibinfo
  {author} {\bibfnamefont {J.~M.}\ \bibnamefont {Perez-Mato}}, \ and\ \bibinfo
  {author} {\bibfnamefont {H.}~\bibnamefont {Wondratschek}},\ }\href@noop {}
  {\bibfield  {journal} {\bibinfo  {journal} {Acta Cryst. A}\ }\textbf
  {\bibinfo {volume} {62}},\ \bibinfo {pages} {115} (\bibinfo {year}
  {2006})}\BibitemShut {NoStop}%
\bibitem [{\citenamefont {Zhang}\ \emph {et~al.}(1988)\citenamefont {Zhang},
  \citenamefont {Gros},\ and\ \citenamefont {a~nd
  H.~Shiba}}]{ZhangGrosandRiceShiba}%
  \BibitemOpen
  \bibfield  {author} {\bibinfo {author} {\bibfnamefont {F.~C.}\ \bibnamefont
  {Zhang}}, \bibinfo {author} {\bibfnamefont {C.}~\bibnamefont {Gros}}, \ and\
  \bibinfo {author} {\bibfnamefont {T.~M.~R.}\ \bibnamefont {a~nd H.~Shiba}},\
  }\href@noop {} {\bibfield  {journal} {\bibinfo  {journal} {Supercond. Sci.
  Tech.}\ }\textbf {\bibinfo {volume} {1}},\ \bibinfo {pages} {36} (\bibinfo
  {year} {1988})}\BibitemShut {NoStop}%
\bibitem [{\citenamefont {Anderson}\ \emph {et~al.}(2004)\citenamefont
  {Anderson}, \citenamefont {Lee}, \citenamefont {Randeria}, \citenamefont
  {Rice}, \citenamefont {Trivedi},\ and\ \citenamefont
  {Zhang}}]{AndersonRice_04}%
  \BibitemOpen
  \bibfield  {author} {\bibinfo {author} {\bibfnamefont {P.~W.}\ \bibnamefont
  {Anderson}}, \bibinfo {author} {\bibfnamefont {P.~A.}\ \bibnamefont {Lee}},
  \bibinfo {author} {\bibfnamefont {M.}~\bibnamefont {Randeria}}, \bibinfo
  {author} {\bibfnamefont {T.~M.}\ \bibnamefont {Rice}}, \bibinfo {author}
  {\bibfnamefont {N.}~\bibnamefont {Trivedi}}, \ and\ \bibinfo {author}
  {\bibfnamefont {F.~C.}\ \bibnamefont {Zhang}},\ }\href@noop {} {\bibfield
  {journal} {\bibinfo  {journal} {J. Phys.: Condens. Matter}\ }\textbf
  {\bibinfo {volume} {16}},\ \bibinfo {pages} {R755} (\bibinfo {year}
  {2004})}\BibitemShut {NoStop}%
\bibitem [{\citenamefont {Edegger}\ \emph {et~al.}(2007)\citenamefont
  {Edegger}, \citenamefont {Muthukumar},\ and\ \citenamefont
  {Gros}}]{EdeggerGros_07}%
  \BibitemOpen
  \bibfield  {author} {\bibinfo {author} {\bibfnamefont {B.}~\bibnamefont
  {Edegger}}, \bibinfo {author} {\bibfnamefont {V.~N.}\ \bibnamefont
  {Muthukumar}}, \ and\ \bibinfo {author} {\bibfnamefont {C.}~\bibnamefont
  {Gros}},\ }\href@noop {} {\bibfield  {journal} {\bibinfo  {journal} {Adv.
  Phys.}\ }\textbf {\bibinfo {volume} {56}},\ \bibinfo {pages} {927} (\bibinfo
  {year} {2007})}\BibitemShut {NoStop}%
\bibitem [{\citenamefont {Hur}\ and\ \citenamefont
  {Rice}(2009)}]{LeHurRice_09}%
  \BibitemOpen
  \bibfield  {author} {\bibinfo {author} {\bibfnamefont {K.~L.}\ \bibnamefont
  {Hur}}\ and\ \bibinfo {author} {\bibfnamefont {T.~M.}\ \bibnamefont {Rice}},\
  }\href@noop {} {\bibfield  {journal} {\bibinfo  {journal} {Ann. Phys.}\
  }\textbf {\bibinfo {volume} {324}},\ \bibinfo {pages} {1452} (\bibinfo {year}
  {2009})}\BibitemShut {NoStop}%
\bibitem [{\citenamefont {Bouhon}\ and\ \citenamefont
  {Black-Schaffer}(2017{\natexlab{a}})}]{ABABS_HHL_2}%
  \BibitemOpen
  \bibfield  {author} {\bibinfo {author} {\bibfnamefont {A.}~\bibnamefont
  {Bouhon}}\ and\ \bibinfo {author} {\bibfnamefont {A.}~\bibnamefont
  {Black-Schaffer}},\ }\href@noop {} {\bibfield  {journal} {\bibinfo  {journal}
  {to appear}\ } (\bibinfo {year} {2017}{\natexlab{a}})}\BibitemShut {NoStop}%
\bibitem [{Note1()}]{Note1}%
  \BibitemOpen
  \bibinfo {note} {The nonsymmorphicity of SG70 leads to double degeneracies at
  the boundaries of the BZ, such that each band crossing the Fermi level is
  connected to an other band that does not cross the Fermi level. Since two
  bands are involved in the line nodal Fermi surface, a minimum of four bands
  must be taken into account.}\BibitemShut {Stop}%
\bibitem [{\citenamefont {Koster}\ \emph {et~al.}(1964)\citenamefont {Koster},
  \citenamefont {Dimmock}, \citenamefont {Wheeler},\ and\ \citenamefont
  {Statz}}]{Koster}%
  \BibitemOpen
  \bibfield  {author} {\bibinfo {author} {\bibfnamefont {G.~F.}\ \bibnamefont
  {Koster}}, \bibinfo {author} {\bibfnamefont {J.~O.}\ \bibnamefont {Dimmock}},
  \bibinfo {author} {\bibfnamefont {R.~G.}\ \bibnamefont {Wheeler}}, \ and\
  \bibinfo {author} {\bibfnamefont {H.}~\bibnamefont {Statz}},\ }\href@noop {}
  {\emph {\bibinfo {title} {Properties of the Thirty-Two Point Groups}}}\
  (\bibinfo  {publisher} {MIT Press Cambridge},\ \bibinfo {year}
  {1964})\BibitemShut {NoStop}%
\bibitem [{\citenamefont {Schnyder}\ \emph {et~al.}(2008)\citenamefont
  {Schnyder}, \citenamefont {Ryu}, \citenamefont {Furusaki},\ and\
  \citenamefont {Ludwig}}]{Schnyder08}%
  \BibitemOpen
  \bibfield  {author} {\bibinfo {author} {\bibfnamefont {A.~P.}\ \bibnamefont
  {Schnyder}}, \bibinfo {author} {\bibfnamefont {S.}~\bibnamefont {Ryu}},
  \bibinfo {author} {\bibfnamefont {A.}~\bibnamefont {Furusaki}}, \ and\
  \bibinfo {author} {\bibfnamefont {A.~W.~W.}\ \bibnamefont {Ludwig}},\
  }\href@noop {} {\bibfield  {journal} {\bibinfo  {journal} {Phys. Rev. B}\
  }\textbf {\bibinfo {volume} {78}},\ \bibinfo {pages} {195125} (\bibinfo
  {year} {2008})}\BibitemShut {NoStop}%
\bibitem [{\citenamefont {Zhao}\ and\ \citenamefont {Wang}(2013)}]{Zhao13}%
  \BibitemOpen
  \bibfield  {author} {\bibinfo {author} {\bibfnamefont {Y.~X.}\ \bibnamefont
  {Zhao}}\ and\ \bibinfo {author} {\bibfnamefont {Z.~D.}\ \bibnamefont
  {Wang}},\ }\href@noop {} {\bibfield  {journal} {\bibinfo  {journal} {Phys.
  Rev. Lett.}\ }\textbf {\bibinfo {volume} {110}},\ \bibinfo {pages} {240404}
  (\bibinfo {year} {2013})}\BibitemShut {NoStop}%
\bibitem [{\citenamefont {Chiu}\ \emph {et~al.}(2016)\citenamefont {Chiu},
  \citenamefont {Teo}, \citenamefont {Schnyder},\ and\ \citenamefont
  {Ryu}}]{Class_sym_review}%
  \BibitemOpen
  \bibfield  {author} {\bibinfo {author} {\bibfnamefont {C.-K.}\ \bibnamefont
  {Chiu}}, \bibinfo {author} {\bibfnamefont {J.~C.~Y.}\ \bibnamefont {Teo}},
  \bibinfo {author} {\bibfnamefont {A.~P.}\ \bibnamefont {Schnyder}}, \ and\
  \bibinfo {author} {\bibfnamefont {S.}~\bibnamefont {Ryu}},\ }\href@noop {}
  {\bibfield  {journal} {\bibinfo  {journal} {Rev. Mod. Phys.}\ }\textbf
  {\bibinfo {volume} {88}},\ \bibinfo {pages} {035005} (\bibinfo {year}
  {2016})}\BibitemShut {NoStop}%
\bibitem [{\citenamefont {Wen}\ and\ \citenamefont
  {Zee}(2002)}]{WenZee_nodal02}%
  \BibitemOpen
  \bibfield  {author} {\bibinfo {author} {\bibfnamefont {X.~G.}\ \bibnamefont
  {Wen}}\ and\ \bibinfo {author} {\bibfnamefont {A.}~\bibnamefont {Zee}},\
  }\href@noop {} {\bibfield  {journal} {\bibinfo  {journal} {Phys. Rev. B}\
  }\textbf {\bibinfo {volume} {66}},\ \bibinfo {pages} {235110} (\bibinfo
  {year} {2002})}\BibitemShut {NoStop}%
\bibitem [{\citenamefont {Volovik}(2007)}]{Volovik_lect_notes07}%
  \BibitemOpen
  \bibfield  {author} {\bibinfo {author} {\bibfnamefont {G.~E.}\ \bibnamefont
  {Volovik}},\ }\href@noop {} {\bibfield  {journal} {\bibinfo  {journal} {Lect.
  Notes Phys.}\ }\textbf {\bibinfo {volume} {718}},\ \bibinfo {pages} {31}
  (\bibinfo {year} {2007})}\BibitemShut {NoStop}%
\bibitem [{\citenamefont {Hughes}\ \emph {et~al.}(2011)\citenamefont {Hughes},
  \citenamefont {Prodan},\ and\ \citenamefont {Bernevig}}]{Bernevig1}%
  \BibitemOpen
  \bibfield  {author} {\bibinfo {author} {\bibfnamefont {T.~L.}\ \bibnamefont
  {Hughes}}, \bibinfo {author} {\bibfnamefont {E.}~\bibnamefont {Prodan}}, \
  and\ \bibinfo {author} {\bibfnamefont {B.~A.}\ \bibnamefont {Bernevig}},\
  }\href@noop {} {\bibfield  {journal} {\bibinfo  {journal} {Phys. Rev. B}\
  }\textbf {\bibinfo {volume} {83}},\ \bibinfo {pages} {245132} (\bibinfo
  {year} {2011})}\BibitemShut {NoStop}%
\bibitem [{\citenamefont {Fang}\ \emph
  {et~al.}(2012{\natexlab{a}})\citenamefont {Fang}, \citenamefont {Gilbert},\
  and\ \citenamefont {Bernevig}}]{Bernevig_point_groups}%
  \BibitemOpen
  \bibfield  {author} {\bibinfo {author} {\bibfnamefont {C.}~\bibnamefont
  {Fang}}, \bibinfo {author} {\bibfnamefont {M.~J.}\ \bibnamefont {Gilbert}}, \
  and\ \bibinfo {author} {\bibfnamefont {B.~A.}\ \bibnamefont {Bernevig}},\
  }\href@noop {} {\bibfield  {journal} {\bibinfo  {journal} {Phys. Rev. B}\
  }\textbf {\bibinfo {volume} {86}},\ \bibinfo {pages} {115112} (\bibinfo
  {year} {2012}{\natexlab{a}})}\BibitemShut {NoStop}%
\bibitem [{\citenamefont {Wang}\ \emph {et~al.}(2016)\citenamefont {Wang},
  \citenamefont {Alexandradinata}, \citenamefont {Cava},\ and\ \citenamefont
  {Bernevig}}]{Bernevig3}%
  \BibitemOpen
  \bibfield  {author} {\bibinfo {author} {\bibfnamefont {Z.}~\bibnamefont
  {Wang}}, \bibinfo {author} {\bibfnamefont {A.}~\bibnamefont
  {Alexandradinata}}, \bibinfo {author} {\bibfnamefont {R.~J.}\ \bibnamefont
  {Cava}}, \ and\ \bibinfo {author} {\bibfnamefont {B.~A.}\ \bibnamefont
  {Bernevig}},\ }\href@noop {} {\bibfield  {journal} {\bibinfo  {journal}
  {Nature}\ }\textbf {\bibinfo {volume} {532}},\ \bibinfo {pages} {189}
  (\bibinfo {year} {2016})}\BibitemShut {NoStop}%
\bibitem [{\citenamefont {Alexandradinata}\ and\ \citenamefont
  {Bernevig}(2016)}]{AlexBernevig_berryphase}%
  \BibitemOpen
  \bibfield  {author} {\bibinfo {author} {\bibfnamefont {A.}~\bibnamefont
  {Alexandradinata}}\ and\ \bibinfo {author} {\bibfnamefont {B.~A.}\
  \bibnamefont {Bernevig}},\ }\href@noop {} {\bibfield  {journal} {\bibinfo
  {journal} {Phys. Rev. B}\ }\textbf {\bibinfo {volume} {93}},\ \bibinfo
  {pages} {205104} (\bibinfo {year} {2016})}\BibitemShut {NoStop}%
\bibitem [{\citenamefont {Alexandradinata}\ \emph {et~al.}(2014)\citenamefont
  {Alexandradinata}, \citenamefont {Dai},\ and\ \citenamefont
  {Bernevig}}]{Alex1}%
  \BibitemOpen
  \bibfield  {author} {\bibinfo {author} {\bibfnamefont {A.}~\bibnamefont
  {Alexandradinata}}, \bibinfo {author} {\bibfnamefont {X.}~\bibnamefont
  {Dai}}, \ and\ \bibinfo {author} {\bibfnamefont {B.~A.}\ \bibnamefont
  {Bernevig}},\ }\href@noop {} {\bibfield  {journal} {\bibinfo  {journal}
  {Phys. Rev. B}\ }\textbf {\bibinfo {volume} {89}},\ \bibinfo {pages} {155114}
  (\bibinfo {year} {2014})}\BibitemShut {NoStop}%
\bibitem [{\citenamefont {Muechler}\ \emph {et~al.}(2016)\citenamefont
  {Muechler}, \citenamefont {Alexandradinata}, \citenamefont {Neupert},\ and\
  \citenamefont {Car}}]{Alex2}%
  \BibitemOpen
  \bibfield  {author} {\bibinfo {author} {\bibfnamefont {L.}~\bibnamefont
  {Muechler}}, \bibinfo {author} {\bibfnamefont {A.}~\bibnamefont
  {Alexandradinata}}, \bibinfo {author} {\bibfnamefont {T.}~\bibnamefont
  {Neupert}}, \ and\ \bibinfo {author} {\bibfnamefont {R.}~\bibnamefont
  {Car}},\ }\href@noop {} {\bibfield  {journal} {\bibinfo  {journal} {Phys.
  Rev. X}\ }\textbf {\bibinfo {volume} {6}},\ \bibinfo {pages} {041069}
  (\bibinfo {year} {2016})}\BibitemShut {NoStop}%
\bibitem [{\citenamefont {Bouhon}\ and\ \citenamefont
  {Black-Schaffer}(2017{\natexlab{b}})}]{ABABS_points}%
  \BibitemOpen
  \bibfield  {author} {\bibinfo {author} {\bibfnamefont {A.}~\bibnamefont
  {Bouhon}}\ and\ \bibinfo {author} {\bibfnamefont {A.~M.}\ \bibnamefont
  {Black-Schaffer}},\ }\href@noop {} {\bibfield  {journal} {\bibinfo  {journal}
  {Phys. Rev. B}\ }\textbf {\bibinfo {volume} {95}},\ \bibinfo {pages}
  {241101(R)} (\bibinfo {year} {2017}{\natexlab{b}})}\BibitemShut {NoStop}%
\bibitem [{\citenamefont {Bouhon}\ and\ \citenamefont
  {Black-Schaffer}()}]{ABABS_lines}%
  \BibitemOpen
  \bibfield  {author} {\bibinfo {author} {\bibfnamefont {A.}~\bibnamefont
  {Bouhon}}\ and\ \bibinfo {author} {\bibfnamefont {A.~M.}\ \bibnamefont
  {Black-Schaffer}},\ }\href@noop {} {\bibinfo  {journal} {arXiv:1710.04871}\
  }\BibitemShut {NoStop}%
\bibitem [{Note2()}]{Note2}%
  \BibitemOpen
\bibfield  {journal} {  }\bibinfo {note} {The phases of the eigenvalues of the
  Wilson loops are very stable and only a few number of points are necessary in
  the discretization of the base loop.}\BibitemShut {Stop}%
\bibitem [{\citenamefont {Hatsugai}(2009)}]{Hatsugai_graphene_chiral}%
  \BibitemOpen
  \bibfield  {author} {\bibinfo {author} {\bibfnamefont {Y.}~\bibnamefont
  {Hatsugai}},\ }\href@noop {} {\bibfield  {journal} {\bibinfo  {journal}
  {Solid State Commun.}\ }\textbf {\bibinfo {volume} {149}},\ \bibinfo {pages}
  {1061} (\bibinfo {year} {2009})}\BibitemShut {NoStop}%
\bibitem [{\citenamefont {Ryu}\ \emph {et~al.}(2010)\citenamefont {Ryu},
  \citenamefont {Schnyder}, \citenamefont {Furusaki},\ and\ \citenamefont
  {Ludwig}}]{RyuSchnyder_10ways}%
  \BibitemOpen
  \bibfield  {author} {\bibinfo {author} {\bibfnamefont {S.}~\bibnamefont
  {Ryu}}, \bibinfo {author} {\bibfnamefont {A.~P.}\ \bibnamefont {Schnyder}},
  \bibinfo {author} {\bibfnamefont {A.}~\bibnamefont {Furusaki}}, \ and\
  \bibinfo {author} {\bibfnamefont {A.~W.~W.}\ \bibnamefont {Ludwig}},\
  }\href@noop {} {\bibfield  {journal} {\bibinfo  {journal} {New J. Phys.}\
  }\textbf {\bibinfo {volume} {12}},\ \bibinfo {pages} {065010} (\bibinfo
  {year} {2010})}\BibitemShut {NoStop}%
\bibitem [{\citenamefont {Hatsugai}(2010)}]{Hatsugai_nonabelian}%
  \BibitemOpen
  \bibfield  {author} {\bibinfo {author} {\bibfnamefont {Y.}~\bibnamefont
  {Hatsugai}},\ }\href@noop {} {\bibfield  {journal} {\bibinfo  {journal} {New
  J. Phys.}\ }\textbf {\bibinfo {volume} {12}},\ \bibinfo {pages} {065004}
  (\bibinfo {year} {2010})}\BibitemShut {NoStop}%
\bibitem [{\citenamefont {Ryu}\ and\ \citenamefont
  {Hatsugai}(2002)}]{RyuHatsugai_BulkEC}%
  \BibitemOpen
  \bibfield  {author} {\bibinfo {author} {\bibfnamefont {S.}~\bibnamefont
  {Ryu}}\ and\ \bibinfo {author} {\bibfnamefont {Y.}~\bibnamefont {Hatsugai}},\
  }\href@noop {} {\bibfield  {journal} {\bibinfo  {journal} {Phys. Rev. Lett.}\
  }\textbf {\bibinfo {volume} {89}},\ \bibinfo {pages} {077002} (\bibinfo
  {year} {2002})}\BibitemShut {NoStop}%
\bibitem [{\citenamefont {Wong}\ \emph {et~al.}(2013)\citenamefont {Wong},
  \citenamefont {Liu}, \citenamefont {Law},\ and\ \citenamefont
  {Lee}}]{Majorana_PLee}%
  \BibitemOpen
  \bibfield  {author} {\bibinfo {author} {\bibfnamefont {C.~L.~M.}\
  \bibnamefont {Wong}}, \bibinfo {author} {\bibfnamefont {J.}~\bibnamefont
  {Liu}}, \bibinfo {author} {\bibfnamefont {K.~T.}\ \bibnamefont {Law}}, \ and\
  \bibinfo {author} {\bibfnamefont {P.~A.}\ \bibnamefont {Lee}},\ }\href@noop
  {} {\bibfield  {journal} {\bibinfo  {journal} {Phys. Rev. B}\ }\textbf
  {\bibinfo {volume} {88}},\ \bibinfo {pages} {060504(R)} (\bibinfo {year}
  {2013})}\BibitemShut {NoStop}%
\bibitem [{\citenamefont {Zak}(1989)}]{Zak_1D_phase}%
  \BibitemOpen
  \bibfield  {author} {\bibinfo {author} {\bibfnamefont {J.}~\bibnamefont
  {Zak}},\ }\href@noop {} {\bibfield  {journal} {\bibinfo  {journal} {Phys.
  Rev. Lett.}\ }\textbf {\bibinfo {volume} {62}},\ \bibinfo {pages} {2747}
  (\bibinfo {year} {1989})}\BibitemShut {NoStop}%
\bibitem [{\citenamefont {King-Smith}\ and\ \citenamefont
  {Vanderbilt}()}]{KingVanderbilt}%
  \BibitemOpen
  \bibfield  {author} {\bibinfo {author} {\bibfnamefont {R.~D.}\ \bibnamefont
  {King-Smith}}\ and\ \bibinfo {author} {\bibfnamefont {D.}~\bibnamefont
  {Vanderbilt}},\ }\href@noop {} {\bibfield  {journal} {\bibinfo  {journal}
  {Phys. Rev. B}\ }\textbf {\bibinfo {volume} {47}},\ \bibinfo {pages}
  {1651(R)}}\BibitemShut {NoStop}%
\bibitem [{\citenamefont {Vanderbilt}\ and\ \citenamefont
  {King-Smith}(1993)}]{VanderbiltKingSmith_surface}%
  \BibitemOpen
  \bibfield  {author} {\bibinfo {author} {\bibfnamefont {D.}~\bibnamefont
  {Vanderbilt}}\ and\ \bibinfo {author} {\bibfnamefont {R.~D.}\ \bibnamefont
  {King-Smith}},\ }\href@noop {} {\bibfield  {journal} {\bibinfo  {journal}
  {Phys. Rev. B}\ }\textbf {\bibinfo {volume} {48}},\ \bibinfo {pages} {4442}
  (\bibinfo {year} {1993})}\BibitemShut {NoStop}%
\bibitem [{\citenamefont {Kariyado}\ and\ \citenamefont
  {Hatsugai}(2013)}]{KariyadoHatsugai}%
  \BibitemOpen
  \bibfield  {author} {\bibinfo {author} {\bibfnamefont {T.}~\bibnamefont
  {Kariyado}}\ and\ \bibinfo {author} {\bibfnamefont {Y.}~\bibnamefont
  {Hatsugai}},\ }\href@noop {} {\bibfield  {journal} {\bibinfo  {journal}
  {Phys. Rev. B}\ }\textbf {\bibinfo {volume} {88}},\ \bibinfo {pages} {245126}
  (\bibinfo {year} {2013})}\BibitemShut {NoStop}%
\bibitem [{\citenamefont {Rhim}\ \emph {et~al.}(2017)\citenamefont {Rhim},
  \citenamefont {Behrends},\ and\ \citenamefont
  {Bardarson}}]{Intercellular_Zak}%
  \BibitemOpen
  \bibfield  {author} {\bibinfo {author} {\bibfnamefont {J.-W.}\ \bibnamefont
  {Rhim}}, \bibinfo {author} {\bibfnamefont {J.}~\bibnamefont {Behrends}}, \
  and\ \bibinfo {author} {\bibfnamefont {J.~H.}\ \bibnamefont {Bardarson}},\
  }\href@noop {} {\bibfield  {journal} {\bibinfo  {journal} {Phys. Rev. B}\
  }\textbf {\bibinfo {volume} {95}},\ \bibinfo {pages} {035421} (\bibinfo
  {year} {2017})}\BibitemShut {NoStop}%
\bibitem [{\citenamefont {Ishikawa}\ and\ \citenamefont
  {Matsuyama}(1986)}]{Ishikawa_86}%
  \BibitemOpen
  \bibfield  {author} {\bibinfo {author} {\bibfnamefont {K.}~\bibnamefont
  {Ishikawa}}\ and\ \bibinfo {author} {\bibfnamefont {T.}~\bibnamefont
  {Matsuyama}},\ }\href@noop {} {\bibfield  {journal} {\bibinfo  {journal} {Z.
  Phys. C}\ }\textbf {\bibinfo {volume} {33}},\ \bibinfo {pages} {41} (\bibinfo
  {year} {1986})}\BibitemShut {NoStop}%
\bibitem [{\citenamefont {Shiozaki}\ and\ \citenamefont
  {Sato}(2014)}]{ShiozakiSato_I}%
  \BibitemOpen
  \bibfield  {author} {\bibinfo {author} {\bibfnamefont {K.}~\bibnamefont
  {Shiozaki}}\ and\ \bibinfo {author} {\bibfnamefont {M.}~\bibnamefont
  {Sato}},\ }\href@noop {} {\bibfield  {journal} {\bibinfo  {journal} {Phys.
  Rev. B}\ }\textbf {\bibinfo {volume} {90}},\ \bibinfo {pages} {165114}
  (\bibinfo {year} {2014})}\BibitemShut {NoStop}%
\bibitem [{\citenamefont {Stone}(1976)}]{Stone_76}%
  \BibitemOpen
  \bibfield  {author} {\bibinfo {author} {\bibfnamefont {A.~J.}\ \bibnamefont
  {Stone}},\ }\href@noop {} {\bibfield  {journal} {\bibinfo  {journal} {Proc.
  R. Soc. Lond. A}\ }\textbf {\bibinfo {volume} {351}},\ \bibinfo {pages} {141}
  (\bibinfo {year} {1976})}\BibitemShut {NoStop}%
\bibitem [{\citenamefont {Wan}\ \emph {et~al.}(2011)\citenamefont {Wan},
  \citenamefont {Turner}, \citenamefont {Vishwanath},\ and\ \citenamefont
  {Savrasov}}]{Vishwanath_weyl0}%
  \BibitemOpen
  \bibfield  {author} {\bibinfo {author} {\bibfnamefont {X.}~\bibnamefont
  {Wan}}, \bibinfo {author} {\bibfnamefont {A.~M.}\ \bibnamefont {Turner}},
  \bibinfo {author} {\bibfnamefont {A.}~\bibnamefont {Vishwanath}}, \ and\
  \bibinfo {author} {\bibfnamefont {S.~Y.}\ \bibnamefont {Savrasov}},\
  }\href@noop {} {\bibfield  {journal} {\bibinfo  {journal} {Phys. Rev. B}\
  }\textbf {\bibinfo {volume} {83}},\ \bibinfo {pages} {205101} (\bibinfo
  {year} {2011})}\BibitemShut {NoStop}%
\bibitem [{\citenamefont {Fang}\ \emph
  {et~al.}(2012{\natexlab{b}})\citenamefont {Fang}, \citenamefont {Gilbert},
  \citenamefont {Dai},\ and\ \citenamefont {Bernevig}}]{Bernevig_Weyl0}%
  \BibitemOpen
  \bibfield  {author} {\bibinfo {author} {\bibfnamefont {C.}~\bibnamefont
  {Fang}}, \bibinfo {author} {\bibfnamefont {M.~J.}\ \bibnamefont {Gilbert}},
  \bibinfo {author} {\bibfnamefont {X.}~\bibnamefont {Dai}}, \ and\ \bibinfo
  {author} {\bibfnamefont {B.~A.}\ \bibnamefont {Bernevig}},\ }\href@noop {}
  {\bibfield  {journal} {\bibinfo  {journal} {Phys. Rev. Lett.}\ }\textbf
  {\bibinfo {volume} {108}} (\bibinfo {year} {2012}{\natexlab{b}})}\BibitemShut
  {NoStop}%
\bibitem [{Note3()}]{Note3}%
  \BibitemOpen
  \bibinfo {note} {Since chiral symmetry is absent, the BdG spectrum is not
  symmetric under $E_n \rightarrow -E_n$ at a given $\protect \bm
  {k}_{\parallel }$. However, particle-hole symmetry still imposes the symmetry
  of the spectrum under $E_n(\protect \bm {k}_{\parallel }) \rightarrow
  -E_n(-\protect \bm {k}_{\parallel }) $.}\BibitemShut {Stop}%
\bibitem [{\citenamefont {Bzdusek}\ \emph {et~al.}(2016)\citenamefont
  {Bzdusek}, \citenamefont {Wu}, \citenamefont {R\"uegg}, \citenamefont
  {Sigrist},\ and\ \citenamefont {Soluyanov}}]{Thomas_line}%
  \BibitemOpen
  \bibfield  {author} {\bibinfo {author} {\bibfnamefont {T.}~\bibnamefont
  {Bzdusek}}, \bibinfo {author} {\bibfnamefont {Q.~S.}\ \bibnamefont {Wu}},
  \bibinfo {author} {\bibfnamefont {A.}~\bibnamefont {R\"uegg}}, \bibinfo
  {author} {\bibfnamefont {M.}~\bibnamefont {Sigrist}}, \ and\ \bibinfo
  {author} {\bibfnamefont {A.~A.}\ \bibnamefont {Soluyanov}},\ }\href@noop {}
  {\bibfield  {journal} {\bibinfo  {journal} {Nature}\ }\textbf {\bibinfo
  {volume} {538}},\ \bibinfo {pages} {75} (\bibinfo {year} {2016})}\BibitemShut
  {NoStop}%
\bibitem [{\citenamefont {Ahn}\ \emph {et~al.}(2017)\citenamefont {Ahn},
  \citenamefont {Mele},\ and\ \citenamefont {Min}}]{Ahn_cyclides}%
  \BibitemOpen
  \bibfield  {author} {\bibinfo {author} {\bibfnamefont {S.}~\bibnamefont
  {Ahn}}, \bibinfo {author} {\bibfnamefont {E.~J.}\ \bibnamefont {Mele}}, \
  and\ \bibinfo {author} {\bibfnamefont {H.}~\bibnamefont {Min}},\ }\href@noop
  {} {\bibfield  {journal} {\bibinfo  {journal} {Phys. Rev. Lett.}\ }\textbf
  {\bibinfo {volume} {119}},\ \bibinfo {pages} {147402} (\bibinfo {year}
  {2017})}\BibitemShut {NoStop}%
\bibitem [{Note4()}]{Note4}%
  \BibitemOpen
  \bibinfo {note} {Contrary to the two-dimensional honeycomb lattice, the
  SU(2)-spin symmetry is fully broken in the hyperhoneycomb lattice since the
  second neighbor hopping processes cannot be coplanar. Also, since the
  spinless Fermi surface lies at general positions of the BZ,
  spin-orbit-coupling gaps out the Fermi surface at half-filling possibly
  leading to a topological insulator.\cite {Kim_HHL_topins}}\BibitemShut
  {NoStop}%
\bibitem [{\citenamefont {Agterberg}\ \emph {et~al.}(2017)\citenamefont
  {Agterberg}, \citenamefont {Brydon},\ and\ \citenamefont
  {Timm}}]{Agterberg_BdGsurface}%
  \BibitemOpen
  \bibfield  {author} {\bibinfo {author} {\bibfnamefont {D.~F.}\ \bibnamefont
  {Agterberg}}, \bibinfo {author} {\bibfnamefont {P.~M.~R.}\ \bibnamefont
  {Brydon}}, \ and\ \bibinfo {author} {\bibfnamefont {C.}~\bibnamefont
  {Timm}},\ }\href@noop {} {\bibfield  {journal} {\bibinfo  {journal} {Phys.
  Rev. Lett.}\ }\textbf {\bibinfo {volume} {118}},\ \bibinfo {pages} {127001}
  (\bibinfo {year} {2017})}\BibitemShut {NoStop}%
\bibitem [{\citenamefont {Bzdusek}\ and\ \citenamefont
  {Sigrist}(2017)}]{Bzdusek_mulitnodes}%
  \BibitemOpen
  \bibfield  {author} {\bibinfo {author} {\bibfnamefont {T.}~\bibnamefont
  {Bzdusek}}\ and\ \bibinfo {author} {\bibfnamefont {M.}~\bibnamefont
  {Sigrist}},\ }\href@noop {} {\bibfield  {journal} {\bibinfo  {journal} {Phys.
  Rev. B}\ }\textbf {\bibinfo {volume} {96}},\ \bibinfo {pages} {155105}
  (\bibinfo {year} {2017})}\BibitemShut {NoStop}%
\bibitem [{Note5()}]{Note5}%
  \BibitemOpen
  \bibinfo {note} {With the BdG spectrum invariant under global U(1) gauge
  transformations, i.e.~$\protect \mathaccentV {hat}05E{\psi } \DOTSB
  \mapstochar \rightarrow \protect \mathrm {e}^{i\theta }\protect \mathaccentV
  {hat}05E{\psi }$ for every fermion field, we call two Hamiltonians equivalent
  if $\protect \mathaccentV {tilde}07E{H}(\protect \bm {k}) = U_{\theta
  }^{\dagger } H(\protect \bm {k}) U_{\theta } $ with $U_{\theta } = \protect
  \mathrm {diag}[\protect \mathrm {e}^{-i \theta },\protect \mathrm {e}^{i
  \theta }] \otimes \protect \mathbb {I}_{4\times 4}$, which we write $\protect
  \mathaccentV {tilde}07E{H} \protect \cong H$. As we define symmetries only up
  to such a gauge transformation, we use the equivalence relation `$\protect
  \cong $' instead of the strict equality `$=$'.}\BibitemShut {Stop}%
\bibitem [{\citenamefont {Lee}\ \emph {et~al.}(2014)\citenamefont {Lee},
  \citenamefont {Bhattacharjee}, \citenamefont {Hwang}, \citenamefont {Kim},
  \citenamefont {Jin},\ and\ \citenamefont {Kim}}]{Kim_HHL_topins}%
  \BibitemOpen
  \bibfield  {author} {\bibinfo {author} {\bibfnamefont {E.~K.-H.}\
  \bibnamefont {Lee}}, \bibinfo {author} {\bibfnamefont {S.}~\bibnamefont
  {Bhattacharjee}}, \bibinfo {author} {\bibfnamefont {K.}~\bibnamefont
  {Hwang}}, \bibinfo {author} {\bibfnamefont {H.-S.}\ \bibnamefont {Kim}},
  \bibinfo {author} {\bibfnamefont {H.}~\bibnamefont {Jin}}, \ and\ \bibinfo
  {author} {\bibfnamefont {Y.~B.}\ \bibnamefont {Kim}},\ }\href@noop {}
  {\bibfield  {journal} {\bibinfo  {journal} {Phys. Rev. B}\ }\textbf {\bibinfo
  {volume} {89}},\ \bibinfo {pages} {205132} (\bibinfo {year}
  {2014})}\BibitemShut {NoStop}%
\end{thebibliography}%

\end{document}